\documentclass[prb,preprintnumbers,amsmath,amssymb,showpacs,superscriptaddress,notitlepage]{revtex4-1} 
\usepackage{epsfig}
\usepackage{amssymb}
\usepackage{amsmath}
\usepackage{amsthm}
\usepackage{bm}
\usepackage{bbold}
\usepackage{color}
\usepackage{times}
\usepackage{braket}
\usepackage{cancel}
\usepackage[colorlinks,bookmarks=false,citecolor=blue,linkcolor=red,urlcolor=blue]{hyperref}
\hypersetup{%
  colorlinks = true,
  linkcolor  = black
} 
\newcommand{\be}{\begin{equation}}
\newcommand{\ee}{\end{equation}}
\newcommand{\bea}{\begin{eqnarray}}
\newcommand{\eea}{\end{eqnarray}}
\newcommand{\ba}{\begin{aligned}}
\newcommand{\ea}{\end{aligned}}

\newcommand{\down}{\downarrow}
\newtheorem{mydef}{Definition}

\def\doi{http://dx.doi.org/}

\def\nn{\nonumber\\}
\def\fr#1{(\ref{#1})}

\pdfoutput=1
\begin{document}

\title{Quench dynamics and relaxation in isolated integrable quantum
  spin chains}
\author{Fabian H.L. Essler}
\affiliation{The Rudolf Peierls Centre for Theoretical Physics, Oxford
University, Oxford OX1 3NP, UK}
\author{Maurizio Fagotti}
\affiliation{D\'epartement de Physique, \'Ecole Normale Sup\'erieure / PSL Research
University, CNRS, 24 rue Lhomond, 75005 Paris, France}

\begin{abstract}
We review the dynamics after quantum quenches in integrable quantum
spin chains. We give a pedagogical introduction to relaxation in
isolated quantum systems, and discuss the description of the steady
state by (generalized) Gibbs ensembles. We then turn to general
features in the time evolution of local observables after the quench,
using a simple model of free fermions as an example. In the second
part we present an overview of recent progress in describing quench
dynamics in two key paradigms for quantum integrable models, the
transverse field Ising chain and the anisotropic spin-1/2 Heisenberg
chain.
\end{abstract}
\maketitle
\tableofcontents

\section{Introduction}
An \emph{isolated} many-particle quantum system is characterized by
the absence of any coupling to its environment. According to the laws
of quantum mechanics its time evolution is unitary and governed
by the time dependent Schr\"odinger equation. In order to specify the
state $|\Psi(t)\rangle$ of the system at a given time $t$, it is then
sufficient to know its Hamiltonian $H$ and its state at an earlier
time
\be
|\Psi(t)\rangle=e^{-iHt}|\Psi(0)\rangle.
\ee
In spite of this purely unitary evolution, macroscopic systems are
expected to eventually ``relax'' in some way and be amenable to
a description by quantum Statistical Mechanics \cite{ETH}. For
many-particle systems it is convenient to focus on the time evolution
of the expectation values of particular observables of interest rather
than the state itself, i.e. one considers
\be
\langle\Psi(t)|{\cal O}|\Psi(t)\rangle\ .
\ee
Historically, studies of many-particle quantum systems such as
electronic degrees of freedom in solids mainly focussed on equilibrium
properties at zero and finite temperature. This is because in the context
of solids, the many-body system of interest is typically coupled to an
``environment'', i.e. other degrees of freedom, the presence of which
is felt after very short time scales. The situation changed
dramatically about a decade ago, when it became possible to
experimentally investigate the non-equilibrium dynamics of clouds of
ultra-cold, trapped
atoms\cite{GM:col_rev02,HL:Bose07,kww-06,hacker10,tetal-11,getal-11,shr-12,cetal-12,langen13,MM:Ising13,FK:mob13,FS:magn13,ronzheimer13,zoran1,jurcevic14,richerme14}
(see also the review by T. Langen, T. Gasenzer and J. Schmiedmayer in this
volume\cite{LGS16}). These 
are by design almost isolated. Moreover 
the natural energy scale underlying the dynamics is incredibly small,
so that there is a long time window (on the order of seconds) for
conducting experiments. The main effect of the coupling to the
environment is particle loss through heating. This eventually becomes
significant, but over an intermediate time window the dynamics is to a
good approximation unitary. It is important to note in view of the
following discussion that finite size effects are often important in
cold atom systems, and as a result of the trapping potential these systems
are not translationally invariant. They are however highly tuneable
both with regards to Hamiltonian parameters and their effective
dimensionality. This was exploited in the seminal \emph{Quantum
Newton's Cradle} experiments by Kinoshita, Wenger and
Weiss\cite{kww-06}. These investigated the non-equilibrium evolution
of one, two and three dimensional Bose gases that were initially
driven out of equilibrium. While the two and three dimensional systems
were seen to relax very quickly towards an equilibrium state, the
behaviour in the one dimensional case was very different. In
Ref.~\onlinecite{kww-06} this was attributed to the presence of
approximate conservation laws. Neglecting the trap, the Hamiltonian is
well approximated by the Lieb-Liniger model\cite{LiebLiniger} 
\be
H_{\rm LL}=-\frac{\hbar^2}{2m}\sum_{j=1}^N\frac{\partial^2}{\partial
  x_j^2} +c\sum_{j<k}\delta(x_j-x_k)\ .
\label{HLL}
\ee
The Hamiltonian \fr{HLL} is integrable and as a result has an infinite
number of conservation laws $I_n$\cite{Korepinbook,davies} such that
\be
[H,I_n]=[I_n,I_m]=0.
\ee
It is intuitively clear that conservation laws will affect the quantum
dynamics, because they impose constraints of the form
\be
\langle\Psi(t)|I_n|\Psi(t)\rangle={\rm const}.
\ee
The difference in behaviours between the one and three dimensional
Quantum Newton's Cradle experiments suggested that the non-equilbirium
dynamics of integrable models is unusual. This was one of the motivations
for the recent intense theoretical efforts aimed at understanding the
non-equilibrium dynamics of integrable quantum many-particle systems. 
Integrable models come in a variety of guises, the two main classes
being 
\begin{itemize}
\item{}Lattice models:

These include non-interacting fermion and boson theories~\cite{CCR_PRL11,GR:quasid12,WRDK:anyons}, models that
can be mapped to free fermions like the transverse-field Ising~\cite{CEF1,CEF2,CEF3,FE_13a,EEF:12,F:13a,qentexc,qexc,efftherm,IR:quench00,SPS:04,S:work,IR:quenchB,FCG:f-d11,RI:sc11,SE:Ising,FCG:dyn12,HPL:dyn13,CC:GGEIs11} and XY
chains~\cite{ARRS99,BMD70,BM71a,BM71b,FC08,CIC_PRE11,mau-prep,CLopenXY,BRI:sc_XY},
spin models like the Heisenberg
chain~\cite{gaplessXXZ,LM:XXZ10,SM13,bonnes14,nonint,FE_13b,Pozsgay:13a,fp-13,la-13,bpgda-10,FCEC_14,GGE_int,QANeel,GGE_XXZ_Amst,p-13a,XXZunglong,XXZung,R:gs14,MC:XXZ10,AC:qa15}
and electronic theories like the Hubbard
model\cite{KE:Mott08,ES12,IMWR:14,ROHM15,BDHM15}.

\item{}Continuum models:

These include free field theories like the
Klein-Gordon\cite{cc-07,Sp_th,Klein-Gordon}  and Luttinger models\cite{C:LL06,IC_PRA09,RS:T-L12,KR:LL12,DBZ:GGE12,NI:Coul13,gaplessXXZ}
(see the review by M. Cazalilla and M.-C. Chung~\cite{RevCaz} in this
volume), conformal field
theories\cite{cc-05,CC:tcf06,cc-07,Cardy14,non-abSpyr,Cardy_non-ab} 
(see the review by P. Calabrese and J. Cardy~\cite{CCrev} in this
volume), massive relativistic field theories like the
sine-Gordon\cite{GD:spectro07,sine-Gordon,sc_sineG},
sinh-Gordon\cite{FM_NJPhys10,sinh-Gordon,Pozsgay11,muss13} and nonlinear sigma
models\cite{sc-O3}, and non-relativistic field theories like the Lieb-Liniger
model\cite{austen,MC_NJPhys12,CSC:13b,KSCCI,releBose,nwbc-13,trappedBose,enttrap,KCC14,quenchLL,GGEqft,rel_int,bound_Bose}. 
In continuum models the spectrum of elementary excitations is
unbounded, which leads to certain differences as compared to lattice models.

\end{itemize}

In this review we focus on the equilibrium dynamics in integrable
lattice models. We moreover restrict our discussion to a particular
protocol for inducing out of equilibrium dynamics, the so-called
\emph{quantum quench}. We note that other protocols have been studied
in the literature. One example are \emph{ramps}\cite{MarioRamp,ramps},
which are of interest in relation to the Kibble-Zurek mechanism
\cite{KibbleZurek}. 

The outline of this review is as follows. In section
\ref{sec:essence} we define what we mean by a quantum quench. This is
followed by a discussion in section \ref{sec:relaxation}, of how
isolated many-particle quantum systems relax, and of how to describe
their late time behaviour. In 
section \ref{sec:simple} we provide a simple example that shows these
ideas at work. Having established a framework for the late time
behaviour after a quantum quench, section \ref{sec:spreading} turns
to the discussion of general features of the evolution of
observables at finite times, such as how correlations spread through
the system. Section \ref{sec:TFIC} is concerned with one of the
key paradigms of quantum quenches, the transverse field Ising
chain (TFIC). This constitutes a non-trivial example, for which it is
nevertheless possible to obtain exact results in closed form. From the
point of view of quantum integrability the TFIC is quite special,
because it can be mapped onto a non-interacting fermionic theory by
means of a (nonlocal) Jordan-Wigner transformation. The case of fully
\emph{interacting} integrable models (defined as having scattering
matrices that are different from $\pm 1$) is discussed in section
\ref{sec:interacting}. We conclude with an outlook on some open
problems of current interest in section \ref{sec:Outlook}.

Apart from the other contributions to this Special Issue, there have
been several previous reviews\cite{ramps,review,Austenreview,Eis_rev,DKPR16}
 on closely related topics. They
differ considerably in perspective, focus, scope and style, and are
therefore largely complementary to ours.

\section{The essence of a global quantum quench}
\label{sec:essence}
Our starting point is an isolated many-particle quantum system
characterized by a time-independent, translationally invariant
Hamiltonian $H(h)$ with only short-range interactions. Here $h$ is a
system parameter such as a magnetic field, or an interaction
strength. An example we will return to frequently throughout this
review is the transverse field Ising chain (TFIC)
\be
H(h)=-J\sum_{j=1}^L \sigma^x_j\sigma^x_{j+1}+h\sigma^z_j\ ,
\label{TFIC}
\ee
where $\sigma_j^\alpha$ are Pauli matrices acting on a spin-1/2 on
site $j$ of a one dimensional ring, and $\sigma^\alpha_{L+1}\equiv
\sigma_1^\alpha$ ($\alpha=x,y,z$). 
The short-ranged nature of $H(h)$ is an essential condition,
and it is known that models with infinite range interactions such as the
BCS problem~\cite{BCS} exhibit very different behaviours when driven
out of equilibrium.
Let us imagine that we somehow prepare our system in the
ground state $|\Psi(0)\rangle$ of $H(h_0)$. This state is highly
non-generic in the sense that it has low entanglement~\cite{vidal03,peschel,arealaw}. At time $t=0$
we then suddenly ``quench'' the system parameter to a new value $h$,
and then consider subsequent unitary time evolution with our new
Hamiltonian $H(h)$. As the change of Hamiltonian is assumed to be
instantaneous, the system remains in state $|\Psi(0)\rangle$ (``sudden
approximation''). At times $t>0$ the state of the system is found by
solving the time-dependent Schr\"odinger equation
\be
|\Psi(t)\rangle=e^{-iH(h)t}|\Psi(0)\rangle.
\ee
We are interested in the cases where in a large, finite volume
the initial state $|\Psi(0)\rangle$ has non-zero overlaps with an
exponentially large number of the eigenstates of $H(h)$. The case
where $|\Psi(0)\rangle$ has non-zero overlaps with only a finite,
system-size independent number of eigenstates is equivalent to the one
discussed in undergraduate Quantum Mechanics courses. In terms of
energy eigenstates 
\be
H(h)|n\rangle=E_n|n\rangle\ , \quad E_n\geq E_0\ ,
\label{energyeigenstates}
\ee
we can express the state of the system at time $t$ as
\be
|\Psi(t)\rangle=\sum_n \langle n|\Psi(0)\rangle\ e^{-iE_nt}|n\rangle\ .
\label{Psit}
\ee
Our objective is the description of \emph{expectation values} of
a certain class of operators ${\cal O}$, that will be defined in
detail later, in the state $|\Psi(t)\rangle$
\be
\langle\Psi(t)|{\cal O}|\Psi(t)\rangle=\sum_{n,m}
\langle\Psi(0)|n\rangle\ \langle m|\Psi(0)\rangle\
\langle n|{\cal O}|m\rangle e^{-i(E_m-E_n)t}\ .
\ee
All the interesting phenomena after a quantum quench arise from the
presence of the oscillatory factors $e^{-i(E_m-E_n)t}$, which induce
quantum mechanical interference effects~\cite{BS_PRL08}, in the double sum over
exponentially (in system size) many terms. We note that focussing
on expectation values is not as restrictive as it may first appear,
as it still allows us to consider the full probability distributions
of the observables we are interested in. 

A crucial property of a global quantum quench is that energy is
conserved at all $t>0$, and that the post-quench energy density is
larger than the ground state energy per site
\be
e=\lim_{L\to\infty}\frac{1}{L}\langle\Psi(t)|H(h)|\Psi(t)\rangle>
\lim_{L\to\infty}\frac{E_0}{L}.
\label{energydensity}
\ee
This means that through the quantum quench we explore a region of
Hilbert space that is \emph{macroscopically} different from the sector
containing the ground state and low-lying excitations.

In practice one often considers more general ``quench protocols'',
where for example $H(h)$ and/or $|\Psi(0)\rangle$ are invariant only
under translations by $2,3,...$ sites~\cite{mau-prep,preT_entropy,FCEC_14}, and initial states that are not
necessarily pure~\cite{Sp_th}, but are described by a density matrix $\rho(0)$.
Considering $\rho(0)$ to be a thermal density matrix provides a simple
way of varying the energy density $e$ ``injected'' into the system. 
It is essential for our purposes that the initial density matrix
$\rho(0)$ has a cluster decomposition property\cite{clusterGGE}
\be
\lim_{|x-y|\to\infty}{\rm Tr}\left[\rho(0){\cal O}(x){\cal
    O}(y)\right]=
{\rm Tr}\left[\rho(0){\cal O}(x)\right]
{\rm Tr}\left[\rho(0){\cal O}(y)\right],
\label{cluster}
\ee
and we elaborate on this point in Appendix \ref{app:cluster}.
Invariance of the initial density matrix under translations is another
key requirement. Inhomogeneous initial conditions have been considered
by many authors in the
literature, see
e.g. Refs~\onlinecite{ARRS99,BAW06a,BAW06b,SC08,CHL08,LM:XXZ10,BW11,HM13,ER13,SM13,RS:Cal14,HM16,releBose,AP:XY03,AB:XY06,CK_PRL12,BD:EnCFT12,BD:nessCFT,MS:NESSLutt,NESSIsing,SM:NESSXXZ,DHB:flow14,C-A:QFT14,LMV:freeqa14,DLSB:K-G15,D:bal15,VSDH:inh,PK:XX07,EKPP:conn},
and the reviews by D. Bernard and B. Doyon~\cite{BDrev} and by
R. Vasseur and Joel E. Moore~\cite{VMrev} in this volume, 
but we will not consider them here.

\section{Relaxation in isolated quantum systems}
\label{sec:relaxation}
Given that we are considering an isolated quantum system, an obvious
question is in what sense it may relax to a stationary state at late
times after we have driven it out of equilibrium. In order to sharpen
the following argument, let us revisit the case discussed in
Section \ref{sec:essence}, namely an isolated system initially
prepared in a pure state $|\Psi(0)\rangle$, that is not an eigenstate
of the (time independent) Hamiltonian $H(h)$ describing the time
evolution after our quantum quench. Let us now consider the following
class of hermitian operators 
\be
{\cal O}_{jk}=|j\rangle\langle k|+|k\rangle\langle j|\ .
\ee
Their expectation values in the state $|\Psi(t)\rangle$ can be
expressed using \fr{Psit} as
\be
\langle\Psi(t)|{\cal O}_{jk}|\Psi(t)\rangle=
e^{i(E_j-E_k)t}\langle\Psi(0)|j\rangle\langle k|\Psi(0)\rangle+{\rm c.c.}
\ee
We see that generically the right hand side exhibits periodic
oscillatory behaviour in time. Hence the observables ${\cal O}_{jk}$
typically do not relax at late times. A crucial point is that ${\cal
  O}_{jk}$ are generally very nonlocal in space. As locality is an important
concept in quantum quenches it is useful to define what we mean by a
local operator.  
\begin{mydef}
Local Operators.
\end{mydef}
In lattice models an operator ${\cal O}$ is called \emph{local} if
in the thermodynamic limit ${\cal O}$ acts non-trivially only on a 
finite number of sites separated by a finite distance. For a quantum
spin-1/2 chain with $L$ sites examples of local operators are  
\be
\sigma^\alpha_j\ ,\quad
\sigma^\alpha_j\sigma^\beta_{j+k} ,\quad
\prod_{j=1}^{k}\sigma_j^{\alpha_j}\ ,\qquad k\text{ fixed},
\ee
where $\sigma_j^\alpha$ ($\alpha=x,y,z$) are Pauli matrices acting on
site $j$. On the other hand operators such as
\be
\sigma^x_1\sigma^x_{L/2}\ ,\qquad \prod_{j=1}^{L/2}\sigma_j^{\alpha_j}
\ee
are nonlocal. Acting with a local operator on a state $|\psi\rangle$
does not change its macroscopic properties (e.g. its energy per
volume in the thermodynamic limit). 
\begin{mydef}
Range of a local operator.
\end{mydef}
The range of a local operator ${\cal O}$ is the size of the largest
interval on which it acts non-trivially. For the examples
above we have 
\be
{\rm range}(\sigma^\alpha_j)=1\ ,\quad
{\rm range}(\sigma^\alpha_j\sigma^\beta_{j+k})=k+1\ ,\quad
{\rm range}\big(\prod_{j=1}^{k}\sigma_j^{\alpha_j}\big)=k\ .
\ee
\subsection{Local Relaxation}
The previous argument shows that an isolated quantum system prepared
in a pure state cannot relax in its entirety. However, it can and does relax
\emph{locally in space}\cite{BS_PRL08,CD:therm08,Cramer_10,CEF2,FE_13a,SK:therm14}. To explain this concept let
us consider the example of a one-dimensional spin system with
Hamiltonian $H$ with only short-range interactions. An example would
be the TFIC \fr{TFIC}. We partition the entire system into an
arbitrary but finite subsystem $B$ and its complement $A$. Eventually
we will take the thermodynamic limit while keeping $B$ fixed. Let us
prepare our spin system at time $t=0$ in some initial density matrix
$\rho(0)$ that is not an eigenstate of $H$. At later times the entire
system is characterized by the density matrix
\be
\rho(t)=e^{-iHt}\rho(0)e^{iHt}.
\ee
The \emph{reduced density matrix} of the subsystem $B$ is obtained by
tracing out the degrees of freedom in $A$
\be
\rho_B(t)={\rm Tr}_A\rho(t).
\ee
\begin{figure}[ht]
\begin{center}
\includegraphics[width=0.6\textwidth]{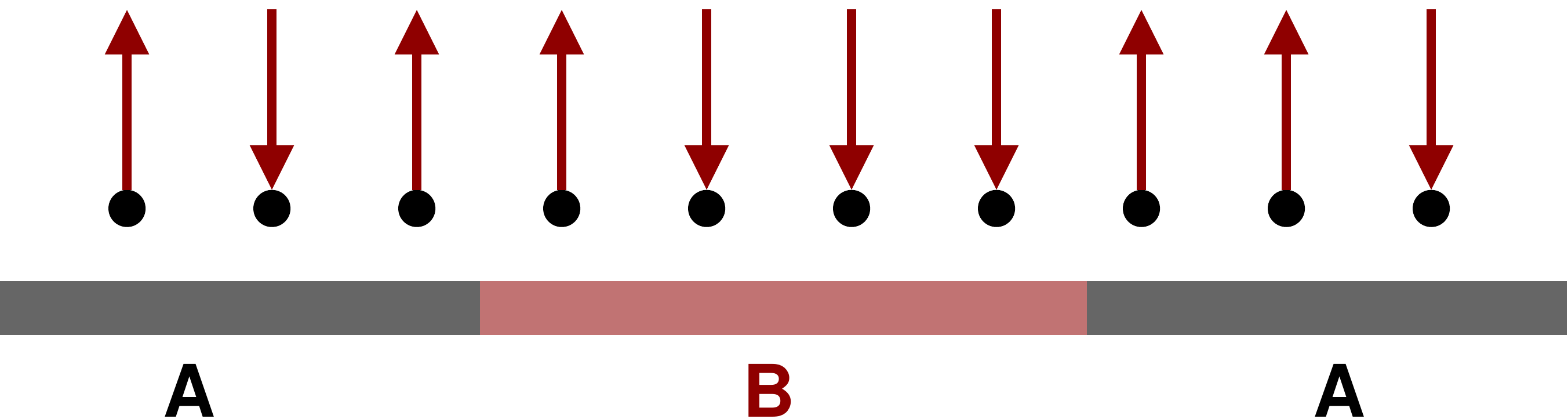}
\caption{Local relaxation in an isolated many-particle quantum system:
we partition the entire system into an arbitrary \emph{finite}
subsystem $B$ and its complement $A$. We then take the thermodynamic
limit while keeping $B$ fixed. Expectation values of all operators
that act non-trivially only in $B$ will relax to stationary values at
late times.}
\label{fig:localrelaxation}
\end{center}
\end{figure}
It is instructive to give an explicit representation of the reduced
density matrix for our spin chain example, \emph{cf.}
Fig.~\ref{fig:localrelaxation}. An orthonormal basis of 
states is given by 
\be
|\sigma_1,\sigma_2,\ldots,\sigma_L\rangle=\mathop{\otimes}_{j=1}^L
|\sigma_j\rangle_j\ ,\quad
\sigma_j=\pm 1,
\ee
where $|\sigma_j\rangle_j$ denote the two eigenstates of $S^z_j$. Let
us take our subsystem $B$ to consist of sites $1,2,\dotsm\ell$ for
simplicity. The reduced density matrix is then
\bea
\rho_B(t)&=&\sum_{\sigma_{\ell+1},\dots,\sigma_L}\left(\mathop{\otimes}_{k=\ell+1}^L{}_k\langle\sigma_k|\right) 
\rho(t)\left(\mathop{\otimes}_{n=\ell+1}^L|\sigma_n\rangle_n\right)\nn
&=&2^{-\ell}\sum_{\alpha_1,\dots,\alpha_\ell=0,x,y,z}{\rm Tr}\left[\rho(t)
\sigma^{\alpha_1}_1\sigma^{\alpha_2}_2\dots
\sigma^{\alpha_\ell}_\ell\right]\ 
\sigma^{\alpha_1}_1\sigma^{\alpha_2}_2\dots
\sigma^{\alpha_\ell}_\ell\ ,
\label{RDMt}
\eea
where we have defined $\sigma^0_j=\mathbb{1}_j$. 
Eqn \fr{RDMt} shows
that the matrix elements of $\rho_B(t)$ are equal to particular
correlation functions of local operators acting non-trivially only on
$B$, and that $\rho_B(t)$ in fact encodes all such correlators.
\begin{mydef}
Local Relaxation.
\end{mydef}
We say that our system relaxes locally if the limit
\be
\lim_{t\to\infty}\lim_{L\to\infty}\rho_B(t)=\rho_B(\infty)
\ee
exists for any finite subsystem $B$.
\begin{mydef}
Stationary State.
\end{mydef}
Consider a system that relaxes locally in the sense just defined.
Then its stationary state is defined as a time-independent density matrix
$\rho^{\rm SS}$ for the full system such that for any
finite subsystem $B$ 
\be
\lim_{L\to\infty}{\rm Tr}_A\left(\rho^{\rm SS}\right)=\rho_B(\infty),
\label{equivalence}
\ee
where $A$ is the complement of $B$. We stress that this equivalence
applies only at the level of finite subsystems in the thermodynamic
limit and not for the density matrices of the full system.

\begin{mydef}
Local equivalence of ensembles.
\end{mydef}
Let $\rho_1$ and $\rho_2$ be two density matrices. We call the
corresponding two ensembles locally equivalent, if in the
thermodynamic limit the reduced density matrices for any finite
subsystem $B$ coincide, i.e.
\be
\lim_{|A|\to\infty}{\rm Tr}_{A}\left(\rho_1\right)=
\lim_{|A|\to\infty}{\rm Tr}_{A}\left(\rho_2\right).
\ee
Here $A$ is the complement of $B$ and $|A|$ denotes its volume. We
denote local equivalence by
\be
\rho_1=_{\rm loc}\rho_2.
\ee
A key problem in the field of non-equilibrium dynamics is the
determination of the stationary state density matrix. A crucial role
is played by symmetries of the Hamiltonian. In translationally
invariant systems there are essentially two paradigms for local
relaxation, and we turn to their respective descriptions next. 
Before doing so an important comment is in order. By design our focus
is on local operators, which in turn leads us to define local
relaxation in the sense formulated above. In practice one might
be interested in observables that are formally not captured by our
framework, and yet might relax to stationary values. An example
would be expectation values of the momentum distribution function,
which is the Fourier transform of the single particle Green's
function\cite{WRDK:anyons,GG15}. The relaxational properties of such
quantities are at present not understood in general, and further
investigation is called for.
\subsection{Thermalization}
As we are dealing with an isolated system, energy is always a
conserved quantity
\be
E={\rm Tr}\left(\rho(t)H\right)
={\rm Tr}\left(e^{-iHt}\rho(0)e^{iHt}H\right)
={\rm Tr}\left(\rho(0)H\right)\ .
\ee
In absence of other conserved quantities isolated systems are believed
to locally relax to thermal equilibrium. This is known as
\emph{thermalization}~\cite{ETH,RS:altET,RDO08}. In the framework we have set up, this means
that our stationary state is described by a Gibbs ensemble
\be
\rho^{\rm SS}=_{\rm loc}\rho^{\rm Gibbs}=\frac{e^{-\beta_{\rm eff}H}}{{\rm
    Tr}\left(e^{-\beta_{\rm eff}H}\right)}\ .
\label{thermalization}
\ee
Here the inverse effective temperature $\beta_{\rm eff}$ is fixed by
the initial value of the energy density
\be
e\equiv \lim_{L\to\infty}\frac{1}{L}{\rm Tr}\left(\rho(0)H\right)=
\lim_{L\to\infty}\frac{1}{L}{\rm Tr}\left(\rho^{\rm Gibbs}H\right)\ .
\label{betaeff}
\ee
We once again stress that \fr{thermalization} implies only that 
in the thermodynamic limit the stationary state is \emph{locally}
indistinguishable from a Gibbs ensemble in the sense that expectation
values of all operators that act non-trivially only in a finite
subsystem are identical to the corresponding thermal expectation
values. The physical picture underlying thermalization is that the
infinitely large complement of our subsystem acts like a heat bath
with an effective inverse temperature $\beta_{\rm eff}$.

We note that $\beta_{\rm eff}$ defined in this way can be either
positive or negative. The meaning of a negative temperature in this
context is as follows. Let us consider a
Hamiltonian with bounded spectrum (e.g. a quantum spin model). Then
the average energy associated with $H$ is
\be
\bar{e}=\lim_{L\to\infty}\frac{1}{L}{\rm Tr}\left(H\right)\ .
\ee
Depending on our initial density matrix we now have two possibilities
\bea
e&<&\bar{e}\Rightarrow \beta_{\rm eff}>0\ ,\nn
e&>&\bar{e}\Rightarrow \beta_{\rm eff}<0\ .
\eea
We note that negative temperature ensembles have been observed
experimentally in Ref.~\onlinecite{negativeT}.

\subsection{Generalized Gibbs Ensembles (GGE)}
We now turn to the case of systems with additional \emph{local
  conservation laws} $I^{(n)}$. These are operators that commute with the
Hamiltonian 
\be
[H,I^{(n)}]=0,
\label{CL}
\ee
and have the property that their densities are local operators. In
most of the cases of interest these conservation laws also commute
with one another 
\be
[I^{(n)},I^{(m)}]=0,
\label{CL2}
\ee
and we will assume this to be the case in the following. We note that
exceptions to \fr{CL2} are
known~\cite{mau-prep,Cardy_non-ab,non-abSpyr} (see also 
Sec.~\ref{ss:locmod}). What we mean by local conservation laws is
best explained by an example. The TFIC \fr{TFIC} has infinitely many local
conservation laws in the thermodynamic limit~\cite{CLIsing} 
\be
\ba
I^{(1,+)}&=H(h)\ , \\
I^{(n,+)}&=-J\sum_j\left(S^{xx}_{j,j+n}+S^{yy}_{j,j+n-2}\right)
+h\left(S^{xx}_{j,j+n-1}+S^{yy}_{j,j+n-1}\right)\equiv \sum_j \mathcal I^{(n,+)}_j\, ,\\
I^{(n,-)}&=-J\sum_j\left(S^{xy}_{j,j+n}-S^{yx}_{j,j+n}\right)\equiv
  \sum_j \mathcal I^{(n,-)}_j \ , 
\ea
\label{In+}
\ee
where $n\geq 1$ and we have defined
\be
S^{\alpha\beta}_{j,j+\ell}=\sigma^\alpha_j\left[
\prod_{k=1}^{\ell-1}\sigma^z_{j+k}\right]\sigma^\beta_{j+\ell}\ ,\quad
S^{yy}_{j,j}=-\sigma^z_j\ .
\ee
The conservation laws themselves are \emph{extensive}, but their
densities $\mathcal I_j^{(n,\pm)}$ involve only finite numbers of
neighbouring sites ($n+1$ sites for $I^{(n,\pm)}$). 
As a consequence of \fr{CL}, expectation values of 
the conservation laws as well as their densities are time independent
\be
\frac{1}{L}{\rm Tr}\left(\rho(t)I^{(n,\pm)}\right)=
{\rm Tr}\left(\rho(t)\mathcal I_j^{(n,\pm)}\right)
=\frac{1}{L}{\rm Tr}\left(\rho(0)I^{(n,\pm)}\right)\equiv 
\frac{E^{(n,\pm)}}{L}\ ,
\label{initial}
\ee
where in the first step we used translational invariance.
An obvious question at this point is why we
insist on conservation laws to be local. The answer lies in the simple
fact that \emph{any} Hamiltonian has at least as many nonlocal
conservation laws as there are basis states in the Hilbert space. Let
us consider one-dimensional projectors on energy eigenstates
\be
H|n\rangle=E_n|n\rangle\ ,\quad P_n=|n\rangle\langle n|.
\ee
Then we clearly have 
\be
[P_n,P_m]=0=[P_n,H]\ ,
\ee
which imply that the expectation values of all $P_n$ are time independent.
Importantly, these conservation laws are not local, and as they exist
for any Hamiltonian they are not expected to have any profound effect.
In contrast, local conservation laws are rather special and, as is
clear from \eqref{initial}, have important consequences for local
relaxation after quantum quenches. 

An immediate consequence of \fr{initial} is that systems with local
conservation laws cannot thermalize after a quantum quench, because
the system retains memory of the initial expectation values of all
conserved quantities at all times.
 The maximum entropy principle~\cite{maxent} then
suggests that the stationary state density matrix should be given by a GGE~\cite{GGE}
\be\label{eq:GGEnaive}
\rho^{\rm GGE}=\frac{e^{-\sum_n \lambda_n I^{(n)}}}{{\rm Tr}
\left(e^{-\sum_n \lambda_n I^{(n)}}\right)}.
\ee
Here $\lambda_n$ are Lagrange multipliers that are fixed by the
initial conditions \fr{initial}, as we require that
\be
\lim_{L\to\infty}\frac{E^{(n,\pm)}}{L}=
\lim_{L\to\infty}\frac{1}{L}{\rm Tr}\left(\rho^{\rm GGE}I^{(n)}\right).
\ee
\subsubsection{Local conservation laws vs mode occupation
  numbers}\label{ss:locmod} 
In free theories GGEs are often formulated using conserved mode
occupation numbers\cite{GGE}. This is usually equivalent to
our formulation in terms of local conservation laws in these cases
\cite{FE_13a} as we now demonstrate for a simple example. Let us
consider a fermionic tight binding model
\be
H=-J\sum_jc^\dagger_jc_{j+1}+c^\dagger_{j+1}c_j-\mu\sum_j c^\dagger_jc_j\ ,\quad
\{c_j,c^\dagger_n\}=\delta_{j,n}.
\label{TBM}
\ee
The Hamiltonian is easily diagonalized by going to momentum space
\be
H=\sum_k \left[-2J\cos(k)-\mu\right] c^\dagger(k)c(k)\ .
\ee
The mode occupation operators $n(k)=c^\dagger(k)c(k)$ commute with the
Hamiltonian and with one another. However, the $n(k)$ are neither local nor
extensive. An equivalent set of local, extensive conserved quantities
is easily found
\bea
I^{(n,+)}&=&2J\sum_k
\cos(nk)\ c^\dagger(k)c(k)=J\sum_jc^\dagger_jc_{j+n}+c^\dagger_{j+n}c_j\ ,\nn
I^{(n,-)}&=&2J\sum_k
\sin(nk)\ c^\dagger(k)c(k)=iJ\sum_jc^\dagger_jc_{j+n}-c^\dagger_{j+n}c_j\ .\label{eq:I+_}
\eea
The crucial point is that these conservation laws are \emph{linearly}
related to the mode occupation numbers.
This implies that the GGE describing the stationary state after a
quantum quench to the Hamiltonian $H$ can be constructed either from
the local conservation laws, or from the mode occupation numbers
\be
\rho^{\rm GGE}=
\frac{e^{-\sum_n \lambda_{n,+}I^{(n,+)}+\lambda_{n,-}I^{(n,-)}}}
{{\rm Tr}\big(e^{-\sum_n \lambda_{n,+}I^{(n,+)}+\lambda_{n,-}I^{(n,-)}}\big)}
=
\frac{e^{-\sum_k \mu_k\ n(k)}}
{{\rm Tr}\big(e^{-\sum_k \mu_k\ n(k)}\big)}.
\label{GGEFF}
\ee
At this point a word of caution is in order. There are cases in which
local conservation laws exist that cannot be expressed in terms of
mode occupation operators. In non-interacting theories this occurs if
the dispersion relation in the finite volume has additional
symmetries. An example is provided by setting $\mu=0$ in our
tight-binding model \fr{TBM}, in which case the dispersion has the
symmetry $\epsilon(p)=-\epsilon(\pi-p)$. In this case the following
operator commutes with the Hamiltonian 
\be
\sum_j(-1)^j\Bigl[c_jc_{j+1}-c^\dagger_{j}c^\dagger_{j+1}\Bigr]=\sum_k
e^{-ik}c(\pi-k)c(k)+{\rm h.c.}\qquad (\mu=0)\, ,
\ee
but is clearly neither expressible in terms of mode occupation
operators, nor does it commute with the latter. In cases like this the
stationary state is not always locally equivalent to a GGE of the form
\eqref{GGEFF} since the constraints arising from the additional local
charges must be imposed as well~\cite{mau-prep,Cardy_non-ab,Doyon}. 

\subsection{Generalized Microcanonical Ensemble (GMC)}
In equilibrium Statistical Physics it is well known that in the
thermodynamic limit the Gibbs ensemble becomes equivalent to the
microcanonical ensemble. In the quench context analogous equivalences
of ensembles hold, and we now turn to their discussion. 

\subsubsection{Generic Systems}
We have stated above that generic systems thermalize when their
stationary state density matrix is equal to a Gibbs ensemble in the
sense of \fr{equivalence}, \fr{thermalization}, where the effective
temperature is fixed by the energy density $e$ given by
\fr{betaeff}. There is strong numerical evidence
\cite{RDO08,rigol09a,rigol09b,BKL_PRL10,RS10a,RS10b,neuenhahn12,steinigeweg13,beugeling1,kid14,beugeling2} 
that an equivalent microcanonical ensemble can be constructed in the form
\be
\rho^{\rm MC}=|n\rangle\langle n|\ ,
\ee
where $|n\rangle$ is \emph{any} energy eigenstate (of the post-quench
Hamiltonian) with energy density $e$. Importantly, no averaging over a
microcanonical energy shell is required. This implies that in generic
systems all energy eigenstates at a given energy density are locally
indistinguishable and thermal. The microcanonical description for the
stationary state of a thermalizing system is then
\be
\rho^{\rm SS}=_{\rm loc}\rho^{\rm MC}\ .
\ee

\subsubsection{Systems with local conservation laws}
If a system has local conservation laws other than energy the above
construction needs to be modified. This was first discussed in the
non-interacting case by Cassidy et.al. \cite{CCR_PRL11} and
subsequently generalized to interacting integrable models by Caux and
Essler \cite{CE_PRL13}. Let us recall that the ``initial data'' for a
(post quench) system with local conservation laws $I^{(n)}$ and
Hamiltonian $H=I^{(1)}$ is
\be
e^{(n)}=\lim_{L\to\infty}\frac{1}{L}{\rm
  Tr}\left(\rho(0)I^{(n)}\right)\ .
\label{initialcond}
\ee
The stationary state is then locally equivalent to the density matrix
\be
\rho^{\rm SS}=_{\rm loc}\rho^{\rm GMC}=|\Phi\rangle\langle\Phi|\ ,
\ee
where $|\Phi\rangle$ is a simultaneous eigenstate of all local
conservation laws such that
\be
\lim_{L\to\infty}\left[\frac{1}{L}I^{(n)}-e^{(n)}\right]|\Phi\rangle=
0\ .
\label{eigenvalues}
\ee
Conditions \fr{eigenvalues} ensure that all $I^{(n)}$ have the
correct expectation values \fr{initialcond} in the stationary
state. An obvious question is how to construct $|\Phi\rangle$ in
practice. For non-interacting lattice models this is often
straightforward as we now discuss. In free theories the Hamiltonian is
diagonalized in terms of mode occupation operators $n(k)=c^\dagger(k)c(k)$ 
\be
H=\sum_k \epsilon(k) n(k)\ ,\quad [n(k),n(q)]=0=[n(k),H].
\ee
As discussed above the $n(k)$ are in one-to-one correspondence
with local conservation laws and can therefore be used to construct
GGEs. The initial conditions \fr{initialcond} therefore fix the
densities $\rho(k)$ of particles with momentum $k$
\be
{\rm  Tr}\left(\rho(0)n(k)\right)=\rho(k)\ .
\label{icond}
\ee
In a large, finite volume $L$, the state $|\Phi\rangle$ can then be
taken as a Fock state 
\be
|\Phi\rangle=\prod_{j=1}^Nc^\dagger(k_j)|0\rangle\ ,\quad
c(q)|0\rangle=0\ .
\label{phi1}
\ee
The $k_j$ need to be chosen to reproduce the correct particle density
$\rho(k)$ in the thermodynamic limit $N,L\to \infty$, $n=N/L$ fixed. Noting that
$\Delta n=\rho(k) \Delta k$ we see that a possible choice is 
\be
k_{j+1}=k_j+\frac{1}{L\rho(k_j)}.
\label{phi2}
\ee
In practice we can determine $\rho(k)$ from \fr{icond}, construct
the state $|\Phi\rangle$ from \fr{phi1}, \fr{phi2}, and then use it to
form $\rho^{\rm GMC}$. In interacting theories the construction
is analogous but considerably more involved and will be discussed in
Section \ref{sec:interacting}.

In contrast to the non-integrable case, energy eigenstates at a given
energy density $e$ are \emph{not} all locally indistinguishable and
thermal. While the \emph{typical} state at a given $e$ is thermal
in this case as well, there exist other, non-thermal, macro-states
at a given $e$. However, their entropies are smaller than the one of
the thermal equilibrium state. A more detailed discussion is given in
Appendix~\ref{app:typical}.

\subsection{Time averaged relaxation and Diagonal Ensemble}
In the literature a different definition of relaxation after quantum
quenches in finite systems is widely used~\cite{review,DKPR16}. Let us
for simplicity consider a one-dimensional system on a ring of length
$L$, which is initially prepared in a state with density matrix
$\rho(0)$. The \emph{time average} of an operator ${\cal O}$ is then
defined as 
\be
\bar{{\cal O}}_L=\lim_{T\to\infty}
\frac{1}{T}\int_0^T dt\ {\rm Tr}\left(\rho(t)\ {\cal O}\right),
\label{Obar}
\ee
where we have introduced the subscript $L$ to indicate that the system
size is kept fixed at $L$, and where we assumed the limit $T\to\infty$
to exist. The operator ${\cal O}$ is then said to relax if its
expectation value is very close to its time average most of the time,
i.e. if 
\be
\frac{1}{T}\int_0^T dt\left[{\rm Tr}\left(\rho(t)\ {\cal O}\right)
-\bar{{\cal O}}_L\right]^2\ll \bar{\mathcal O}_{L}^2.
\ee
Physically this way of thinking about relaxation is very different
from ours. This is most easily understood for a non-interacting
system, which by construction features stable particle and hole
excitations. Let us denote their maximal group velocity by $v_{\rm
  max}$. In time averaged relaxation one considers times $t$ such that
$v_{\rm max} t\gg L$, i.e. the stable excitations traverse the entire
system many times. In contrast, our way of defining relaxation is
based on taking the thermodynamic limit first, and then considering
late times. 

A natural question to ask is what statistical ensemble describes the
averages $\bar{\cal O}$. To see this let us consider time evolution
with a Hamiltonian $H$ that has non-degenerate eigenvalues
$E_n$. Expanding the density matrix in the basis of energy eigenstates gives
\be
\rho(t)=\sum_{n,m} \langle
n|\rho(0)|m\rangle\ e^{-i(E_n-E_m)t}|n\rangle\langle m| \ .
\ee
Substituting this back into \fr{Obar} and using that the energy
eigenvalues are non-degenerate we arrive at
\be
\bar{\cal O}_L=\sum_n\langle n|\rho(0)|n\rangle\ \langle n|{\cal O}|n\rangle.
\label{DE}
\ee
This shows that time averages involve only the diagonal elements (in
the energy eigenbasis) of the density matrix. The description \fr{DE}
is known as \emph{diagonal ensemble}. Is there a relation between the
subsystem view of relaxation and the time-averaged one? It is
believed that for local operators ${\cal O}$ the two viewpoints are in
fact equivalent in the sense that
\be
\lim_{L\to \infty} \bar{\cal O}_L={\rm Tr}\left[\rho^{SS}{\cal
    O}\right]\ .
\ee
In the case of the TFIC this has been shown in Appendix E of
Ref.~\onlinecite{CEF2} and elaborated on in Ref.~\onlinecite{F:13a}.

\subsection{Symmetry restoration}
An interesting question in the quantum quench context concerns
problems where the Hamiltonian is invariant under a symmetry
operation, but the initial state after the quench is not. If we
denote the symmetry operation by $U$, then we have
\be
[H,U]=0\ ,\qquad U|\Psi(0)\rangle\neq|\Psi(0)\rangle.
\ee
The question is then, whether or not the symmetry associated with $U$
is restored in the stationary state
\be
[\rho^{\rm SS},U]\overset{?}{=}0.
\ee
The answer to this is quite straightforward for the stationary states
we have considered here, which are described by (generalized) Gibbs
ensembles
\be
\rho^{\rm SS}=_{\rm loc}\rho_{\rm GGE}=\frac{1}{Z_{\rm
    GGE}}e^{-\sum_n\lambda_nI^{(n)}}\ .
\ee
Clearly, if $U$ is a symmetry of all local conservation laws
\be
[I^{(n)},U]=0\ ,\quad \forall n\ ,
\ee
the symmetry associated with $U$ will be restored in the stationary
state. Conversely, if there is at least one conservation law
for which $[I^{(s)},U]\neq 0$, then the symmetry will remain broken in
the stationary state. It is useful to consider some explicit examples.
\subsubsection{Spin-flip symmetry}
The TFIC Hamiltonian \fr{TFIC} is invariant under rotations in spin
space around the z-axis by 180 degrees 
\be\label{eq:spin-flip}
U\sigma_\ell^{\alpha} U^\dag= -\sigma_{\ell}^{\alpha}\ ,\quad\alpha=x,y,\qquad
U\sigma_\ell^{z} U^\dag= \sigma_{\ell}^{z}\ .
\ee
A quantum quench of the transverse field starting in the ordered phase
$h<1$ leads to an initial state that breaks this symmetry. However,
the local conservation laws \fr{In+} that characterize the stationary
state are invariant with respect to spin reflection. As a result the
spin flip symmetry is restored in the stationary state~\cite{FE_13a}, 
as will be shown in Section \ref{sec:TFIC}.

\subsubsection{Parity symmetry}
The Hamiltonian of the tight-binding model~\eqref{TBM} is invariant
under the parity transformation (reflection across a link)
\be\label{eq:parity}
U c_\ell^\dag U^\dagger= c^\dag_{1-\ell}\, .
\ee
On the other hand, the set of charges $I^{(n,-)}$ in \eqref{eq:I+_} is
odd under $U$, and hence a quantum quench from a state that is not
parity even does not generally result in a parity-symmetric
stationary state. As an example we may consider the time evolution of
the ground state $|\Psi(0)\rangle$ of the following Hamiltonian 
\be
H_0=\sum_\ell \frac{i}{4}\left( 3c^\dag_\ell
c_{\ell+1}-3c^\dag_{\ell+1}c_{\ell} 
-c^\dag_\ell c^\dag _{\ell+1}+c_{\ell+1}c_\ell \right)+\mu_0 c^\dag_\ell c_\ell\, .
\ee
For $|\mu_0|<\sqrt{2}$, $|\Psi(0)\rangle$ is not parity-symmetric and
the expectation value of $I^{(n,-)}$ is nonzero
\be
\braket{\Psi(0)|I^{(n,-)}|\Psi(0)}
=J
\frac{1-(-1)^n}{2}\frac{\cos\Bigl(n\arcsin(\frac{|\mu_0|}{\sqrt{2}})\Bigr)}{\pi
  n}\, \neq 0.
\label{nonzero}
\ee
As $I^{(n,-)}$ are conserved we must have
\be
\lim_{L\to\infty}\frac{1}{L}{\rm Tr}\big(\rho^{\rm SS}\ I^{(n,-)}\big)
=\lim_{L\to\infty}\frac{1}{L}\braket{\Psi(0)|I^{(n,-)}|\Psi(0)},
\ee
and consequently the stationary state cannot be not symmetric under
\eqref{eq:parity}. We note that the above symmetry argument does not
guarantee a non-zero value for the expectation value in \fr{nonzero},
as there could be other symmetries that force it to vanish.
\subsubsection{Translational symmetry}
If we allow the initial state to partially break translational symmetry
the situation can become significantly more complicated. An interesting
example is provided by the quantum XY model~\cite{LSM:XY}
\be
H_{\rm
  XY}=J\sum_{\ell=1}^L\frac{1+\gamma}{4}\sigma_\ell^x\sigma_{\ell+1}^x
+\frac{1-\gamma}{4}\sigma_\ell^y\sigma_{\ell+1}^y\ ,
\ee
where we take $L$ to be even.
As long as we consider quenches from translationally invariant initial
states, the stationary behaviour is described by a GGE constructed
from local conservation laws $Q^{(n,\pm)}$ along the lines described
above. The situation changes for initial states $|\Psi_2(0)\rangle$
that are invariant only under translations by two sites. One might
naively expect that translational symmetry gets restored in 
the stationary state, but this is in fact not the case~\cite{mau-prep}. This can be
understood by noting that $H$ has local conservation laws that are not
translationally invariant, an example being
\be
J_1^{+}=\sum_{\ell=1}^L(-1)^\ell
\Bigl[\frac{1+\gamma}{4}\sigma_\ell^x\sigma_{\ell+1}^x-\frac{1-\gamma}{4}\sigma_\ell^y\sigma_{\ell+1}^y\Bigr]\,
,\quad
[J_1^+,H_{\rm XY}]=0.
\ee
Generically one will have $\langle\Psi_2(0)|J_1^+|\Psi_2(0)\rangle\neq
0$, and as $J_1^+$ is conserved, translational invariance must remain
broken at all times. The construction of a GGE in this case is
complicated by the fact that $J_1^+$ does not commute with all
$Q^{(n,\pm)}$, and involves an initial state dependent selection of
mutually commuting local conservation laws.

\subsection{Truncating Generalized Gibbs Ensembles}
As we have seen above, GGEs in integrable models involve infinite
numbers of conservation laws in the thermodynamic limit. A natural
question to ask is whether all of these are equally important for the
description of the stationary state, or whether certain classes are
more important than others. A general framework for investigating this
question was developed in Ref.~\onlinecite{FE_13a} and then applied
to the specific example of quenches in the TFIC. The main findings are
conjectured to apply much more generally to quenches in integrable
models. A key measure of importance of a given conservation law is its
\emph{degree of locality}\cite{FE_13a} $D_{\rm loc}$.  
This is straightforward to quantify for conservation laws like \fr{In+},
which have densities that act non-trivially only on a fixed number of
$n+1$ neighbouring sites, 
\be
I^{(n)}=\sum_{j}{\cal I}^{(n)}_j\ ,\quad {\rm range}({\cal
  I}^{(n)}_j)=n+1\qquad \Rightarrow
D_{\rm loc}\big(I^{(n)}\big)=n+1.
\label{DoL}
\ee
The idea of Ref.~\onlinecite{FE_13a} was to select various finite subsets
$\{J_m|m=1,\dots,y\}\subset\{I^{(n)}\}$, and ask how well the density
matrices 
\be\label{pGGE}
\rho^{\rm pGGE,y}=\frac{e^{-\sum_{n=1}^y \lambda_n^{(y)} J_n}}{{\rm Tr}
\left(e^{-\sum_{n=1}^y \lambda_{n}^{(y)} J_n}\right)}
\ee 
approximate the full GGE density matrix. Here the Lagrange multipliers
$\lambda_n^{(y)}$ are fixed by imposing the appropriate initial conditions
\be
\lim_{L\to\infty}\frac{\langle\Psi(0)|J_n|\Psi(0)\rangle}{L}=
\lim_{L\to\infty}\frac{1}{L}{\rm Tr}\left(\rho^{\rm
  pGGE,y}J_n\right)\ ,\quad n=1,\dots,y.
\ee
A useful way of comparing the full $\rho^{\rm GGE}$ and partial
$\rho^{\rm pGGE,y}$ GGEs is through a distance 
${\cal D}(\rho_\ell,\rho'_\ell)$ on the space of reduced density
matrices on an interval of length $\ell$. In order for
$\rho^{\rm pGGE,y}$ to be a good approximation to $\rho^{\rm GGE}$ it
should be possible to make the distance between their respective
reduced density matrices arbitrarily small
\be
{\cal D}\big(\rho^{\rm GGE}_\ell,\rho^{\rm
  pGGE,y}_\ell\big)\overset{!}{<}\epsilon\ ,\quad \forall\ell.
\label{compare}
\ee
The findings of Ref.~\onlinecite{FE_13a} show that if even a single
conservation law with small degree of locality is excluded, the
approximation \fr{pGGE} immediately becomes very poor, and the only
way to achieve \fr{compare} is by retaining all conservation laws with
the smallest degrees of locality \fr{DoL}, i.e. by considering
``truncated GGEs'' 
of the form
\be\label{tGGE}
\rho^{\rm tGGE,y}=\frac{e^{-\sum_{n=1}^y \lambda_n^{(y)} I^{(n)}}}{{\rm Tr}
\left(e^{-\sum_{n=1}^y \lambda_{n}^{(y)} I^{(n)}}\right)}\ .
\ee 
The value of $y$ is then determined by the required degree of accuracy
$\epsilon$ in \fr{compare}. A consequence of these findings is that
the full GGE can in fact be \emph{defined} as the limit of truncated
GGEs 
\be\label{eq:GGElim}
\rho^{\rm GGE}\equiv \lim_{y\rightarrow\infty} \rho^{\rm tGGE,y}\ .
\ee
The limiting procedure \eqref{eq:GGElim} was originally introduced in
a non-interacting model (the transverse-field Ising)\cite{FE_13a}, 
and has since proved very useful for the construction of GGEs and the
calculation of the stationary state values of local observables
in interacting integrable models\cite{FE_13b,Pozsgay:13a,richerme14,GGE_int,A:GGE15}.  

\begin{figure}[ht]
\begin{center}
\includegraphics[width=0.45\textwidth]{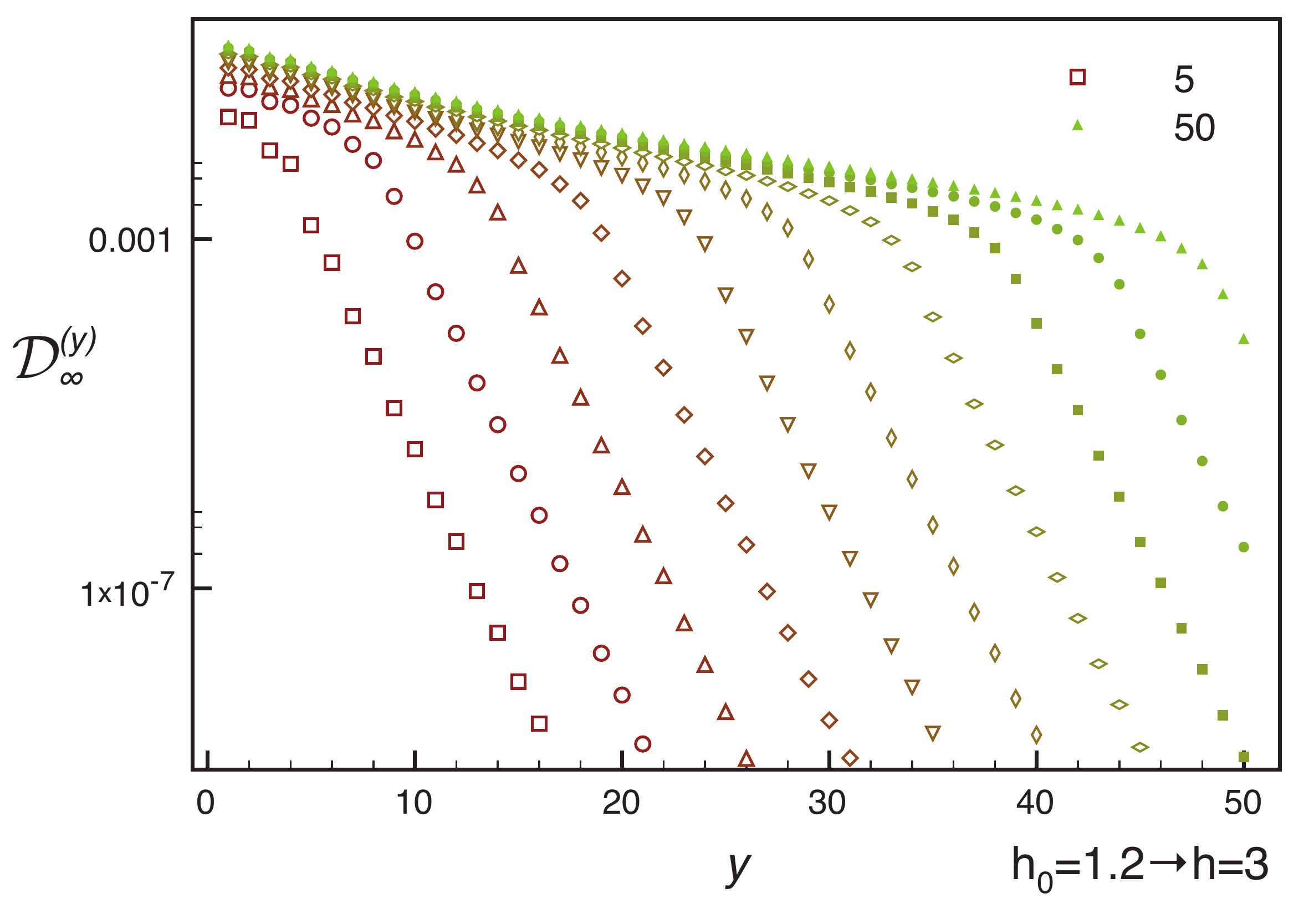}
\caption{Distance $\mathcal D_\infty^{(y)}=\mathcal
  D(\rho^{\rm GGE}_{\ell},\rho^{\rm tGGE,y}_\ell)$, as defined in
\eqref{eq:distance}, between the GGE and the 
truncated GGEs obtained by taking into account local conservation laws
with densities involving at most $y+1$ consecutive sites. The quench is
from $h_0=1.2$ to $h=3$ and the subsystem size ranges from
$\ell=5$ to $\ell=50$. Colors and sizes change gradually as a function
of the size $\ell$. For $y>\ell$, the distance starts decaying
exponentially in $y$, with an $\ell$-independent decay rate.
\emph{[Figure taken from Ref.~\onlinecite{FE_13a}]}}
\label{fig:local}
\end{center}
\end{figure}

To exhibit the above ideas more explicitly it is useful to consider the
example of a transverse field quench in the disordered phase of the
TFIC\cite{FE_13a}. Fig.~\ref{fig:local} shows results for an
appropriately defined distance
between the reduced density matrices of the full and truncated GGEs as
a function of the number $y$ of conservation laws retained in \fr{tGGE}.
Ten different subsystem sized $\ell=5,10,\dots,50$ are shown.

We see that for a given subsystem size $\ell$ the distance
${\cal D}\big(\rho^{\rm GGE}_\ell,\rho^{\rm pGGE,y}_\ell\big)$ starts
decaying exponentially in $y$ above an $\ell$-dependent value. This
means that conservation laws $I^{(n)}$ with $n\gg \ell$ play a
negligible role in describing stationary state properties in a
subsystem of size $\ell$. The main message is then as follows:
\begin{quote}
\emph{The smaller the degree of locality of a conservation law
is, the more important it is for describing the stationary
properties of local observables.}
\end{quote}

\subsection{Dynamical properties in the stationary state}
It is clearly of interest to go beyond the equal time correlators we
have discussed so far and consider dynamical (non equal time)
correlations. In particular one can envisage using them to
characterize the stationary state in the same way they are used in
thermal equilibrium. An example would be the measurement of
dynamical response functions in the stationary state by e.g.
photoemission spectroscopy\cite{spscopy}. For such purposes the
objects of interest are of the form
\be
\lim_{t\to\infty}
\braket{\Psi(t)|\mathcal O_1(t_1)\cdots \mathcal O_n(t_n)|\Psi(t)}\, ,\qquad \mathcal O_j(t)=e^{i H t}\mathcal O_j e^{-i H t}\, ,
\ee
where $\mathcal O_j$ are local observables. The problem one is faced
with is that the descriptions of the stationary state by statistical
ensembles (GGE, GMC) a priori hold only at the level of finite
subsystems in the thermodynamic limit. On the other hand, time
dependent operators act by construction non-trivially on the entire
system, which moves them beyond the remit of applicability of the
framework set out above. This issue was addressed in
Ref.~\onlinecite{EEF:12}, which established that the stationary state
density matrix $\rho^{\rm SS}$ that describes the local relaxation in
fact provides a correct description of dynamical correlations as well,
i.e. 
\be\label{eq:infdyn}
\lim_{t\rightarrow\infty}\braket{\Psi(t)|\mathcal O_1(t_1)\cdots \mathcal O_n(t_n)|\Psi(t)}=\mathrm{Tr}[\rho^{\rm SS}\mathcal O_1(t_1)\cdots \mathcal O_n(t_n)]\, .
\ee
The proof of this statement is based on the Lieb-Robinson bound (see
Sec.~\ref{ss:L-R}) and more specifically on a theorem by Bravyi,
Hastings, and Verstraete~\cite{bravyi06}, who showed that
time-evolving operators (in the Heisenberg picture) are well
approximated by local operators with a range that increases
linearly in time. We stress that \eqref{eq:infdyn} does not depend on
which statistical ensemble describes the local relaxation in the
stationary state, i.e. for generic systems $\rho^{\rm SS}$ would be a
Gibbs or a microcanonical density matrix, while for integrable systems
it would be that of a GGE or GMC.

The issue of whether or not a fluctuation dissipation
relation (FDR) holds in the stationary state was
investigated in Refs~\onlinecite{FCG:f-d11,FCG:dyn12, EEF:12,KPSR:f-d}.  
In thermal equilibrium, the FDR connects the imaginary part of the
linear response function $\chi_{AB}(\omega,\vec q|\rho)$ of two observables
$A$, $B$ to the corresponding spectral function
$S_{AB}(\omega,\vec{q}|\rho)$ 
\bea
\label{eq:FD_eq}
-\frac{1}{\pi}{\rm Im}\
\chi_{AB}\big(\omega,\vec{q}\ |\rho^{\rm Gibbs}(\beta)\big)
=(1-e^{-\beta \omega})S_{AB}\big(\omega,\vec{q}\ |\rho^{\rm Gibbs}(\beta)\big)\, .
\eea
Here we have defined
\bea
\chi_{AB}(\omega,\vec q\ |\rho)&=&-\frac{i}{L}\sum_{j,\ell}
\int_0^\infty\mathrm dt\ e^{i\omega t-i\vec q\cdot (\vec r_j-\vec
  r_\ell)}\mathrm{Tr}\Big(\rho\ [A_j(t),B_\ell]\Big)\, ,\nn
S_{AB}(\omega,\vec q\ |\rho)&=&
\frac{1}{L}\sum_{j,\ell}\int_{-\infty}^\infty\frac{\mathrm dt}
{2\pi}\ e^{i\omega t-i\vec q\cdot (\vec r_j-\vec r_\ell)}
\mathrm{Tr}\big(\rho\ A_j(t)B_\ell\Big)\, .
\eea
In the derivation of \fr{eq:FD_eq} one uses that in thermal
equilibrium the spectral function for negative and positive
frequencies are related in a simple way
\be
\label{eq:Sbaab}
S_{BA}\big(-\omega,-\vec q\ |\rho^{\rm Gibbs}(\beta)\big)
=e^{-\beta \omega}S_{AB}\big(\omega,\vec q\ |\rho^{\rm Gibbs}(\beta)\big).
\ee
Let us now turn to FDRs in the steady state after quantum
quenches. Clearly, if the system thermalizes~\eqref{thermalization}, 
the equilibrium FDR~\eqref{eq:FD_eq} with inverse temperature
$\beta_{\rm eff}$ applies. If on the other hand the system locally
relaxes to a GGE, the imaginary part of the linear response function ceases to be
proportional to the spectral function\cite{FCG:f-d11,FCG:dyn12,
  EEF:12}. This can be traced back to the absence of a simple
relationship between the spectral functions at positive and negative
frequencies. However, the basic form of the FDR still holds\cite{EEF:12}
\be
-\frac{1}{\pi}{\rm Im}\ \chi_{AB}\big(\omega,\vec q\ |\rho^{\rm
  GGE}\big)=
S_{AB}\big(\omega,\vec q\ |\rho^{\rm GGE}\big)
-S_{BA}\big(-\omega,-\vec q\ |\rho^{\rm GGE}\big)\, .
\ee


\section{A simple example}
\label{sec:simple}
In order to see the ideas presented above at work, we now consider the
specific example of a one dimensional fermionic pairing model
\be\label{eq:Hpar}
H(\Delta,\mu)=-J\sum_{j=1}^Lc^\dagger_jc_{j+1}+c^\dagger_{j+1}c_j-\mu\sum_{j=1}^L
c^\dagger_jc_j
+\Delta\sum_{j=1}^L
c^\dagger_jc^\dagger_{j+1}+c_{j+1}c_j\ .
\ee
Here $c^\dagger_j$, $c_j$ are canonical fermion creation
and annihilation operators at site $j$ and
$\{c_j,c^\dagger_n\}=\delta_{j,n}$. In momentum space we have
\be
H(\Delta,\mu)=\sum_k \epsilon_0(k)\ c^\dagger(k)c(k)-i\Delta
\sin(k)\big(c^\dagger(k)c^\dagger(-k)-c(-k)c(k)\big)\ ,
\label{Hdm}
\ee
where we have defined $\epsilon_0(k)=-2J\cos(k)-\mu$ and
\be
c_j=\frac{1}{\sqrt{L}}\sum_k e^{-ikj} c(k)\ .
\ee
The Hamiltonian \fr{Hdm} is diagonalized by a Bogoliubov
transformation
\bea
\begin{pmatrix}
\alpha(k)\\
\alpha^\dagger(-k)
\end{pmatrix}
&=&\begin{pmatrix}
\cos(\Theta_k/2) & -i\sin(\Theta_k/2)\\
-i\sin(\Theta_k/2) & \cos(\Theta_k/2)
\end{pmatrix}
\begin{pmatrix}
c(k)\\
c^\dagger(-k)
\end{pmatrix},\quad k\neq 0,
\eea
where
\bea
\epsilon(k)&=&\sqrt{(2J\cos(k)+\mu)^2+4\Delta^2\sin^2(k)}\ ,\quad
e^{i\Theta_k}=\frac{-2J\cos(k)-\mu+2i\Delta\sin(k)}{\epsilon(k)}.
\eea
Defining $\alpha(0)=c^\dagger(0)$ we find
\be
H(\Delta,\mu)=\sum_{k}
\epsilon(k)\ \alpha^\dagger(k)\alpha(k)+{\rm const}\ .
\ee
Let us now implement a quantum quench by initially preparing our
system in the ground state of $H(\Delta,\mu)$, and at $t=0$ quenching the
pairing amplitude from $\Delta$ to zero. The initial state is the
Bogoliubov fermion vacuum 
\be
|\Psi(0)\rangle=|0\rangle\ ,\quad \alpha(k)|0\rangle=0\ \forall k.
\ee
The time evolution of the fermion annihilation operators is obtained
by solving the Heisenberg equations of motion
$\frac{d}{dt}c(k,t)=i[H(0,\mu),c(k,t)]=-i\epsilon_0(k)c(k,t)$, which gives
\be
c(k,t)=e^{-i\epsilon_0(k)t}c(k)=e^{-i\epsilon_0(k)t}\left[
\cos(\Theta_k/2)\ \alpha(k) -i\sin(\Theta_k/2)\ \alpha^\dagger(-k)\right].
\ee
The fermion two-point functions at $t>0$ are thus equal to
\bea
\langle\Psi(t)|c^\dagger(k) c(q)|\Psi(t)\rangle&=&
\langle 0|c^\dagger(k,t) c(q,t)|0\rangle=\delta_{k,q}
\sin^2\big(\Theta_k/2\big)\ ,\nn
\langle\Psi(t)|c(k)
c(q)|\Psi(t)\rangle&=&\delta_{k,-q}\frac{i}{2}\sin\Theta_k\ e^{-2i\epsilon_0(k)t}.
\eea
In position space we obtain 
\bea
\langle\Psi(t)|c^\dagger_{j+\ell} c_{j}|\Psi(t)\rangle&=&
\frac{1}{L}\sum_ke^{ik\ell}\sin^2\big(\Theta_k/2\big)=f_L(\ell)\ ,\nn
\langle\Psi(t)|c_{j+\ell} c_{j}|\Psi(t)\rangle&=&
\frac{1}{L}\sum_ke^{-ik\ell}\frac{i}{2}\sin\Theta_k\ e^{-2i\epsilon_0(k)t}
=g_L(\ell,t)\ .
\label{fLgL}
\eea
Importantly, multi-point correlation functions can be calculated by
Wick's theorem, e.g.
\be
\langle\Psi(t)|c^\dagger_{j} c^\dagger_k c_{n}c_m|\Psi(t)\rangle=
g_L^*(k-j,t)g_L(n-m,t)-f_L(j-n)f_L(k-m)+f_L(k-n)f_L(j-m).
\ee
In the limit $L\to\infty$ we can turn the sums in \fr{fLgL} into
integrals, which at late times can be evaluated by a stationary phase
approximation. At infinite times we obtain
\bea
\lim_{L\to\infty}f_L(\ell)&=&\int_0^{2\pi}\frac{dk}{2\pi}\ e^{ik\ell}\ \sin^2\big(\Theta_k/2\big)\ ,\nn
\lim_{t\to\infty}\lim_{L\to\infty}g_L(\ell,
t)&=&0\ .
\label{2pointfns}
\eea
Importantly the ``anomalous'' average $\langle\Psi(t)|c_{j+\ell}
c_{j}|\Psi(t)\rangle$ vanishes in this limit. We can now immediately
conclude that our system relaxes locally to some steady state: any
operator ${\cal O}$ acting non-trivially only on a given, finite 
subsystem $B$ can be expressed in terms of  fermionic creation and
annihilation operators acting only on sites in $B$. We then can use
Wick's theorem to express the expectation value of
$\langle\Psi(t)|{\cal O}|\Psi(t)\rangle$ in terms of the functions
$f_L(\ell)$ and $g_L(\ell,t)$. After taking the infinite volume
limit $L\to\infty$, the limit $t\to\infty$ of the resulting expression
exists. This argument shows that the steady state is  completely
characterized by the two-point functions \fr{2pointfns} and the fact
that Wick's theorem holds. 

\subsection{Generalized Gibbs Ensemble}
According to our previous discussion the steady state should be
described by the GGE \fr{GGEFF}, where the Lagrange multipliers are
fixed by
\be
{\rm Tr}\big[\rho^{\rm GGE}c^\dagger(k)c(k)\big]=
\frac{1}{1+e^{\mu_k}}=
\langle\Psi(0)|c^\dagger(k)c(k)|\Psi(0)\rangle=
\sin^2\big(\Theta_k/2\big)\ .
\label{fixmu}
\ee
As the GGE density matrix is Gaussian a Wick's theorem holds, and
as a consequence of \fr{fixmu} the two-point functions in the GGE
coincide with those of our steady state. As we have a Wick's theorem
in the steady state as well, the GGE correctly reproduces \emph{all}
multi-point
correlation functions. This proves that the steady state in
our example is locally equivalent to the GGE \fr{GGEFF}, \fr{fixmu}.

\section{Spreading of correlations after a quantum quench}
\label{sec:spreading}
\begin{figure}[ht]
\begin{center}
(a)\includegraphics[width=0.4\textwidth]{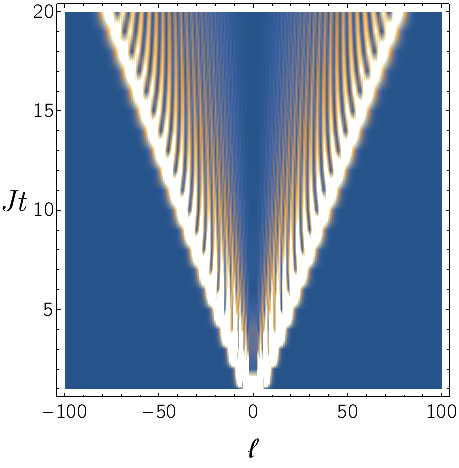}\qquad
(b)\includegraphics[width=0.4\textwidth]{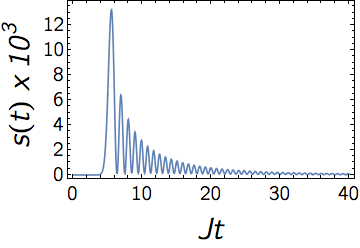}
\caption{(a) Time dependent part $S_\infty(\ell,t)+|f_\infty(\ell)|^2$ of
  the connected density-density correlator after a quantum quench
  where the system is initialized in the ground state of
  $H(\Delta=2J,\mu=-J)$, and time evolved with $H(0,\mu=-J)$. A
  light cone effect is clearly visible. (b) 
$s(t)=S_\infty(\ell=20,t)+|f_\infty(20)|^2$ as a function of time.}
\label{fig:lightcone}
\end{center}
\end{figure}

\begin{figure}[ht]
\begin{center}
\includegraphics[width=0.7\textwidth]{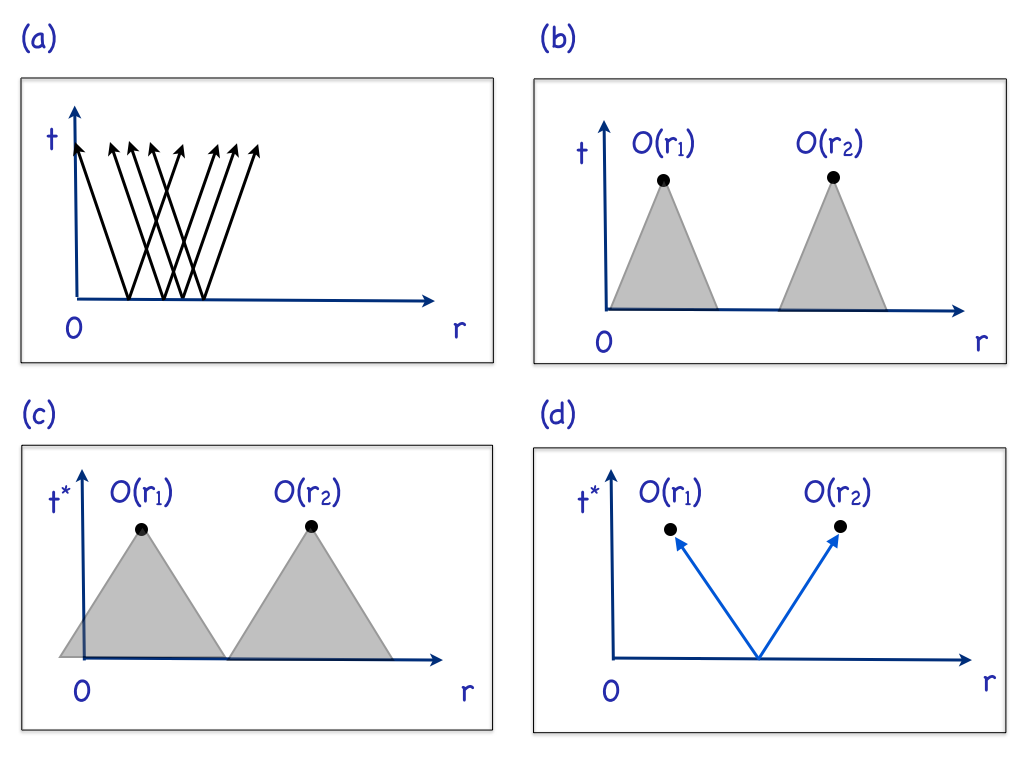}
\caption{``Quasi-particle picture'' for the light cone effect
\cite{cc-05}: (a) At
time $t=0$ the quantum quench creates quasi-particle excitations
throughout the system. (b) At time $t$ quasi-particles from within
the ``backward light cone'' $[r_j-v_{\rm max}t,r_j+v_{\rm max}t]$
will affect a measurement at position $r_j$. This leads to de-phasing
of 1-point functions $\langle\Psi(t)|{\cal
  O}(r_j)|\Psi(t)\rangle$. (c) At time $t^*=\frac{|r_2-r_1|}{2v_{\rm
max}}$ the backwards light cones touch, and measurements at $r_1$
and $r_2$ become correlated. (d) Connected correlations are induced by
quasi-particle pairs created at time $t=0$ and propagating with group
velocities $v_{\rm max}$ in opposite directions.}
\label{fig:QPP}
\end{center}
\end{figure}

Let us now turn to the time dependence of the expectation values of
local operators after a quantum quench. As an example we consider the
connected correlation function of the fermionic density 
$n_j=c^\dagger_jc_j$ in our example
of section~\ref{sec:simple} 
\be
S_L(\ell,t)=\langle\Psi(t)|n_{j+\ell}n_j|\Psi(t)\rangle-
\langle\Psi(t)|n_{j+\ell}|\Psi(t)\rangle
\langle\Psi(t)|n_{j}|\Psi(t)\rangle\ .
\label{SL}
\ee
Application of Wick's theorem gives
\be
S_L(\ell,t)=|g_L(-\ell,t)|^2-|f_L(\ell)|^2.
\ee
It is convenient to isolate the time dependent part $|g_L(-\ell,t)|^2$,
which is shown in Fig.~\ref{fig:lightcone} for a quantum quench, where
we start in the ground state of $H(\Delta=2J,\mu=-J)$ and time evolve
with $H(0,\mu=-J)$.
At a fixed value of $\ell$ the connected correlator is exponentially
small (in $\ell$) until a time
\be
t_F=\frac{\ell}{2v_{\rm max}}\ ,
\ee
where in our example $v_{\rm max}=2J$ is the maximal group velocity of
elementary particle and hole excitations of our post-quench
Hamiltonian $H(0,\mu=-J)$. At
$t\approx t_F$ the connected correlator increases substantially, goes
through a maximum, and then decays in an oscillating fashion.
A physical explanation for the light cone effect was provided by
Calabrese and Cardy\cite{cc-05,cc-07}. For a non-interacting,
non-relativistic theory like the one in our example it goes as
follows, \emph{cf.} Fig.~\ref{fig:QPP}. We focus on
connected correlation functions of local operators, e.g.  
\be
G_{\cal O\cal O}(r_1,r_2;t)=\langle\Psi(t)|{\cal O}(r_1){\cal
  O}(r_2)|\Psi(t)\rangle -\langle\Psi(t)|{\cal O}(r_1)|\Psi(t)\rangle 
\langle\Psi(t)|{\cal O}(r_2)|\Psi(t)\rangle\ .
\ee
This is the average of the simultaneous measurement of the observables
${\cal O}(r_{j})$ minus the product of the averages of separate
measurements of ${\cal O}_j$ at time $t$ after the quench.
The initial state in our case is characterized by a finite correlation
length $\xi$
\be
G_{\cal O\cal O}(r_1,r_2;t=0)\propto e^{-|r_1-r_2|/\xi} \ ,
\ee
and is therefore extremely small at large spatial separations.
At time $t=0$ the quantum quench generates a finite density
of stable quasiparticle excitations throughout the system. Their
dispersion relation is $\epsilon_0(p)$ as our post-quench Hamiltonian
is simply $H(0,\mu)=\sum_k\epsilon_0(k)n(k)$. The maximal group
velocity of these free fermionic excitations is 
\be
v_{\rm max}={\rm max}_p\frac{d\epsilon_0(p)}{dp}=2J.
\ee
At times $t>0$ the quasi-particles created by the quench propagate
through the system. A measurement at $r_j$ will be influenced by
quasi-particles from within the ``backwards light cone'' $[r_j-v_{\rm
    max}t,r_j+v_{\rm     max}t]$. At all times $t>0$ this will affect
the value of the 1-point functions $\langle\Psi(t)|{\cal
  O}(r_j)|\Psi(t)\rangle$, but the effect cancels in the
\emph{connected} two-point function. 
At time $t^*=\frac{|r_2-r_1|}{2v_{\rm max}}$ the backwards light cones
emanating from $r_1$ and $r_2$ touch, and the average of measurements
at $r_1$ and $r_2$ becomes correlated. These connected correlations
are induced by quasi-particle pairs created at time $t=0$ and
propagating with group velocities $v_{\rm max}$ in opposite directions.

Since the work of Calabrese and Cardy light cone effects after quantum
quenches have been
analyzed in a number of lattice models
\cite{dechiara06,laeuchli08,FC08,manmana09,CEF1,CEF2,CEF3,carleo14}
and observed in experiments 
on systems of ultra-cold atomic gases~\cite{cetal-12,langen13} and
trapped ions~\cite{jurcevic14,richerme14}. The experimental work
raises the poignant theoretical issue of which velocity underlies the
observed light cone effect in non-relativistic systems at finite
energy densities. Here there is no unique velocity of light, and
quasi-particles in interacting systems will generally have finite life
times depending on the details of the initial density matrix.
For the case of the spin-1/2 Heisenberg XXZ chain, an integrable
model, it was shown in Ref.~\onlinecite{bonnes14} that the light cone
propagation velocity in general depends on the energy density of the
initial state, and an explanation for this effect in terms of
properties of stable excitations at finite energy densities was put
forward. More recent theoretical works address the influence of
long-range interactions on the spreading of
correlations~\cite{schachenmayer13,hauke13,eisert13,vodola15,regemortel15,buyskikh16}. 
Sufficiently long-range interactions lead to a destruction of light cone
effects. Non-relativistic continuum models are also known to exhibit
modifications to light cone behaviour \cite{KCC14}. 

It is useful to contrast the above discussion to the spreading of
correlations in equilibrium and after ``local quantum quenches''
\cite{localquenches}. In the latter context one is concerned with the
spreading of a local perturbation that has been imposed on an
equilibrium state. Light cone effects are observed in such situations
as well, but the spreading occurs at the maximum group velocity of
elementary excitations over the equilibrium state that is being
considered. In other words, unlike for global quantum quenches, there
is no factor of two.

\subsection{Relation to Lieb-Robinson bounds}
\label{ss:L-R}
As shown by Lieb and Robinson~\cite{LR72,SN10}, the velocity of
information transfer in quantum spin chains is effectively
bounded. More precisely, there exists a causal structure in
commutators of local operators at different times
\be
\|[{\cal O}_A(t),{\cal O}_B(0)]\|\leq c\ {\rm min}(|A|,|B|)\ \|{\cal O}_A\|\
\|{\cal O}_B\|\ e^{-\frac{L-vt}{\xi}}\ .
\label{LRB}
\ee
Here ${\cal O}_A$ and ${\cal O}_B$ are local operators acting
non-trivially only in two subsystems $A$ and $B$ that are spatially
separated by a distance $L$, $\|.\|$ denotes the operator norm and
$|A|$ the number of sites in subsystem $A$. Finally, $c$, $v$ and
$\xi$ are constants. More recently, the Lieb-Robinson bounds have been
refined~\cite{juneman13,kliesch13} and extended to mixed state
dynamics in open quantum systems~\cite{poulin10,kliesch13}.

The Lieb-Robinson bound has important consequences  for
quantum quenches starting in initial states with finite
correlation lengths, and time evolving under a short-ranged
Hamiltonian. It was shown in Ref.~\onlinecite{bravyi06} that
\fr{LRB} implies a bound on the connected two-point correlation
functions after such quenches
\be
\langle\Psi(t)|{\cal O}_A\ {\cal O}_B|\Psi(t)\rangle_{\rm conn}<
\bar{c}\big(|A|+|B|\big)\ e^{-\frac{L-2vt}{\chi}}\ .
\label{LRB2}
\ee
Here $\bar{c}$, $v$ and $\chi$ are constants. The bound \fr{LRB2}
shows that connected two-point functions of local operators after
quantum quenches in spin chains are exponentially small up to times
$t=L/2v$. This tallies very nicely with the light cone effects
discussed above. We note that the bound does not provide values for
the velocity $v$ or the length $\chi$.

\subsection{Finite-size effects}
Throughout our discussion we have stressed that we are ultimately
interested in taking the thermodynamic limit. In a large but finite
system local observables can never truly relax. There always will be
\emph{recurrences}\cite{Bocc57} such that the return amplitude
${\cal F}(t)=|\langle\Psi(0)|\Psi(t)\rangle|$ is arbitrarily close to
$1$ 
\be
|1-{\cal F}(t)|<\epsilon\ .
\ee
However, in many-particle systems these typically occur only at
astronomically late times. The exception to this rule are cases in
which the spectrum of the post-quench Hamiltonian in a (large) finite
volume for some reason has a highly commensurate structure. A very
simple example are Hamiltonians with equidistant energy
levels $E_n=E_0+n\delta$, for which we have \mbox{${\cal F}(2\pi j/\delta)=1$}.

A more relevant effect in practice are \emph{revivals}, which
refers to situations where ${\cal F}(t)={\cal O}(1)$. This again
requires the finite size energy spectrum to have a certain regularity,
as is the case for example in conformal field theories \cite{Cardy14,CCrev}.

\begin{figure}[ht]
\begin{center}
(a) \includegraphics[width=0.4\textwidth]{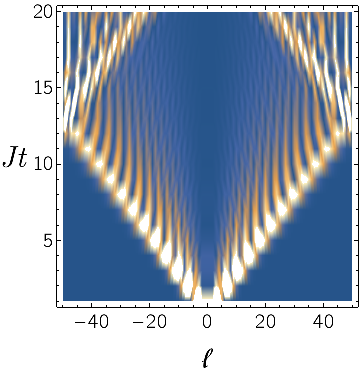}
\qquad
(b) \includegraphics[width=0.4\textwidth]{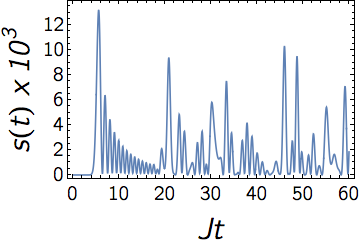}
\caption{(a) Same as Fig.~\ref{fig:lightcone} but for a finite system
with $L=100$ sites and periodic boundary conditions. At times
$Jt>12.5$ the light cone has traversed the system and eventually
causes a revival of the connected correlation function at a given
separation $\ell$. This is strictly a finite-size effect and has no
analog in the thermodynamic limit.
(b) $s(t)=S_L(\ell=20,t)+|f_L(20)|^2$ for $L=100$ as a function of time.
}
\label{fig:lightconeL100}
\end{center}
\end{figure}

\begin{figure}[ht]
\begin{center}
\includegraphics[width=0.4\textwidth]{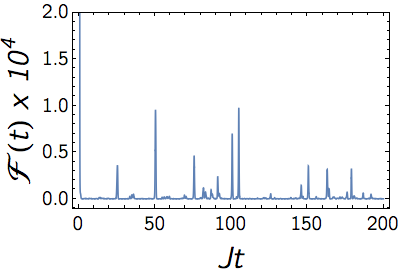}
\caption{The return probability for the same quench as in
  Fig.~\ref{fig:lightcone} with $L=100$ sites and periodic boundary
  conditions. The overlap with the initial state remains negligible at
  any time $t>0$.  
}
\label{fig:returnprob}
\end{center}
\end{figure}

A different finite-size effect affects all observables and is related
to the light cone repeatedly traversing the system~\cite{CSC:13b,releBose,IR:quenchB,BRI:sc_XY}. We refer to this
effect as a \emph{traversal}. As an example we consider the connected
two point function \fr{SL} on a finite ring. A density plot is shown
for a system of size $L=100$ in Fig.~\ref{fig:lightconeL100}.  
We see that the light cone traverses the system and induces a signal
in $S_L(\ell,t)$ at a time $(L-|\ell|)/2v_{\rm max}$ after the
light cone first reaches. Clearly at times $t>(L-|\ell|)/2v_{\rm max}$
correlation functions of the finite system look very different from
the ones in the thermodynamic limit. As is shown in
Fig.~\ref{fig:returnprob}, the traversal in not associated with any
revival, because the return amplitude remains exponentially small in
the system size. 

\section{Transverse-field Ising chain (TFIC)}
\label{sec:TFIC}
We now turn our attention to a key paradigm for quantum quenches,
the transverse-field Ising chain
\be\label{eq:HTFIC} 
H(h)=-J\sum_{\ell=1}^L \sigma_\ell^x\sigma_{\ell+1}^x+h\sigma_\ell^z\, .
\ee
Here we impose periodic boundary conditions $\sigma^\alpha_{L+1}\equiv
\sigma^\alpha_1$, $L$ even, $h\geq 0$ and $J>0$. We note that the signs
of $h$ and $J$ can be reversed by unitary transformations with
respectively 
\be
U_1=\prod_{\ell=1}^L\sigma_\ell^x\ ,\quad
U_2=\prod_{\ell=1}^{L/2}\sigma_{2\ell-1}^x\sigma_{2\ell}^y.
\ee
The Hamiltonian \fr{eq:HTFIC} has a $\mathbb{Z}_2$ symmetry of
rotations by $\pi$ around the z-axis. The ground state phase diagram
of the TFIC features a paramagnetic (for $h>1$) and a ferromagnetic (for
$h<1$) phase, in which the $\mathbb{Z}_2$ symmetry is spontaneously
broken. The two phases are separated by a quantum critical point at
$h=1$, which is described by the Ising conformal field theory with
central charge $c=1/2$, \emph{cf.} Ref.~\onlinecite{sachdevbook}. 

It is well known that the TFIC admits a representation in terms of
non-interacting fermions. However, the Jordan-Wigner transformation
between spins and fermions is nonlocal. This renders the TFIC an ideal
testing ground for relaxation ideas, in particular in relation to the
crucial role played by locality.

The study of non-equilibrium dynamics in the TFIC was initiated in a
seminal paper by Barouch, McCoy and Dresden in 1970\cite{BMD70}. They
analyzed the time evolution of the transverse magnetization
$\langle\Psi(t)|\sigma^z_\ell|\Psi(t)\rangle$ and observed that it
relaxes to a non-thermal value at late times. The focus of research on
the TFIC and its two-dimensional classical counterpart then shifted to
the determination of equilibrium properties, and it took thirty years
before the issue of relaxation after quantum quenches returned to
centre stage\cite{IR:quench00,SPS:04}. A combination of experimental
advances in cold atom systems and the theoretical insights gained through
the study of quantum quenches in conformal field
theories\cite{cc-05,CC:tcf06,cc-07} revitalized the quest for obtaining a 
complete understanding of the quench dynamics in the TFIC. Exact
closed form expressions for the time evolution of the entanglement
entropy after a global quench were obtained in
Ref.~\onlinecite{FC08}. By combining free-fermion techniques with
numerical methods important insights on thermalization issues were
gained\cite{efftherm}, and finite-size effects like traversals were
analyzed in detail\cite{IR:quenchB}. Exact results for the time
evolution of order parameter correlations were finally obtained in
Refs~\onlinecite{CEF1,CEF2,CEF3,SE:Ising}, and it was demonstrated
that the stationary state is described by a generalized Gibbs ensemble.
The time evolution of reduced density matrices was studied in
Ref.~\onlinecite{FE_13a}, and the question of which conservation
laws are most important for characterizing the stationary state with a
given accuracy was resolved. Refs~\onlinecite{RI:sc11,BRI:sc_XY}
introduced a semi-classical approach to quantum quenches by
generalizing a method developed by Sachdev in the context of
equilibrium dynamics\cite{sachdevbook}. Dynamical (non-equal time)
correlations were studied in Refs~\onlinecite{RI:sc11,BRI:sc_XY,EEF:12} 
in relation to the question whether the stationary state fulfils a
fluctuation-dissipation theorem.
Progress on experimental studies of non-equilibrium evolution in TFICs
has been more limited. Ref.~\onlinecite{MM:Ising13} reported results on
quantum quenches in a cold atom system described by one dimensional
Bose Hubbard chains, which map onto the TFIC in a particular limit.
Refs~\onlinecite{VD:QED13} proposed a realization of the TFIC with a
time-dependent magnetic field  in the framework of circuit QED.
\subsection{Fermionic form of the Hamiltonian}
Spin chains that can be mapped to free fermions differ in two
important aspects from free fermion models like the one we considered
in Section \ref{sec:simple}:
\begin{itemize}
\item[-] two-point functions of spin operators map onto $n$-point
correlation functions of fermions, where $n$ is related to the
distance between the two spins and can be arbitrarily large; 
\item[-] The ground state of the TFIC in the ordered phase is not a
Fock state (as a result of spontaneous symmetry breaking).
\end{itemize} 
The Hamiltonian \eqref{eq:HTFIC} can be mapped to a fermionic theory
by a Jordan-Wigner transformation 
\bea
\sigma^z_\ell=ia_{2\ell}a_{2\ell-1}\ ,\quad
\sigma^x_\ell=\left(\prod_{j=1}^{\ell-1}(ia_{2j}a_{2j-1})\right) a_{2\ell-1}\ ,\quad
\sigma^y_\ell=\left(\prod_{j=1}^{\ell-1}(ia_{2j}a_{2j-1})\right) a_{2\ell},
\label{majorana}
\eea
Here $a_\ell$ are Majorana fermions satisfying the anti-commutation relations
$\{a_\ell,a_n\}=2\delta_{\ell n}$. The usual spinless fermions are
obtained by taking linear combinations $c^\dag_\ell=(a_{2\ell-1}+i
a_{2\ell})/2$. It is now straightforward to see that spin-spin
correlation functions map onto expectation values of strings of
fermions, e.g.  
\be\label{eq:egEV}
\braket{\sigma_\ell^x}=(-i)^{\ell-1} \braket{\prod_{j=1}^{2\ell-1}a_j} \qquad \braket{\sigma_\ell^x\sigma_{\ell+n}^x}=(-i)^n \braket{\prod_{j=2\ell}^{2\ell+2n-1}a_j }\, .
\ee
Application of the Jordan-Wigner transformation to the TFIC
Hamiltonian \fr{eq:HTFIC} results in a fermion Hamiltonian of the form 
\bea
\label{eq:nonint}
H&=&\frac{\mathrm I-e^{i\pi\mathcal N}}{2}H_{\rm R}+\frac{\mathrm
  I+e^{i\pi\mathcal N}}{2}H_{\rm NS}\, ,\nn 
H_{\rm NS/R}&=&iJ\sum_{j=1}^{L-1}a_{2j}\left[a_{2j+1}-h a_{2j-1}\right]
-iJa_{2L}\left[ha_{2L-1}\mp a_{1}\right].
\eea
Here $e^{i\pi\mathcal N}$ is the fermion parity operator with
eigenvalues $\pm 1$
\be
e^{i\pi\mathcal
  N}=\prod_{\ell=1}^L\sigma_\ell^z=(-i)^L\prod_{j=1}^{2L}a_j\ ,\qquad
e^{i\pi\mathcal  N}a_j=-a_je^{i\pi\mathcal  N} \, .
\ee
$H_{\rm R,NS}$ commute with the fermion parity operator, and the full
Hamiltonian \fr{eq:nonint} is therefore block-diagonal: $H_{\rm R}$
($H_{\rm NS}$) describes the action on states with an odd (even)
number of fermions. We note that the free fermion Hamiltonians $H_{\rm
  NS/R}$ are closely related to the pairing model \eqref{Hdm}
considered earlier
\be
H_{\mathrm R}=-H(-J,-2h J)+J h  L\ .
\ee
\subsubsection{Ground states}
The Hamiltonians $H_{\rm NS/R}$ can be diagonalized by Bogoliubov
transformations to canonical momentum space fermion operators $b_p$
(details can be found in e.g. Appendix A of Ref.~\onlinecite{CEF1}) 
%
\be
H_{\rm a}(h)=\sum_{p\in{\rm a}}
\varepsilon_h(p)
\left(b^\dagger_pb_p-\frac{1}{2}\right),\quad
{\rm a=R,\ NS}\ ,
\ee
where the single-particle energy is given by
\be
\varepsilon_h(k)=2J\sqrt{1+h^2-2h\cos k}.
\label{eq:energy}
\ee
The difference between R and NS sectors enters via the allowed values
of the momenta, which are $p=\frac{\pi n}{L}$, where $n$ are
even/odd integers for R and NS fermions respectively. The ground
states of $H_{\rm R,NS}(h)$ are the fermionic vacua 
\be
b_p|GS\rangle_{\rm a}=0\ \forall p\in {\rm a}\ ,\quad {\rm a =R,NS}.
\ee
These vacuum states are also eigenstates of the fermion parity
operator
\be
e^{i\pi \mathcal N}\ket{GS}_{\rm NS}=\ket{GS}_{\rm NS}\ ,\qquad
e^{i\pi \mathcal N}\ket{GS}_{\rm R}=\mathrm{sgn}(h-1)\ket{GS}_{\rm
  R}\, .
\label{vacuumparity}
\ee
From \eqref{eq:nonint} it follows that in the ferromagnetic phase
$h<1$ both fermion vacua are eigenstates of the full Hamiltonian
$H$. Their respective energies are exponentially (in system size)
close, and they become degenerate in the thermodynamic
limit. Spin-flip symmetry then gets spontaneously broken, and the
ground state is either the symmetric or the antisymmetric combination
of the two vacuuum states. In the paramagnetic phase $h>1$ the ground
state of $H$ is given by the NS vacuum state. In summary, we have
\be
\ket{GS}=\left\{\begin{array}{ll}
\frac{\ket{GS}_{\rm NS}\pm\ket{GS}_{\rm R}}{\sqrt{2}}& h<1\\
\ket{GS}_{\rm NS}& h>1\, .
\end{array}\right.
\label{groundstates}
\ee 
This shows that for $h<1$ the ground state of $H$ is not a Fock state.

\subsection{Quantum quench of the transverse field} 
We now consider the following quench protocol. We prepare the system
in the ground state of $H(h_0)$, and at time $t=0$ quench the
transverse field to a new value $h$. At times $t>0$ we time evolve
with the new Hamiltonian $H(h)$. All local operators in spin basis can
be classified according to their fermion parity, and this turns out to
be very useful. We have
\be
e^{i\pi \mathcal N}{\cal O}_{e/o}e^{-i\pi \mathcal N}=\pm 
{\cal O}_{e/o}.
\label{evenodd}
\ee

\subsubsection{Quenches originating in the paramagnetic phase}
For a quench starting in the paramagnetic phase it follows from
\fr{groundstates}, \fr{eq:nonint} and \fr{vacuumparity} that 
the state of the system at times $t>0$ is given by
\be\label{eq:equiv}
|\Psi(t)\rangle=
e^{-i H t}\ket{GS}\equiv e^{-i H_{\rm NS} t}\ket{GS}_{\rm NS}\, .
\ee
This is even under fermion parity. Hence expectation values of odd
operators must vanish 
\be
\langle\Psi(t)|O_o |\Psi(t)\rangle=0.
\ee
By virtue of the simple form of both the time evolution
operator $H_{\rm NS}$ and the initial state $|{\rm GS}\rangle_{\rm
  NS}$ in \fr{eq:equiv}, expectation values of even operators can be
calculated by applying Wick's theorem. This allows the reduction of 
expectation values of strings of fermion operators to Pfaffians
involving only the two-point functions $\braket{a_i a_j}$
\cite{LSM,BM71a,BM71b}. For example we have   
\bea
\label{eq:egWick}
\braket{\sigma_\ell^x\sigma_{\ell+2}^x}&=&
-\braket{a_{2\ell}a_{2\ell+1}a_{2\ell+2}a_{2\ell+3}}\nn
&=&-\braket{a_{2\ell}a_{2\ell+1}}\braket{a_{2\ell+2}a_{2\ell+3}}+\braket{a_{2\ell}a_{2\ell+2}}\braket{a_{2\ell+1}a_{2\ell+3}}-\braket{a_{2\ell}a_{2\ell+3}}\braket{a_{2\ell+1}a_{2\ell+2}}\, .
\eea
\subsubsection{Quenches originating in the ferromagnetic phase}
As a result of spontaneous symmetry breaking in the ground state, the
situation for quenches originating in the ferromagnetic phase 
$h_0<1$ is more complicated. The time evolved initial state is 
\be
|\Psi(t)\rangle=\frac{e^{-i H_{\rm NS} t}\ket{GS}_{\rm NS}\pm e^{-i
    H_{\rm R} t}\ket{GS}_{\rm R}}{\sqrt{2}}\, . 
\ee
The expectation values of even operators can be expressed in the form
\bea
\label{eq:Oeven}
\langle\Psi(t)|\mathcal O_e|\Psi(t)\rangle&=&
\frac{1}{2}\sum_{{\rm a}\in\{{\rm R, NS}\}}
{}_{\rm a}\braket{GS|e^{i H_{\rm NS} t}\mathcal O_e e^{-i H_{\rm NS}
    t}|GS}_{\rm a}\nn
&\underset{L\to\infty}{\longrightarrow}
& {}_{\rm NS}
\braket{GS|e^{i H_{\rm NS} t}\mathcal O_e e^{-i H_{\rm NS} t}|GS}_{\rm NS}\, .
\eea
In the last line we have used that the expectation values in the R and
NS sectors become equal in the thermodynamic limit. The last line in
\fr{eq:Oeven} can again be evaluated by application of Wick's theorem.

In contrast, expectation of odd operators cannot be simplified in this
way. Instead we have 
\be
\label{eq:Oodd}
\braket{GS|e^{i H t}\mathcal O_o e^{-i H t}|GS}=\pm \mathrm{Re}[{}_{\rm NS}\!\braket{GS|e^{i H_{\rm NS} t}\mathcal O_o e^{-i H_{\rm R} t}|GS}_{\rm R}]\, .
\ee
Wick's theorem does not apply here, and in order to proceed one
commonly resorts to one of the following methods\cite{Isingusual}:
\begin{enumerate}
\item  Using the cluster decomposition property we can obtain
  \fr{eq:Oodd} from the expectation value of an even operator by
  considering the limit
\be
|\braket{GS|e^{i H t}\mathcal O_o e^{-i H t}|GS}|=\lim_{d\rightarrow\infty} \sqrt{{}_{\rm NS}\!\braket{GS|e^{i H_{\rm NS} t}\mathcal \mathcal O_o(r) \mathcal O_o(r+d) e^{-i H_{\rm NS} t}|GS}_{\rm NS}}\, .
\ee
\item One imposes open boundary conditions on the spins (in that case
the Hamiltonian is mapped into a purely quadratic form of fermions)
and considers the expectation value of $\mathcal O_o$ asymptotically
far away from the boundaries.
\end{enumerate}
An alternative method that applies more generally to integrable models
was developed in Refs~\onlinecite{CEF1,CEF2,CEF3}. It is based on the
observation that the initial state after the quench can be expressed
in a squeezed state form, \emph{e.g.} 
\be
\ket{\Psi(0)}=\exp\Bigl[i\sum_{0<p\in {\rm NS}}K(p)b^\dag_p b^\dag_{-p}\Bigr]\ket{GS}_{\rm NS}\, .
\ee
Here $K(p)$ is a function that depends on the quench (\emph{cf.}
\eqref{eq:K2}),  $b^\dag_p$ are the aforementioned momentum space
Bogoliubov fermion operators. Two point correlation functions can then
be written in a Lehmann representation based on energy eigenstates.
The matrix elements (``form factors'') in the Lehmann representation
are known\cite{Isingformfactors}, and it is possible to obtain
explicit results in the framework of a low-density expansion. We refer
the reader to Refs~\onlinecite{CEF2,CEF3} for further details and the
explicit calculations. Here we only review and discuss the main results. 
\subsection{Stationary state properties}
Stationary state properties were analyzed in detail in
Ref.~\onlinecite{CEF3}. An important simplification at late times is
that expectation values of odd operators go to zero. As discussed
above, for quenches originating in the paramagnetic phase they vanish
identically because the $\mathbb{Z}_2$ symmetry remains unbroken. On the
other hand, for quenches originating in ferromagnetic phase ($h_0<1$)
their expectation value is generally nonzero
(\emph{cf}. \eqref{eq:Oodd}). However, as shown in
Ref.~\onlinecite{FE_13a}, expectation values of all odd local
operators decay exponentially in time to zero. This leaves us with
expectation values for even operators, which can be analyzed by
standard free fermion methods. The basic object is the fermion
two-point function, which in the thermodynamic limit can be written in
the form (\emph{cf.} \eqref{fLgL}) 
\be
\label{eq:aat}
i \braket{a_{\ell}a_{n}}\Bigr|_{\ell\neq
  n}=\int_{-\pi}^\pi\frac{\mathrm d k}{2\pi}
\left[ A_{\ell-n}(k)+B_{\ell-n}(k)e^{2i\varepsilon_h(k)t}+B^*_{\ell-n}(k)e^{-2i\varepsilon_h(k)t}\right]\, .
\ee
Here $A(k)$ and $B(k)$ are smooth functions (we are considering
noncritical quenches) that depend on the quench details and
$\varepsilon_h(k)$ is the dispersion relation \eqref{eq:energy}. 
By the Riemann-Lebesgue lemma, in the infinite time limit the fermion two-point functions approach stationary values  
\be\label{eq:aainf}
\lim_{t\to\infty}i \braket{a_{\ell}a_{n}}\Bigr|_{\ell\neq
  n}= \int_{-\pi}^\pi\frac{\mathrm d k}{2\pi}
A_{\ell-n}(k)\, . 
\ee
Since the expectation value of any even local operator can always be
written as a finite sum of finite products of fermion two-point
functions (\emph{cf.} \eqref{eq:egWick}), the infinite time limit
exists and is obtained by replacing \eqref{eq:aat} with
\eqref{eq:aainf}. Courtesy of Wick's theorem, the stationary properties
of all other even operators can be expressed in terms of
\fr{eq:aainf}.

\subsubsection{Description of the steady state by a GGE}
Since expectation values of odd operators vanish at infinite times,
the steady state can be constructed in complete analogy to our fermionic
example considered in Sec.~\ref{sec:simple}. The appropriate GGE
density matrix is of the form
\be
\rho_{\rm GGE}=\frac{e^{-\sum_{j=1}^\infty \lambda_n^+
    I^{(n,+)}+\lambda_n^- I^{(n,-)}}}{Z_{\rm GGE}}\, , 
\ee
where the conservation laws $I^{(n,\pm)}$ have been reported earlier in
\fr{In+}. The Lagrange multipliers for transverse field quenches were
determined in Ref.~\onlinecite{FE_13a}
\be\label{eq:Lmult}
\lambda_j^+=\Bigl\{\frac{\mathrm{sgn}[(h-1)(h_0-1)]}{\varepsilon_h(0)}+\frac{(-1)^j}{\varepsilon_h(\pi)}\Bigr\}\frac{2}{j}\ ,\qquad
\lambda_j^-=0.
\ee
The local equivalence of the steady state density matrix to this GGE
was demonstrated in Refs~\onlinecite{CEF1,FE_13a}
\be
\lim_{t\to\infty}\ket{\Psi(t)}\bra{\Psi(t)}=_{\rm loc}
\rho_{\rm GGE}.
\ee
The GGE can be interpreted as a Gibbs ensemble at inverse temperature
$\beta=J^{-1}$ for an effective ``GGE Hamiltonian'', defined as 
\be
\label{eq:HGGE}
H_{\rm GGE}\equiv J\sum_{j=1}^\infty \lambda_j^+ I^+_j\, .
\ee
By virtue of \fr{eq:Lmult} this ``Hamiltonian'' is long-ranged. It can
be diagonalized by combined Jordan-Wigner and Bogoliubov
transformations \cite{FE_13a}, which for a quench from $h_0$ to $h$
results in
\be
H_{\rm GGE}= \int_{-\pi}^\pi\frac{dp}{2\pi}
J \log\big( K^2(p)\big)\left[b^\dagger(p)b(p)-\frac{1}{2}\right],
\label{eq:epsGGE}
\ee
where
\be\label{eq:K2}
K^2(p)=\frac{\sin^2(p)\ (h-h_0)^2}{(\varepsilon_{h_0}(p)
\varepsilon_{h}(p)\big(2J\big)^{-2}+1+hh_0-(h+h_0)\cos(p))^2}\, .
\ee
The ``dispersion relation'' $J \log K^2(p)$ diverges logarithmically
at momenta zero and $\pi$. This is related to the fact that the mode
occupation numbers at these momenta are independent of $h$, and
ultimately produces the algebraic decay \eqref{eq:Lmult} of the
Lagrange multipliers. 
These logarithmic singularities do not compromise the cluster
decomposition properties of the steady state.
\subsubsection{Connected spin-spin correlation functions}
In the stationary state the connected two-point correlators decay
exponentially with distance\cite{CEF2}
\be
\rho^{\alpha\alpha}_c(\ell)=\braket{\sigma_j^\alpha\sigma_{j+\ell}^\alpha}-\braket{\sigma_j^\alpha}\braket{\sigma_{j+\ell}^\alpha}
\simeq C^\alpha(\ell) e^{-\ell/\xi_\alpha}\, .
\label{static2point}
\ee
The correlation lengths are given by\cite{CEF2}
\bea
\label{eq:xiz}
\xi_z^{-1}&=&|\ln h_0|+\min(|\ln h_0|,|\ln h|)\ ,\\
\xi_x^{-1}&=&\theta_H(h-1)\theta_H(h_0-1)\ln\left[\min(h_0,h_1)\right]-
\ln\Bigl[x_++x_-+\theta_H((h-1)(h_0-1))\sqrt{4 x_+x_-}\Bigr]\, ,
\label{xiasy}
\eea
where $\theta_H(x)$ is the Heaviside step function and
\bea\label{eq:h1}
x_\pm=\frac{[{\rm min}(h,h^{-1})\pm 1][{\rm min}(h_0,h_0^{-1})\pm 1]}4
\,, \quad  
h_1=\frac{1+h h_0+\sqrt{(h^2-1)(h_0^2-1)}}{h+h_0}\, .
\eea
The prefactors $C^\alpha(\ell)$ depend on the details of
the quench and are listed in Appendix~\ref{app:statics}.

\subsection{Time dependence}
Having established the stationary behaviour of spin correlations, we
now turn to their dynamics at late times. The first question of
interest is how they relax towards their stationary values. 

For even operators we can use Wick's theorem to express
spin-spin correlators in terms of the fermion two-point functions
\eqref{eq:aat}. At late times the resulting expression can be evaluated
by a stationary phase approximation, which gives 
\be
\label{eq:Opeven}
\braket{\mathcal O_e}\simeq \braket{\mathcal O_e}_{\rm
  GGE}+O(t^{-\frac{n[\mathcal O_e]}{2}})\, .
\ee
Here the exponent $n[\mathcal O_e]$ is an integer that depends on the
particular even operator under consideration.

Expectation values of odd operators were argued in
Ref.~\onlinecite{FE_13a} to decay exponentially in time 
\be
\braket{\mathcal O_o}\simeq O(e^{-t/\tau[\mathcal O_o]})\, ,
\ee
where $\tau[\mathcal O_o]$ denotes a relaxation time. Having
established the gross structure of the late time dynamics, we now turn
to a more quantitative description.

\subsubsection{One-point functions}

\begin{figure}[ht]
\begin{center}
\includegraphics[width=0.45\textwidth]{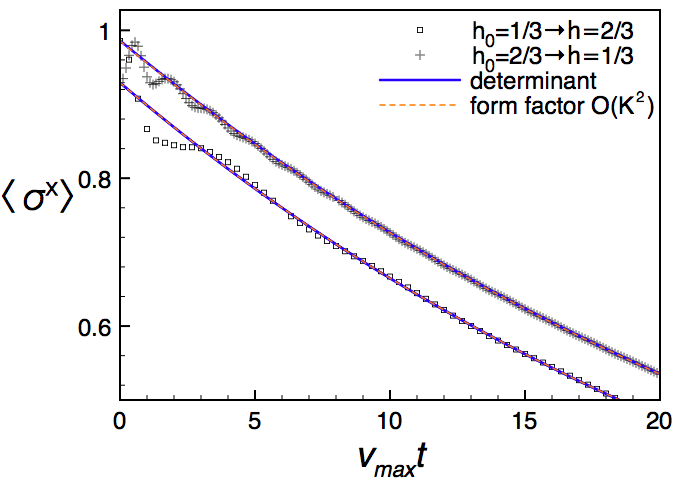}
\includegraphics[width=0.45\textwidth]{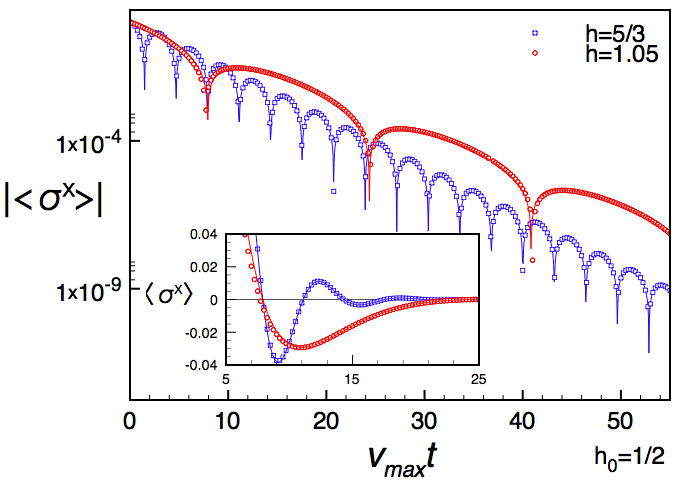}
\caption{
Expectation value of the order parameter after two quenches within the
ferromagnetic phase (left) and two quenches across the critical point
(right). Numerical results obtained in the thermodynamic limit are
compared with the asymptotic predictions \eqref{eq:sx1} (labelled as
``determinant''). In the left panel analytic results obtained by form
factor methods\cite{CEF2} are shown as well (labelled as ``form
factor''). The agreement in all cases is excellent even at the short
times depicted.
\emph{[Figures taken from Ref.~\onlinecite{CEF2}]} 
}
\label{fig:ordpar}
\end{center}
\end{figure}

\paragraph{Longitudinal spin operator.}
The longitudinal spin operator $\sigma_j^x=(-i)^{\ell-1}
\prod_{j=1}^{2\ell-1}a_j $ is the simplest and most important example
of an odd operator. Its expectation value is the order parameter in
the ferromagnetic phase. As we are dealing with an odd operator, its
expectation value is identically zero for quenches originating
in the paramagnetic phase. For quenches from the ferromagnetic phase
($h_0<1$) it was shown in Ref.~\onlinecite{CEF1,CEF2} that
\be
\label{eq:sx1}
\langle\Psi(t)|\sigma_j^x|\Psi(t)\rangle
\simeq C^x(t)e^{-t/\tau_x}\, ,\qquad h_0<1\ ,
\ee
where the inverse decay time is given by
\be
\tau_x^{-1}=\int_{0}^\pi
  \frac{\mathrm d k}{\pi} \varepsilon'_h(k) 
\ln\left |\frac{1-K^2(k)}{1+K^2(k)}\right|.
\ee
Here the function $K^2(k)$ has been previously defined in \eqref{eq:K2}.
The prefactor $C^x(t)$ was calculated in Ref.~\onlinecite{CEF2}
\be\label{eq:Cxt1}
C^x(t)=\left\{
\begin{array}{ll}
\sqrt{\frac{1-h h_0+\sqrt{(1-h^2)(1-h_0^2)}}{2\sqrt{1-h h_0}(1-h_0^2)^{\frac{1}{4}}}}&h<1\\
\Bigl[\frac{h\sqrt{1-h_0^2}}{h+h_0}\Bigr]^\frac{1}{4}\left[1+\cos(2\varepsilon_h(k_0)t+\alpha)+\ldots\right]^\frac{1}{2}&h>1\, .
\end{array}
\right.
\ee
The oscillatory behaviour for $h>1$ can be related to the presence
of a gapless mode with momentum $k_0$ in the GGE Hamiltonian
\eqref{eq:HGGE} ($K^2(k_0)=1$). It was noted in Ref.~\onlinecite{HPL:dyn13} 
that the period of the oscillations in \fr{eq:Cxt1} coincides with
cusps in the time evolution of the logarithm of the return probability
per unit length
$f(t)=-\lim_{L\rightarrow\infty}\frac{1}{L}\log|\braket{\Psi(0)|e^{-i
    H t}|\Psi(0)}|^2$ after quenches between the phases.
The result \fr{eq:sx1} has been derived in the late time limit $Jt\gg
1$. However, it gives an excellent account of the full answer except
at very short times. This can be shown by comparing \fr{eq:sx1} to a
numerical solution based on free fermion methods\cite{CEF2}. The
latter works directly in the thermodynamic limit and does not suffer
from finite-size effects. Fig.~\ref{fig:ordpar} shows such comparisons
for two different quenches within the ferromagnetic phase, and for two
quenches from the ferromagnetic to the paramagnetic phase. The
agreement is visibly excellent even at moderate times.

At first sight the exponential decay \fr{eq:sx1} of the order
parameter for quenches within the ferromagnetic phase may look
surprising. Even a very small quench will lead to the eventual
disappearance of the order parameter. A simple way of understanding
this is to note that the ferromagnetic order persists only at zero
temperature $T=0$, and melts for any $T>0$. By means of our quantum
quench we deposit a finite energy density into the system, which is
very similar to imposing a finite temperature. This consideration provides an
intuitive explanation for why even small quenches wipe out the long
range order present in the initial state. We note that this behaviour
is specific to one dimensional systems, where discrete symmetries can
be spontaneously broken only at $T=0$. In higher dimensional systems
we expect order parameters to be generally stable when subjected to
sufficiently small quantum quenches.

Finally we note that the expectation value $\braket{\sigma_j^y(t)}$
can be obtained in a simple way by considering the Heisenberg equations of
motion for $\sigma^x_j$, and is given by
\be
\langle\Psi(t)|\sigma_j^y|\Psi(t)\rangle=
\frac{\partial}{\partial t}
\frac{ \langle\Psi(t)|\sigma_j^x|\Psi(t)\rangle}{2J h}\, .
\ee

\paragraph{Transverse spin operator.}
The spin operator $\sigma_j^z=i a_{2j} a_{2j-1}$ is even and its
expectation value can be straightforwardly calculated with
free-fermion techniques \cite{BMD70,BM71b}. It decays like a $t^{-3/2}$ power
law in time towards its stationary value $\braket{\cdot}_{\rm GGE}$ 
\be
\langle\Psi(t)|\sigma_l^z|\Psi(t)\rangle\Big|_{J t\gg 1}=\braket{\sigma_l^z}_{\rm GGE}+\frac{h_0-h}{4\sqrt{\pi}(2 h J t)^{\frac{3}{2}}}\Bigl[\frac{\sin(4 J |1-h|t+\pi/4)}{\sqrt{|1-h|}|1-h_0|}-\frac{\sin(4 J (1+h)t-\pi/4)}{\sqrt{1+h}(1+h_0)}\Bigr]+O((J t)^{-\frac{5}{2}})\, .
\ee
The relaxation to the stationary value is only algebraic, in agreement with
\eqref{eq:Opeven}.  
\subsubsection{Spin-spin correlators in the ``space-time scaling limit''} 
A particularly useful way of describing the time dependence of
two-point functions after a quantum quench is by considering an
asymptotic expansion around the so-called \emph{space-time scaling
  limit}\cite{CEF1,CEF2}. The latter refers to the behaviour along a
particular ray in space-time
\be
t,\ell\to\infty\ ,\quad \frac{v_{\rm max}t}{\ell}=\kappa={\rm fixed}.
\ee
Here $v_{\rm max}={\rm max}_k\frac{d\varepsilon_h(k)}{dk}$ is the
maximal group velocity of elementary excitations of the post-quench
Hamiltonian.

 
\paragraph{Transverse spin-spin correlator}
In the space-time scaling limit, the asymptotic behavior of
$\rho^{zz}_c(\ell,t)$ can be evaluated by means of
Wick's theorem, followed by a stationary
phase approximation. The leading behaviour is a $t^{-1}$ power-law
decay, and the subleading corrections are power laws as well\cite{CEF2}
\be
\rho_c^{zz}\big(\ell= \frac{v_{\rm max}t}{\kappa},t\big)
\sim\frac{D^z(t)}{\kappa^2t}+o\big(t^{-1}\big)\ .
\label{STSR}
\ee
Here $D^z(t)$ is the sum of a constant contribution and oscillatory
terms with constant amplitudes.
\paragraph{Longitudinal spin-spin correlator}  
In the space-time scaling limit the order parameter two-point function
$\rho^{xx}(\ell,t)$ takes the form
\be\label{eq:predictionintr}
\rho^{xx}(\ell,t)\simeq {\cal C}^x(\ell, t)
\exp\Big[ \int_0^\pi \frac{\mathrm d k}{\pi}\ln\left|\frac{1-K^2(k)}{1+K^2(k)}\right|
\min\big(2\varepsilon'_h(k)t,\ell\big)\Big]\, .
\ee
The function ${\cal C}^x(\ell, t)$ has been determined in
Ref.~\onlinecite{CEF2}.
\begin{enumerate}
\item For quenches within the ferromagnetic phase, $h_0,h<1$, ${\cal
  C}^x(\ell, t)$ equals the constant denoted by ${\cal C}^x_{\rm FF}$ in
  \eqref{eq:CxFF}. For times smaller than the Fermi time
\be\label{eq:Fermi_t}
t_F=\frac{\ell}{2 v_{\rm max}}\, ,
\ee
\eqref{eq:predictionintr} equals the square of the one-point function
\eqref{eq:sx1}.
Thus, in the space-time scaling limit,  connected correlations {\it
  vanish identically} for times $t<t_F$ and begin to form only after
the Fermi time. We stress that this does not imply that the
connected correlations are exactly zero for $t<t_F$: in any model,
both on the lattice or in the continuum there are exponentially
suppressed terms (in $\ell$), which however vanish in the scaling limit.   
\item For quenches from the ferromagnetic phase to the paramagnetic
phase the prefactor is given by
\be
{\cal C}^x(\ell, t)={\cal C}^x_{\rm FP}\Bigl[1+\theta_H(t_F-t)\Bigl(\cos(2\varepsilon_h(k_0)t+\alpha)+\dots\Bigr)\Bigr]\, ,
\ee
where  ${\cal C}^x_{\rm FP}$ is the constant defined in \eqref{eq:CxFP}, while $k_0$ and $\alpha$ are the constants appearing in the one-point function~\eqref{eq:Cxt1}. 
For $t<t_F$, \eqref{eq:predictionintr} is simply the square of
the corresponding one-point function, which ensures that connected
correlations vanish for $t<t_F$ in the space-time scaling regime. We
note that the expression for $t<t_F$ is a conjecture\cite{CEF2}.
\item For quenches within the paramagnetic phase one has
\be
{\cal C}^x(\ell, t) \simeq {\cal C}^x_{\rm PP}(\ell)+
(h^2-1)^\frac{1}{4}\sqrt{4J^2h}\int_{-\pi}^\pi\frac{dk}{\pi}\frac{K(k)}{\varepsilon_k}
\sin(2t\varepsilon_k-k\ell)+\ldots \, ,
\label{cx}
\ee
where ${\cal C}^x_{\rm PP}(\ell)$ is the function defined in
\eqref{eq:CxPP}.  Eq. \fr{cx} constitutes the leading order in a
low-density expansion computed within the form-factor formalism. The
exact expression for a generic (not small) quench is not known.
\item For quenches from the paramagnetic to the ferromagnetic phase,
for $t>t_F$, ${\cal C}^x(\ell, t)$ is independent of time and is
given by ${\cal C}^x_{\rm PF}(\ell)$ of \eqref{CPFcos}. For $t<t_F$
the correlator is exponentially small and, to the best of our
knowledge, there are no analytic predictions for its behaviour. 
\end{enumerate}

\subsubsection{Long time asymptotics of connected spin-spin
  correlators at a fixed separation $\ell$} 
The late time asymptotics of spin-spin correlation functions at a
fixed separation $\ell$ between the spin operators was analyzed in
Ref.~\onlinecite{CEF2}. 
\begin{enumerate}
\item In the late time regime at fixed, large $\ell$, 
$\rho_c^{zz}(\ell,t)$ decays in a power law fashion to its stationary
value 
\be
\rho^{zz}_c(\ell,t)\sim 
\rho^{zz}_c(\ell,\infty)+\frac{E^z(t) \ell
  e^{-\ell/\tilde\xi_z}}{t^{3/2}}
+o\big(t^{-3/2}\big)\ .
\label{t_to_infty}
\ee
Here
$E^z(t)=\sum_{q=0,\pi}A_q\cos\bigl(2t\varepsilon_h(q)+\varphi_q\bigr)$
and the steady state value is exponentially small in~$\ell$:
$\rho^{zz}_c(\ell,\infty)\propto e^{-\ell/\xi_z}$. Crucially
one has 
(\emph{cf.} \eqref{eq:xiz})
\be
\tilde \xi_z^{-1}=\min(|\log h_0|,|\log h|)<\xi_z^{-1}\, .
\ee
This implies that the time scale after which the stationary behaviour
becomes apparent is in fact exponentially large in the separation
$\ell$. This makes the stationary behaviour difficult to observe in practice.
\item{} 
For quenches originating in the ferromagnetic phase, the stationary
value of $\rho^{xx}(\ell,t)$ emerges at a time scale 
\be 
\tau^{\rm x}_{\rm F}\sim v_{\rm max}^{-1}\ell^{4/3}\, ,
\ee 
where $v_{\rm max}$ is the maximal velocity at which information propagates.
This makes the approach to the steady state straightforward to observe.
\item{} For quenches within the paramagnetic phase, 
$\rho^{xx}_c(\ell,t)$ exhibits an oscillatory power-law decay in time
towards its stationary value, which is exponentially small in
$\ell$. Hence, in complete analogy to the case of the transverse
two-point function, the time scale $\tau^{\rm x}_{\rm PP}$ after which
the stationary behavior reveals itself is exponentially large 
\be
\tau^{\rm xx}_{\rm PP}\propto e^{2\ell/3\xi_x},
\ee
and very difficult to observe in practice.
\end{enumerate}


\subsection{Reduced density matrices}
As we have seen above, the quench dynamics of one and two point
functions of quantum spins is rich and interesting. However, these are 
nonetheless very special observables. 
Ideally one would like to have access to the full reduced density
matrix for a given subsystem size, as its matrix elements encode the
time evolution of all correlation functions, 
\emph{cf.} \fr{RDMt}. In practice this is only possible in very simple
non-interacting examples, or for very small subsystem sizes. An
example are quenches in the disordered phase of the
TFIC\cite{CE_PRL13,FE_13a}. Here the reduced density matrix on
the interval $[1,\ell]$ is given by
\be
\rho_\ell(t)=\frac{1}{Z} \exp\Bigl(\frac{1}{4}\sum_{\ell, n}a_l W_{lm} a_m\Bigr)\ ,
\label{quad}
\ee
where $a_{2n}$ and $a_{2n-1}$ are the Majorana fermion operators
\fr{majorana} and the factor $Z$ ensures ${\rm Tr}(
\rho_\ell(t))=1$. The matrix $W$ is related to the two-point function
of Majorana fermions by\cite{vidal03,peschel03,peschel} 
\be
\label{Gamma}
\tanh\frac{W}2=\Gamma\ ,\qquad
\Gamma_{jk} =\langle\Psi(t)|a_ka_j|\Psi(t)\rangle-\delta_{j,k}= -\Gamma_{kj},
\ee
where the time evolved initial state $|\Psi(t)\rangle$ is given by
\fr{eq:equiv}. The matrix elements of the correlation matrix are
simple (single) integrals and can be found in Ref.~\onlinecite{CEF2}.

\begin{figure}[ht]
\begin{center}
\includegraphics[width=0.45\textwidth]{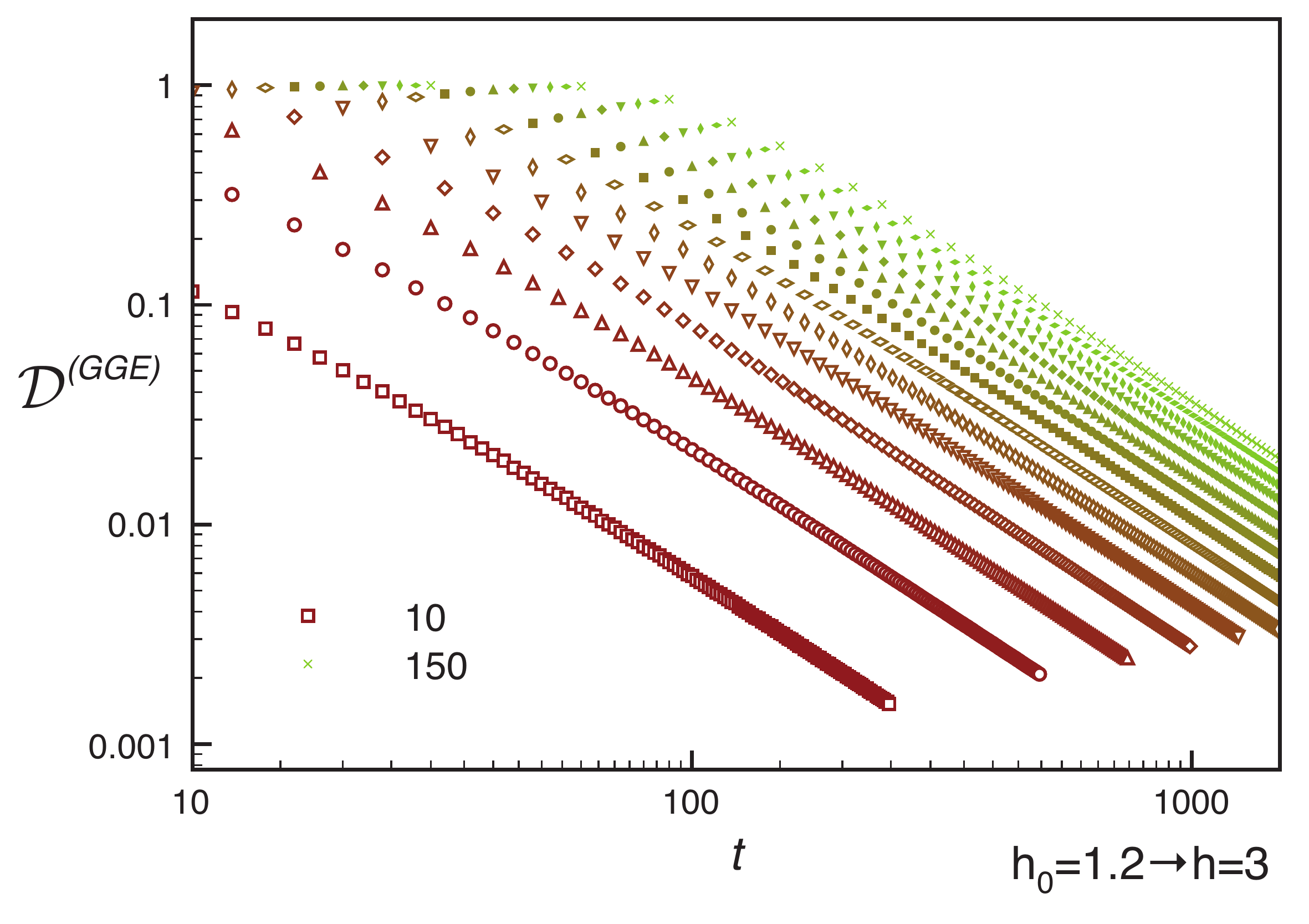}
\includegraphics[width=0.45\textwidth]{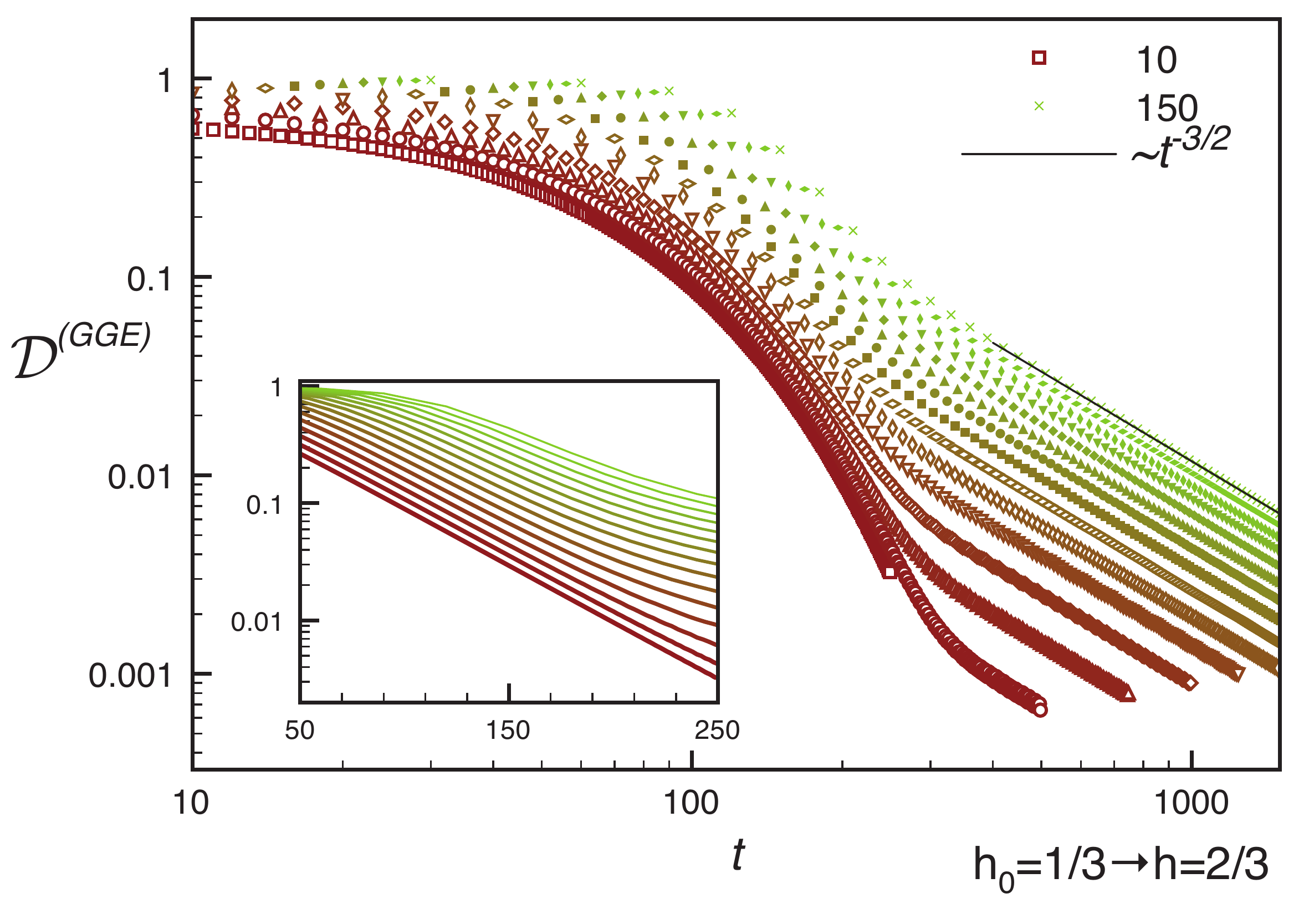}
\caption{
Normalized distance ${\mathcal D}^{(\rm GGE)}=\mathcal
D(\rho_\ell(t),\rho^{\rm GGE}_\ell)$ after a quench within the
paramagnetic (left) and ferromagnetic (right) phase for
subsystem sizes $\ell=10,20,\dots,150$. As $\ell$ increases, the color
changes from brown to green, the symbols become smaller and the curves
narrower. At late times $D(\rho_\ell(t),\rho^{(GGE)}_\ell)$ tends to zero 
in a universal power-law fashion $\propto (Jt)^{-3/2}$. For quenches
in the ordered phase there is an intermediate time regime, in which
the distance decays exponentially (inset). This stems from
the non-vanishing spontaneous magnetization in the initial state for
this quench. \emph{[Figures taken from Ref.~\onlinecite{FE_13a}]}}
\label{fig:para_gge}
\end{center}
\end{figure}

As we have argued above, at late times after global quantum quenches
isolated quantum systems relax locally towards some steady states
$\rho^{\rm SS}$. How quickly this relaxation occurs can be efficiently
measured by considering the distance of the time evolving reduced
density matrix $\rho_B(t)$ from its steady state value $\rho^{\rm
  SS}_B$, where $B$ is a subsystem of a given size. This diagnostic
can be implemented quite generally numerically as long as the
subsystem size is small\cite{strongweakt}. For models that can be
mapped to non-interacting theories, it is possible to go considerably
further. An example is the TFIC, which was considered in
Ref.~\onlinecite{FE_13a}. The first step is to introduce a
measure of distance on the space of RDMs on a given
subsystem. A convenient choice  was introduced
in Ref.~\onlinecite{FE_13a}
\be\label{eq:distance}
\mathcal
D(\rho_1,\rho_2)=
\sqrt{\frac{\mathrm{Tr}[(\rho_1-\rho_2)^2]}
{\mathrm{Tr}[(\rho_1)^2]+\mathrm{Tr}[(\rho_1)^2]}}\, . 
\ee
Fig.~\ref{fig:para_gge} shows results for the distance between
the RMDs for the time-evolving and stationary states for quantum
quenches within the disordered and ordered phases in the
TFIC. Subsystems consisting of $\ell$ neighbouring sites are
considered, where $\ell$ ranges from 10 to 150.  
In both cases the distance is seen to eventually decay as a power law in time
\be
\mathcal D(\rho_\ell(t),\rho_\ell^{\rm GGE})\propto \ell^2(J t)^{-\frac{3}{2}}\, .
\ee
The power-law decay of the distance in time can be interpreted by relating
it to expectation values of local operators in the subsystem. Using
\eqref{RDMt} one can show that
\be\label{eq:D}
\mathcal D(\rho_1,\rho_2)=\big(\overline{R(\mathcal O)^2}\big)^{\frac{1}{2}}\ ,\qquad
R(\mathcal O)=\frac{|\mathrm{Tr}[(\rho_1-\rho_2)\mathcal
    O]|}{\sqrt{\mathrm{Tr}[\rho_1 \mathcal O]^2+\mathrm{Tr}[\rho_2
      \mathcal O]^2}}\, ,
\ee
where the bar denotes an average on the space of operators acting
on the spins in the subsystem, taken with respect to the probability
distribution 
\be\label{eq:dist}
P(\mathcal O)=\frac{\mathrm{Tr}[\rho_1\mathcal
    O]^2+\mathrm{Tr}[\rho_2\mathcal O]^2}{\sum_{\mathcal
    O'}\mathrm{Tr}[\rho_1\mathcal O']^2+\mathrm{Tr}[\rho_2\mathcal
    O']^2}\ .
\ee
Here the sum is over all operators
$\sigma_1^{\alpha_1}\sigma_2^{\alpha_2}\dots
\sigma_\ell^{\alpha_\ell}$, where $\alpha_j=0,x,y,z$ and where we have
assumed for simplicity that the subsystem is the interval
$[1,\ell]$. The contribution from a given operator ${\cal O}$ to the
distance is weighted by the square of its expectation value. This
shows that \eqref{eq:D} measures a mean relative difference between
the expectation values of local operators in the two states.  

As $\overline{R(\mathcal O)}\leq (\overline{R(\mathcal
  O)^2})^{1/2}=\mathcal D(\rho_\ell(t),\rho_\ell^{\rm GGE})$ we may
use \fr{eq:D} to identify a time scale $t^*_{\rm rms}$ associated with
the relaxation of the ``typical'' operator (with respect to the
probability distribution \eqref{eq:dist})
\be
Jt^*_{\rm rms}\sim \ell_B^{\frac{4}{3}}\, .
\ee
The time scale $t^*_{\rm rms}$ is very different from the ones
governing the time evolution of the two-point functions of spin
operators. 

\subsection{Entanglement entropy}

\begin{figure}[ht]
\begin{center}
\includegraphics[width=0.75\textwidth]{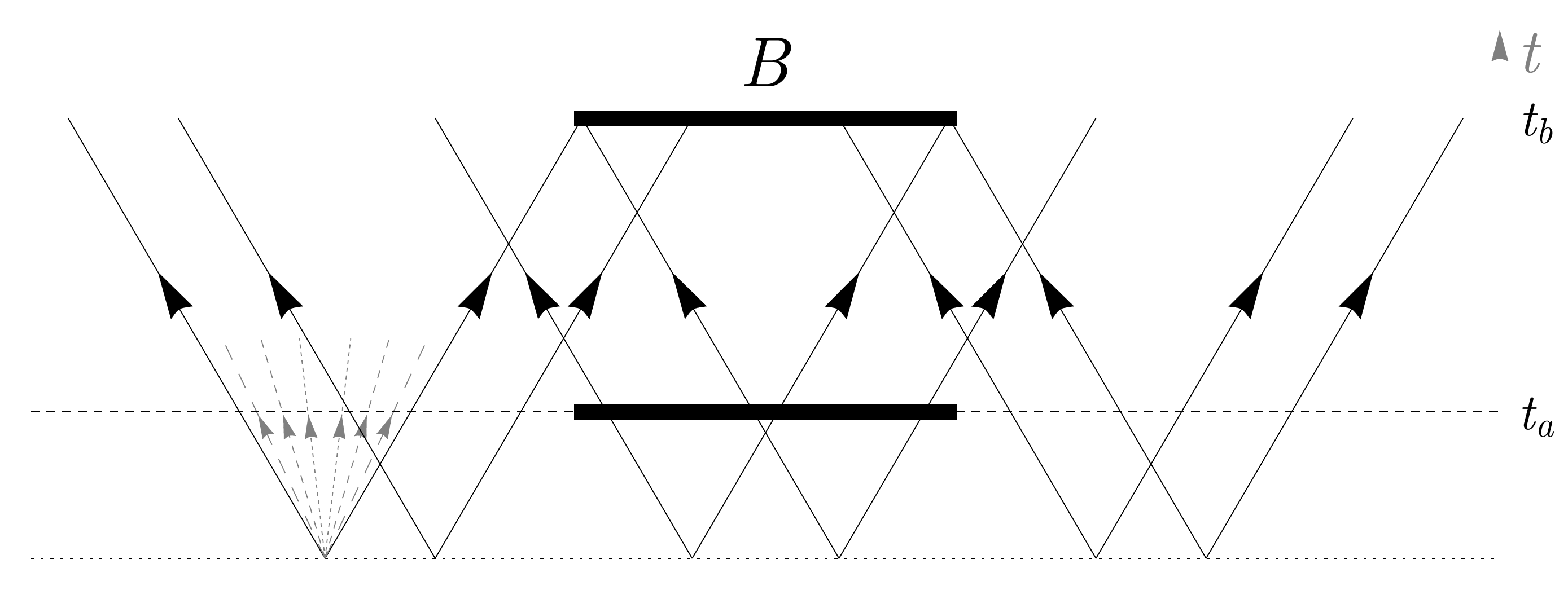}
\caption{
Space-time picture illustrating the semiclassical interpretation
of entanglement entropy growth after a global quantum
quench\cite{cc-05}. Quasi-particles moving at the maximal group
velocity are indicated by thick black arrows, and are initially
generated throughout the system by the quantum quench. Slower
quasi-particles are shown only on the left (short dashed gray
arrows). At time $t_a$ the entanglement entropy of $B$ is still
increasing linearly in time, as there are still trajectories such that
maximum velocity quasi-particle pairs are incident inside $B$.
At times $t=t_b$ the entanglement entropy starts saturating because
\emph{any} maximum velocity quasi-particle incident in $B$ generates
entanglement with the rest of the system.}\label{fig:SvN}
\end{center}
\end{figure}

The von Neumann entropy (also known as entanglement entropy) of a
density matrix $\rho $ is defined as 
\be
S_{\rm vN}[\rho]=-\mathrm{Tr}[\rho\log \rho]\, .
\ee
If $\rho$ is a reduced density matrix in a system that is in a pure
state, $S_{\rm vN}$ measures the entanglement between the subsystem
and its complement. Entanglement entropies have become a standard
diagnostic for detecting and identifying quantum phase transitions.
In the context of quantum quenches the time evolution of the von
Neumann entropy and other entanglement measures provides very useful
information about the spreading of correlations\cite{cc-05,
  FC08,dechiara06,laeuchli08,jurcevic14,preT_entropy,enttrap,qentexc,ISL:entdis12,TT:spectrum14,CTC:neg14,NR:harm14}. A
key result obtained in Ref.~\onlinecite{FC08} is that after quenches
to conformal field theories the von Neumann entropy of a subsystem
of length $\ell$ increases linearly in time until it eventually
saturates (see the review by P. Calabrese and J. Cardy~\cite{CCrev} in this volume)
\be
S_{\rm vN}[\rho]\sim\begin{cases}
\frac{\pi c vt}{6\epsilon}&vt<\frac{\ell}{2}\\
\frac{\pi c \ell}{12\epsilon}&vt\gtrsim\frac{\ell}{2}\, .
\end{cases}
\label{SCFT}
\ee
Here $c$ is the central charge of the CFT, $\epsilon$ is a constant
with dimensions of length that depends on the initial state and $v$ is
the speed of light. The two behaviours in \fr{SCFT} connect smoothly
over a region $|vt-\frac{\ell}{2}|\sim\epsilon$.  
A physical interpretation of the result \fr{SCFT} is provided by the
Calabrese-Cardy quasi-particle picture\cite{cc-05} we already
encountered in Section~\ref{sec:spreading}. Its application to
the time evolution of the entanglement entropy in integrable models
initialized in squeezed states with finite correlation lengths proceeds
as follows. The idea is that in a squeezed state correlations spread
via the propagation of pairs of quasi-particles with equal but
opposite momenta. At time $t=0$ the quantum quench generates such
quasi-particles pairs throughout the system. Correlations between
quasi-particles produced at a distance larger than the correlation
length in the initial state can be neglected. The entanglement between
a given region $B$ and its complement is generated by quasi-particle
pairs. The entanglement entropy is interpreted as a measure of the
number of correlated pairs such that, at a given time, one
quasi-particle is inside $B$ and one outside, see
Fig.~\ref{fig:SvN}. Entanglement is initially generated 
at the boundaries of $B$, and the entangled region spreads outwards in
the form of two light cones. This picture suggests the following
semiclassical expression for the von Neumann entropy
\be\label{eq:Svn-sc}
S_{\rm v N}[\rho_B]\overset{\rm sc}{=}\int \mathrm d k\  f(k)\min\Bigl(\ell_B, 2 |v(k)|t\Bigr)\, .
\ee
Here $v(k)$ is the semiclassical group velocity $v(k)=\frac{\mathrm d
  \varepsilon(k)}{\mathrm d k}$, $\varepsilon(k)$ is the dispersion
relation of the quasiparticles and $f(k)$ is an unknown function that
contains information on the initial state.

This behaviour has been observed in a variety of lattice and
continuum
models~\cite{cc-05,FC08,dechiara06,laeuchli08,preT_entropy,enttrap,qentexc,NR:harm14,ISL:entdis12},
including non-integrable cases~\cite{laeuchli08}, in which the
quasi-particle picture does not apply by virtue of the finite
quasi-particle life time. The semiclassical interpretation has been
generalized to prethermalized regimes in models with weak
integrability breaking\cite{preT_entropy}. 

The linear entanglement growth after quantum quenches has important
ramifications. It is a crucial limiting factor for applying
matrix-product state methods such as t-DRMG~\cite{tDMRG} and i
iTEBD~\cite{iTEBD} algorithms to the computation of the dynamics after
global quantum quenches. We note that translational invariance is essential
for the quasi-particle picture to hold, and the time evolution of the
entanglement entropy in e.g. disordered models \cite{ISL:entdis12}
is very different.

Exact results for the evolution of the entanglement entropy after
quenches in the transverse field Ising chain\cite{FC08} are in
accordance with the structure \fr{eq:Svn-sc} suggested by the
quasi-particle picture. In the limit $1\ll \ell,J t$, the entanglement
entropy of a block of $\ell$ neighbouring spins is
\begin{equation}\label{eq:SvNfree}
S_{\rm v N}[\rho_B]\simeq \int_{\pi}^\pi \frac{\mathrm d k}{2\pi}\ w(\braket{\Psi(0)|n(k)|\Psi(0)})\ \min\Bigl(\ell, 2 |\varepsilon'(k)|t\Bigr)+o(\ell)\, ,
\end{equation}
where $w(x)=-x\log x-(1-x)\log(1-x)$ is the entropy per site
(\emph{cf.} Appendix~\ref{app:typical}), and
$\braket{\Psi(0)|n(k)|\Psi(0)}$ is the (conserved) density of elementary
excitations of the post-quench Hamiltonian $H$ with momentum k at
times $t>0$. It is given by
\be
\braket{\Psi(0)|n(k)|\Psi(0)}=\frac{K^2(k)}{1+K^2(k)}\, ,
\ee
where $K^2(k)$ was defined previously in \eqref{eq:K2}. It follows from
\eqref{eq:SvNfree} that the entanglement entropy increases linearly
until the Fermi time \eqref{eq:Fermi_t}, and then slowly approaches
its stationary value set by the GGE. The latter equals the entropy per
site of the GGE for the \emph{entire} system~\cite{enttrap} 
\be
\int_{\pi}^\pi \frac{\mathrm d k}{2\pi}\
w(\braket{\Psi(0)|n(k)|\Psi(0)})=\lim_{L\rightarrow\infty}
\frac{1}{L}S_{vN}[\rho^{\rm GGE}] 
\, .
\ee 
These observations persist for quantum quenches in the TFIC
starting in excited states~\cite{qentexc}.  

Interestingly, the stationary value of the entropy density in the
diagonal ensemble differs from that in the GGE\cite{F:13a, G:14}. This
is not a problem, because the entropy per site is a global property of
the system, while the equivalence between ensembles only holds for
(finite) subsystems in the thermodynamic limit. A detailed explanation
of the origin of this difference for the case of the TFIC was provided
in Ref.~\onlinecite{F:13a}. 

\subsection{Dynamical spin-spin correlation functions}
We now turn to dynamical correlation functions
\be
\rho^{\alpha\alpha}(\ell,t+t_1,t+t_2)=\braket{\Psi(t)|\sigma_{j+\ell}^\alpha(t_1)
\sigma_{j}^\alpha(t_2)|\Psi(t)}\, .
\label{dyncorr}
\ee
The transverse correlator ($\alpha=z$ in \fr{dyncorr}) can be
calculated by elementary means~\cite{BMD70}, as $\sigma^z_j$ is
quadratic in Jordan-Wigner fermions. The order parameter correlator 
($\alpha=x$ in \fr{dyncorr}) is much more difficult to evaluate. In
Ref.~\onlinecite{EEF:12} it was determined by means of a
generalization of the form factor methods developed in 
Ref.~\onlinecite{CEF2}, and by exploiting exact results in
particular limits\cite{CEF3}. These methods are so far restricted to
quenches within either the paramagnetic or the ferromagnetic phase,
and lead to answers of the form\cite{EEF:12}
\be
\rho^{xx}(\ell,t+\tau,t+t_2)\simeq C^{x}(\ell,\tau,t)\exp\Bigg[\int_0^\pi\frac{\mathrm d k}{\pi}\log\Bigl|\frac{1-K^2(k)}{1+K^2(k)}\Bigr|\min\left\{\max\Bigl[\varepsilon'_h(k)\tau,\ell\Bigr],\varepsilon'_h(k)(2t+\tau)\right\}\Bigg]\, .
\ee
Here the functions $\epsilon_h(k)$ and $K^2(k)$ are given in
\fr{eq:energy} and \fr{eq:K2} respectively, and $h_0$ and $h$ are the
initial and final values of the transverse field. The function
$C^{(x)}(\ell,\tau,t)$ depends on the phase in which the quench is
performed:
\begin{enumerate}
\item In the ferromagnetic phase, $h_0, h<1$, $C^{x}(\ell,\tau,t)$
equals  the constant ${\cal C}^x_{\rm FF}$ in \eqref{eq:CxFF}. 
\item In the paramagnetic phase ($h_0, h>1$) $C^{x}(\ell,\tau,t)$ is
given (to leading order in the form factor expansion) by\cite{EEF:12}
\bea
C^{x}(\ell,\tau,t)&=&\int_{-\pi}^\pi\frac{\mathrm d k}{2\pi}\frac{J e^{i\ell
    k}}{\varepsilon_h(k)}\Bigl[e^{-i\varepsilon_h(k)t}+2 i
  K(k)\cos(2\varepsilon_h(k)(2
  t+\tau))\mathrm{sgn}(\ell-\varepsilon_h'(k)\tau)\Bigr]\nn
&\times&\frac{h h_0-1+\sqrt{(h^2-1)(h_0^2-1)}}
{\sqrt{h_0-1/h}\sqrt[4]{h_0^2-1}}\ ,\qquad \ell<v_{\rm max}(2t+\tau).
\eea
In the complementary regime $v_{\rm max}(2t+\tau)<\ell$ the correlator
is exponentially small and this expression no longer applies.  
\end{enumerate}
The form factor result gives an excellent approximation to the exact
answer (which can be computed numerically using free fermion
techniques) for ``small'' quenches\cite{EEF:12}. These are defined as
being characterized by having low densities of excitations in the
initial state.

It is possible to obtain some of these results in an alternative way
by generalizing the semiclassical approach of Ref.~\onlinecite{RI:sc11} 
to the non-equal time case, and then elevating it using exact limiting
results derived in Refs~\onlinecite{CEF1,CEF2}. This method fails to reproduce
the result for quenches in the disordered phase outside the light cone
$v_{\rm max}\tau<\ell$, but is significantly simpler.


\section{Relaxation in interacting integrable models}
\label{sec:interacting}
We now turn to \emph{interacting} integrable models that are solvable
by the Bethe Ansatz\cite{Korepinbook,book,Gaudin}. By ``interacting''
we mean theories in which the scattering matrix is momentum
dependent. Most of our discussion will focus on the example of the
spin-1/2 Heisenberg chain.

\subsection{The ``initial state problem''}\label{ssec:initialstate}
Integrability allows the construction of a basis of simultaneous
eigenstates of the Hamiltonian and all its conservation laws. Unlike
in the non-interacting case these states have a very complicated
structure described by the Bethe
Ansatz\cite{Korepinbook,book,Gaudin}. 
When we consider a quench between two integrable
Hamiltonians $H(h_0)\rightarrow H(h)$, we are thus faced with
the problem of how to translate between the eigenbases of the two
integrable theories. This is a difficult
undertaking\cite{SFM:Z-F12,ini_state}, and no general formalism for
achieving it is currently known. Progress has however been made in cases
where the initial state has a simpler structure, in particular for
(matrix) product states in either
position\cite{FE_13b,FCEC_14,GGE_XXZ_Amst,XXZung,QANeel,XXZunglong} or 
momentum/rapidity
space\cite{FM_NJPhys10,sine-Gordon,KSCCI,nwbc-13,quenchLL,rel_int,bound_Bose}. 
There are two main methods for encoding the relevant information
contained in the initial state.
\begin{enumerate}
\item{} Let us denote the eigenstates of the post-quench Hamiltonian $H(h)$ by
$|n\rangle$. One way to implement the initial conditions is via the
\emph{overlaps} $\langle n|\Psi(0)\rangle$\cite{p-13a}. If these are known, the
initial state can be translated into the eigenbasis of the time
evolution operator. This method is used in the Quench Action
Approach\cite{CE_PRL13} (see the review by J.-S. Caux~\cite{Crev} in this volume).
\item{} Let us denote the local conservation laws of $H(h)$ by
$\{I^{(n)}\}$. If the set $\{I^{(n)}\}$ is in some sense complete\cite{Pozs_qB,GGEqft,GA:fail14},
then the initial conditions can be encoded in the constraints
\be
\lim_{L\to\infty}\frac{\langle\Psi(0)|I^{(n)}|\Psi(0)\rangle}{L}=
\lim_{L\to\infty}\frac{{\rm Tr}\big(\rho^{\rm SS}I^{(n)}\big)}{L}\ ,
\label{constraints}
\ee
where $\rho^{\rm SS}$ is one of the density matrices (GGE, GMC,
diagonal ensemble) that describes the local properties of the
stationary state.  
\end{enumerate}
In the following we will discuss implementations of the second approach.
\subsection{On mode occupation operators}
In free theories a convenient way for constructing the GGE is by
exploiting the linear relation between the local conservation laws and the
mode occupation operators, \emph{cf.} \fr{eq:I+_}. In interacting
integrable models the situation is different. Like free theories they
feature stable excitations. In the thermodynamic limit these can be
described by creation and annihilation operators
$Z^\dagger_a(\lambda)$, $Z_a(\lambda)$ (the index $a$ labels different
particle species) fulfilling the Faddeev-Zamolodchikov
algebra\cite{Zam,Fadd}   
\bea
Z_{a}(\lambda_1)Z_{b}(\lambda_2)&=&
S_{ab}^{cd}(\lambda_1,\lambda_2)Z_{d}(\lambda_2)Z_{c}(\lambda_1),\nn
Z_{a}(\lambda_1)Z^\dagger_{b}(\lambda_2)&=&2\pi\delta(\lambda_1-\lambda_2)
\delta_{a,b}+S_{bc}^{da}(\lambda_2,\lambda_1)Z^\dagger_{d}(\lambda_2)Z_{c}(\lambda_1).
\eea
Here $\lambda$ parametrizes the momenta $p_a(\lambda)$ and
$S_{ab}^{cd}(\lambda_1,\lambda_2)$ is the purely elastic two-particle
S-matrix. The generalized mode occupation operators
$N_a(\lambda)=Z^\dagger_a(\lambda)Z_a(\lambda)$ then indeed provide a
set of mutually commuting conserved charges. The problem is that, due
to the interacting nature of the stable excitations, there is no
simple way of defining such operators in the finite
volume\cite{sine-Gordon}, which is the standard way of
making the theory well defined (working directly in the thermodynamic
limit requires the regularization of very complicated
singularities\cite{EK:finiteT,SE:Ising,sine-Gordon}, which appears 
impractical in general). The problem lies in the nature of the
quantization conditions in the finite volume, which on a ring of
length $L$ read  
\be
e^{iLp_a(\lambda^{(a)}_j)}=-\prod_{b,k}S_{ab}(\lambda^{(a)}_j,\lambda^{(b)}_k)\ .
\label{quantcond}
\ee
The solutions to this complicated system of coupled equations are
such that the possible values of $\lambda^{(a)}_j$, and hence
$p_a(\lambda^{(a)}_j)$, depend in a very sensitive way on all the other
particles present in a given excitation. This is fundamentally
different from the non-interacting case, where the momenta are simply
given by
\be
p_a(\lambda^{(a)}_j)=\frac{2\pi}{L}\times \text{integer},
\ee
and are \emph{independent} of the particle content of a given
excitation. This makes it clear that defining finite volume analogues of
$N_a(\lambda)$ is difficult.

\subsection{The spin-1/2 Heisenberg model}
Our paradigm for an interacting integrable model will be the spin-1/2
Heisenberg XXZ chain. Its Hamiltonian on a ring with $L$
sites is
\be
H_{\rm XXZ}=\frac{J}{4}\sum_{j=1}^L \sigma^x_j\sigma^x_{j+1}+\sigma^y_j\sigma^y_{j+1}+\Delta
\left[\sigma^z_j\sigma^z_{j+1}-1\right]\ ,
\label{HXXZ}
\ee
where we will assume for definiteness that 
\be
\Delta=\cosh(\eta)\geq 1.
\ee
From now on we set $J=1$.

\subsubsection{Generalized microcanonical ensemble}
\label{GMC_XXZ}
For interacting integrable models the GMC is easier to work with than
the GGE. It is based on working with macro-states obtained by taking
the thermodynamic limit of eigenstates constructed from the Bethe
Ansatz. This procedure is an essential ingredient of the Thermodyamic
Bethe Ansatz and is reviewed in several
monographs\cite{Takahashibook,book}. A very brief summary is given in
Appendix~\ref{app:TBA}. The upshot is that macro-states in integrable
models are characterized by an (infinite) set of densities
$\{\rho_{n,p}(\lambda)|n=1,2,\dots\}$, where $n$ labels all distinct
stable species of excitations in the model. A given macro-state
corresponds to a set of micro-states $|\Phi\rangle$, called
\emph{representative states} in Ref.~\onlinecite{CE_PRL13}. 
These are by construction simultaneous eigenstates of all local
conservation laws. For the macro-state describing the stationary state
after our quench, they satisfy the initial conditions
\be
\lim_{L\to\infty}\frac{\langle\Phi^{\rm SS}|I^{(n)}|\Phi^{\rm
    SS}\rangle}{L}=
\lim_{L\to\infty}\frac{\langle\Psi(0)|I^{(n)}|\Psi(0)\rangle}{L}\ .
\label{repstate}
\ee
The GMC density matrix is then defined\cite{CE_PRL13} in terms of a
single such ``representative'' micro-state $|\Phi^{\rm SS}\rangle$ 
\be
\rho^{\rm GMC}=|\Phi^{\rm SS}\rangle\langle\Phi^{\rm SS}|\ .
\label{rhoGMC}
\ee
Here we have assumed that the stationary state is given in terms of a
single macro-state constructed from the Bethe Ansatz. In principle it
is possible that the steady state has a more complicated structure and
requires a description in terms of a sum of several density matrices
of the form \fr{rhoGMC}. 

\subsubsection{Transfer matrix and ``ultra-local'' conservation laws}
\begin{figure}[ht]
\begin{center}
(a)\includegraphics[width=0.22\textwidth]{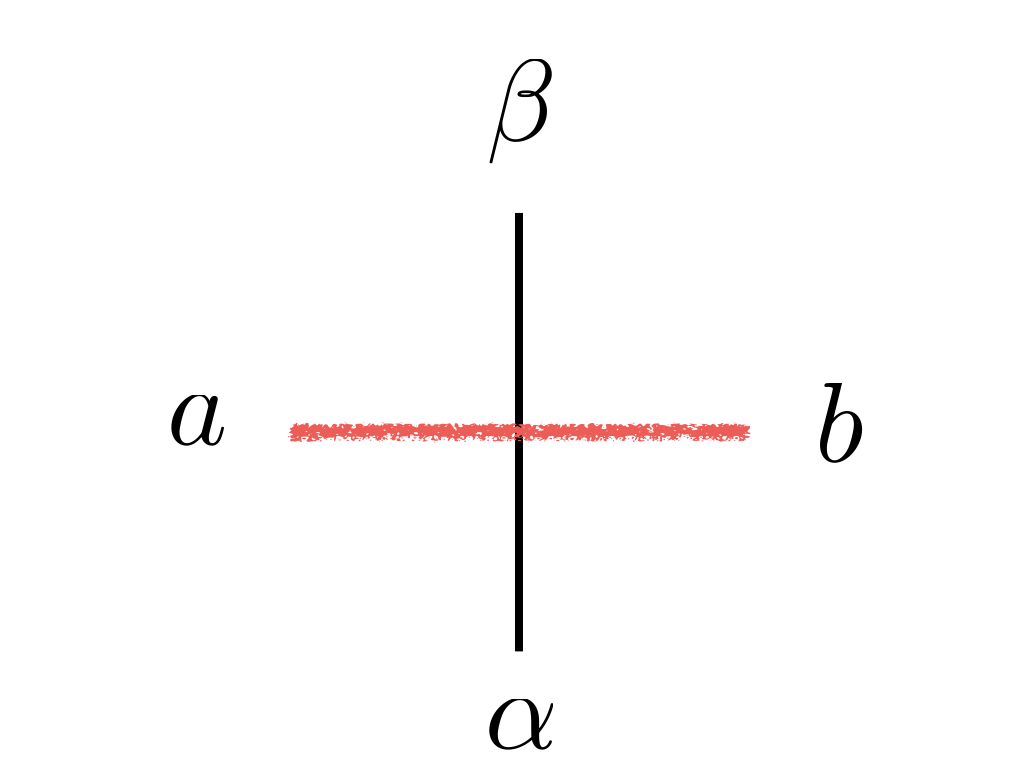}\qquad
(b)\includegraphics[width=0.7\textwidth]{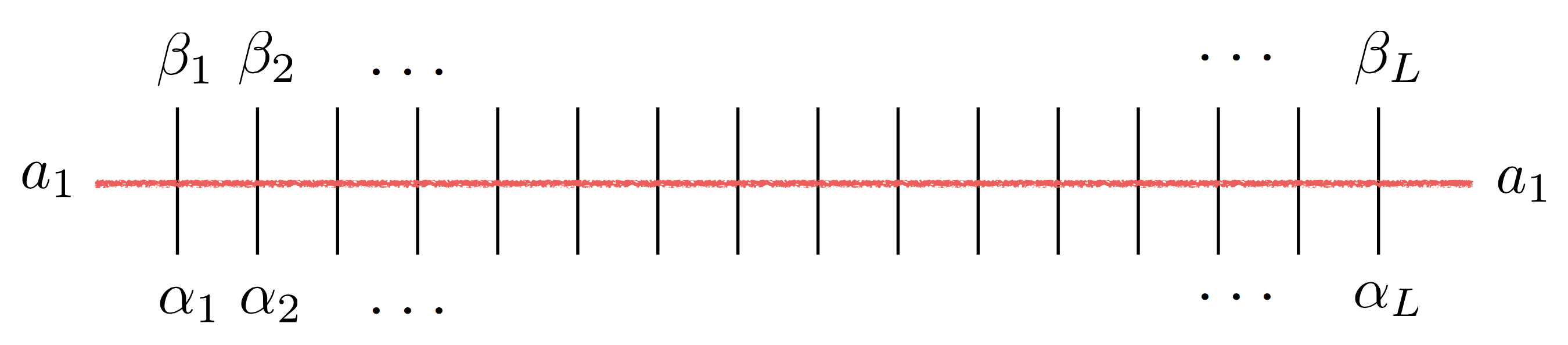}
\caption{(a) Vertex with weight
$\left[L_n(\lambda)\right]_{\alpha\beta}^{ab}$. The horizontal and
vertical lines are associated with the ``auxiliary'' and ``quantum''
spaces respectively. (b) Transfer matrix element
$\tau_{\frac{1}{2}}(\lambda)^{\beta_1\dots\beta_L}_{\alpha_1\dots\alpha_L}$. 
}
\label{fig:6-vertex}
\end{center}
\end{figure}

According to our general discussion, local observables should relax to
an appropriate GGE after quenches to the XXZ chain. In order to
construct this GGE, we need to know the required set of local
conservation laws of \fr{HXXZ}. One family of conservation laws has
been known for a long time and is most conveniently constructed by
exploiting the relation of the Heisenberg Hamiltonian to the transfer
matrix of the six-vertex model \cite{6vertex,FT84,Korepinbook}.
The fundamental building block of the six-vertex model is the L-operator
\be
L_n(\lambda)=\frac{1}{\sinh(\eta+i\lambda)}
\Big[\sinh(\frac{\eta}{2}+i\lambda) \cosh{(\frac{\eta}{2})}
+\cosh(\frac{\eta}{2}+i\lambda) \sinh{(\frac{\eta}{2})}\tau^z\sigma_n^z
+\sinh (\eta) (\tau^{-}\sigma_n^{+} +\tau^{+}\sigma_n^{-})\Big],
\ee
which acts on the tensor product $\mathbb{C}^2\otimes\mathbb{C}^2$
of ``auxiliary'' and ``quantum'' spaces through the Pauli matrices
$\tau^\alpha$ and $\sigma^\alpha$ respectively. Matrix elements 
in the auxiliary/quantum spaces are denotes by Roman/Greek letters
respectively, e.g. 
\be
\left(\tau^-\sigma_n^+\right)^{ab}_{\alpha\beta}=
\tau^-_{ab}\big(\sigma_n^+\big)_{\alpha\beta}.
\ee
The vertex
weights of the six-vertex model are obtained by taking matrix elements
$L(\lambda)^{ab}_{\alpha\beta}$ and have a graphical representation as
shown in Fig.~\ref{fig:6-vertex} (a).
The row-to-row transfer matrix is obtained as shown in
Fig.~\ref{fig:6-vertex} (b)
\be
\big(\tau_{\frac{1}{2}}(\lambda)\big)^{\beta_1\dots\beta_L}_{\alpha_1\dots\alpha_L}
=\big(L_1(\lambda)\big)^{a_1a_2}_{\alpha_1\beta_1}
\big(L_2(\lambda)\big)^{a_2a_3}_{\alpha_2\beta_2}\dots
\big(L_L(\lambda)\big)^{a_La_1}_{\alpha_L\beta_L}\ .
\label{tau}
\ee
The partition function of the 6-vertex model on an $L\times M$ rectangular
lattice with periodic boundary conditions is then 
\be
Z_{\rm 6-vertex}={\rm Tr}\left[\big(\tau_{\frac{1}{2}}(\lambda)\big)^M\right] ,
\ee
where the trace is over the quantum space. As a consequence of the
Yang-Baxter relation for the L-operators~\cite{Korepinbook}, the
transfer matrices form a commuting family 
\be
[\tau_\frac{1}{2}(\lambda),\tau_\frac{1}{2}(\mu)]=0.
\label{commuting}
\ee
The Heisenberg Hamiltonian is related to the transfer matrix by taking
a logarithmic derivative
\be
H_{\rm XXZ}=-i\frac{\sinh\eta}{2}\frac{\partial}{\partial\lambda}\Big|_{\lambda=0}
\ln\left[\tau_{\frac{1}{2}}(\lambda)\right].
\ee
By virtue of the commutation relations \fr{commuting} it is clear that
a set of mutually commuting operators can be obtained by taking higher
derivatives, i.e.
\be
H^{(\frac{1}{2},k)}=i\left(-\frac{\sinh\eta}{2}\frac{\partial}{\partial\lambda}
\right)^k \Big|_{\lambda=0} \ln\left[\tau_{\frac{1}{2}}(\lambda)\right].
\label{higher}
\ee
Crucially, these conservation laws have the form
\be
H^{(\frac{1}{2},k)}=\sum_j H^{(\frac{1}{2},k)}_{j,j+1,\dots,j+k}\ ,
\ee
where the densities $H^{(\frac{1}{2},k)}_{j,j+1,\dots,j+k}$ act
non-trivially only on the $k+1$ consecutive sites
$j,j+1,\dots,j+k$. These conservation laws are sometimes referred to
as \emph{ultra-local}. They have been studied extensively in the
literature \cite{Korepinbook,GM:chain,higherCL,takahashi}. We note
that the above construction is not restricted to Heisenberg models,
but works much more generally \cite{Korepinbook}.
\subsubsection{``Ultra-local'' GGE}
According to our general discussion, the GGE describing the steady
state after a quench to the Heisenberg model should contain all of the
conservation laws \fr{higher}. An important question is whether these
conservation laws are also sufficient. This was investigated in
Refs~\onlinecite{Pozsgay:13a,FE_13b,FCEC_14}. The basic idea is as
follows. One considers time evolution induced by the Hamiltonian \fr{HXXZ}
starting from an initial state $|\Psi(0)\rangle$. The quantities of
interest are the matrix elements of the reduced density matrix on a
short interval in the steady state, \emph{cf.} \fr{RDMt}
\be
g_{\alpha_1,\dots,\alpha_n}=\lim_{t\to\infty}\lim_{L\to\infty}
\langle\Psi(t)|\sigma^{\alpha_1}_1\sigma^{\alpha_2}_{2}\dots\sigma^{\alpha_n}_{n}
|\Psi(t)\rangle \ .
\ee
The question is whether these expectation values can be obtained, to a
given accuracy, from a GGE density matrix of the form
\be
\rho^{(y)}_{\rm ulGGE}=\frac{1}{Z_{\rm ulGGE}}
\exp\big(\displaystyle{-\sum_{k=1}^y\lambda^{(y)}_{k}H^{(k,\frac{1}{2})}}\big)\ ,
\label{rhoy}
\ee
which takes into account the first $y$ conservation laws in the series
\fr{higher}. The Lagrange multipliers $\lambda^{(y)}_k$ are in
principle fixed by the requirements
\be
\langle\Psi(0)|H^{(k,\frac{1}{2})}|\Psi(0)\rangle={\rm Tr}\left(
\rho^{(y)}_{\rm ulGGE}H^{(k,\frac{1}{2})}\right)\ ,\quad
k=1,\dots,y.
\label{initialvalues}
\ee
In practice it is very difficult to determine the Lagrange multipliers
from these conditions, even for very simple initial states
$|\Psi(0)\rangle$. A method to circumvent this problem
was developed in Ref.~\onlinecite{FE_13b}.  The idea is to define a
generating function for the initial values \fr{initialvalues}
\bea
\Omega_{\Psi(0)}^{(\frac{1}{2})}(\lambda)&=&
\lim_{L\to\infty}\frac{i}{L}
\langle\Psi(0)|\tau'_\frac{1}{2}(\lambda)\tau^{-1}_\frac{1}{2}(\lambda)|\Psi(0)\rangle\nn
&=&\lim_{L\to\infty}\frac{1}{L}\sum_{k=1}\left(-\frac{2}{\sinh\eta}\right)^k\frac{\lambda^{k-1}}{(k-1)!}
\frac{\langle\Psi(0)|H^{(k,\frac{1}{2})}|\Psi(0)\rangle}{L}.
\label{Omega}
\eea
For large $L$ (and real $\lambda$) the inverse of the transfer
matrix becomes
\be
\tau^{-1}_\frac{1}{2}(\lambda)\simeq\left(\tau_\frac{1}{2}(\lambda)\right)^\dagger
=\left[\frac{\sinh\big(-i\lambda\big)}
{\sinh\big(\eta-i\lambda\big)}\right]^L
\tau_{\frac{1}{2}}(\lambda+i\eta).
\ee
Using the expression \fr{tau} of the transfer matrix as a product of
L-operators, the generating function can thus be expressed in the form
\be
\Omega_{\Psi(0)}^{(\frac{1}{2})}(\lambda)=
\lim_{L\to\infty}\frac{i}{L}\frac{\partial}{\partial
  x}\Big|_{x=\lambda} {\rm Sp}\langle\Psi(0)|V_L(x,\lambda)\dots
V_1(x,\lambda)|\Psi(0)\rangle, 
\label{V}
\ee
where $V_n(x,\lambda)$ are $4\times 4$ matrices with entries
$\left(V_n(x,\lambda)\right)^{ab}_{cd}$ that are operators acting on
the 2-dimensional quantum space at site $n$, and ${\rm Sp}$ denotes
the usual trace for $4\times 4$ matrices. The explicit expression is
\be
\left[\left(V_n(x,\lambda)\right)^{ab}_{cd}\right]_{\alpha_n\beta_n}=
\frac{\sinh\big(-i\lambda\big)}
{\sinh\big(\eta-i\lambda\big)}\sum_{\gamma_n}
\big(L_n(x)\big)^{ab}_{\alpha_n\gamma_n}\big(L_n(\lambda+i\eta)\big)^{cd}_{\gamma_n\beta_n}\ .
\ee
The advantage of representation \fr{V} is that it can be efficiently
evaluated for initial states $|\Psi(0)\rangle$ of  matrix product form
\cite{FE_13b,FCEC_14}. To understand the principle behind this let us
consider a translationally invariant product state 
\be
|\Psi(0)\rangle=\otimes_{j=1}^L|\psi\rangle_j\ .
\ee
In this case the generating function is obtained from the eigenvalues
of the $4\times 4$ matrix $U(x,\lambda)=
{}_1\langle\psi|V_1(x,\lambda)|\psi\rangle_1$ as
\be
\Omega_{\Psi(0)}^{(\frac{1}{2})}(\lambda)=
\lim_{L\to\infty}\frac{i}{L}\frac{\partial}{\partial
  x} \Big|_{x=\lambda} {\rm  Sp}\left(U(x,\lambda)^L\right).
\ee
An efficient algorithm for calculating
$\Omega_{\Psi(0)}^{(\frac{1}{2})}(\lambda)$ for matrix-product states
was given in Ref.~\onlinecite{FCEC_14}.
Having encoded the ``initial data'' of our quantum quench in the
generating function $\Omega_{\Psi(0)}^{(\frac{1}{2})}(\lambda)$, we
now move on to the calculation of expectation values of local
operators in the state given by \fr{rhoy}. A very useful observation
is that $\rho^{(y)}_{\rm ulGGE}$ can be viewed as a Gibbs ensemble for
the ``Hamiltonian'' 
\be
H_{\rm eff}=\sum_{k=1}^y\lambda_k^{(y)}H^{(k,\frac{1}{2})}\ 
\label{Heff}
\ee
at an effective inverse temperature $\beta=1$. As all
$H^{(k,\frac{1}{2})}$ commute and are obtained by taking logarithmic
derivatives of the transfer matrix, finite temperature properties of 
\fr{Heff} can be studied by standard methods\cite{higherCL,KS:2002}. 
The \emph{Quantum Transfer Matrix approach} (QTM)\cite{QTM} is particularly
useful in this regard, as it provides an efficient way to obtain
explicit results for thermal averages of local
operators\cite{GKS:2004,wuppertal} (see also Ref.~\onlinecite{Smirnov}). 
Refs~\onlinecite{Pozsgay:13a,FE_13b,FCEC_14} employed the QTM
approach to the calculation of steady state properties of the
density matrix \fr{rhoy} for quenches from simple initial states.
By employing the generating function \fr{Omega}, it is possible to
take into account \emph{all} ultra-local conservation laws, and arrive
at explicit results for local observables without having to determine
the Lagrange multipliers $\lambda_k^{(\infty)}$\cite{FE_13b}. In the QTM
approach the stationary state is described in terms of the solution of
a system of coupled, nonlinear integral equations. Remarkably, the
information on the initial state enters this system only \emph{via}
the function $\Omega^{(\frac{1}{2})}_{\Psi(0)}(\lambda)$. Results
for spin correlators obtained in this way were compared to t-DMRG
computations for quenches from a variety of initial states in
Ref.~\onlinecite{FCEC_14}, and found to be compatible within the
limitations of the numerical analysis. 

The subsequent application of the \emph{Quench Action
  Approach}\cite{CE_PRL13} (reviewed by J.-S. Caux~\cite{Crev} in this volume) to
the same problem revealed that the ultra-local GGE in fact does not
correctly describe the steady state for quenches to the spin-1/2
Heisenberg chain\cite{GGE_XXZ_Amst,XXZung,QANeel,XXZunglong,R:gs14}, although
it does provide a very good approximation for e.g. quenches from the
N\'eel state. This suggested the existence of hitherto unknown conservation
laws in the Heisenberg chain, which need to be taken into account in
the construction of the GGE. 
\subsubsection{``Quasi-local'' GGE}
The ``missing'' conservation laws for the spin-1/2 Heisenberg XXZ
chain were discovered in
Refs~\onlinecite{quasilocal_XXX,GGE_int} (see the review by
E. Ilievski, M. Medenjak, T. Prosen, and L. Zadnik~\cite{Prosrev} in
this volume). Their structure is quite 
different from that of the ultra-local conservation laws discussed
above: their densities are not local in the sense that they act
non-trivially only on a finite number of neighbouring sites, but
\emph{quasi-local}. Similar conservation laws had been identified
earlier in relation to transport properties of the Heisenberg
chain\cite{qlXXZ}. In order to define the concept of quasi-locality
one introduces an inner product on the space of operators by
\be
(A,B)=\langle A^\dagger B\rangle_\infty\ ,\qquad \langle
A\rangle_\infty=\frac{1}{2^L}{\rm Tr}\big(A\big).
\ee
\begin{mydef}
Quasi-local operators\cite{quasilocal_XXX}.
\end{mydef}
Let us consider an operator $Q$ and expand it in terms of mutually
orthogonal local operators $q_{j,r}$ of range $r$ 
\be
Q=\sum_{j}\sum_r q_{j,r}.
\ee
$Q$ is called \emph{quasi-local} if it fulfils the following three
conditions: 
\bea
{\rm (QL1):}&\qquad& \lim_{L\to\infty}\frac{1}{L}
(Q-\langle Q\rangle_\infty,Q-\langle Q\rangle_\infty)={\rm const};\nn
{\rm (QL2):}&\qquad& \lim_{L\to\infty}(Q,B_k)\ \text{exists}\ ,\nn
{\rm (QL3):}&\qquad&(q_{j,r},q_{j,r})<Ce^{-r/\xi}\, ,
\eea
where $B_k$ is any operator that acts non-trivially only on a fixed
number of $k$ sites, and $\xi$ and $C$ are positive constants.

In the anisotropic Heisenberg chain quasi-local
charges can be constructed as follows\cite{quasilocal_XXX,GGE_int}.
It is well known that the six-vertex model transfer matrix \fr{tau}
is part of a much larger family, built from the L-operators
\bea
L^{(S)}_n(\lambda)=\frac{1}{\sinh\big(\frac{\eta}{2}(1+s)+i\lambda)\big)}
&\Big[&\sinh(\frac{\eta}{2}+i\lambda) \cosh{(\eta{\cal S}^z)}
+\cosh(\frac{\eta}{2}+i\lambda) \sinh{(\eta {\cal
    S}^z)}\sigma_n^z\nn
&& +\sinh (\eta) ({\cal S}^{-}\sigma_n^{+} +{\cal S}^{+}\sigma_n^{-})\Big),
\label{LS}
\eea
which again acts on the tensor product of auxiliary and
quantum spaces, but now the auxiliary space is $2S+1$
dimensional. Here $S$ is an arbitrary half integer. The operators
${\cal S}^\alpha$ obey a q-deformed SU(2) algebra 
\be
[{\cal S}^{+},{\cal S}^{-}]=[2{\cal S}^{z}]_{q}\ ,\qquad
[{\cal S}^{z},{\cal S}^{\pm}]=\pm {\cal S}^{\pm}\ ,
\ee
where $[x]_q =\sinh(\eta x)/\sinh(\eta)$, and act on a $q$-deformed
spin-S representation as
\begin{equation}
{\cal S}^{z}\ket{k}=k\ket{k},\qquad {\cal S}^{\pm}\ket{k}
=\sqrt{[S+1\pm k]_{q}[S\mp k]_{q}}\ket{k\pm 1}\ ,\quad
k=-S,\dots,S.
\end{equation}
A family of row-to-row transfer matrices is then obtained as
\be
\big(\tau_{S}(\lambda)\big)^{\beta_1\dots\beta_L}_{\alpha_1\dots\alpha_L}
=\big(L^{(S)}_1(\lambda)\big)^{a_1a_2}_{\alpha_1\beta_1}
\big(L^{(S)}_2(\lambda)\big)^{a_2a_3}_{\alpha_2\beta_2}\dots
\big(L^{(S)}_L(\lambda)\big)^{a_La_1}_{\alpha_L\beta_L}\ .
\label{tauS}
\ee
All $\tau_S(\lambda)$ are operators on the same quantum space (a
tensor product of $L$ spin-1/2's), and as a consequence of the
Yang-Baxter relation form a commuting family
\be
[\tau_S(\lambda),\tau_{S'}(\mu)]=0.
\label{commutingS}
\ee

By virtue of the commutation relations \fr{commuting} it is clear that
a set of mutually commuting operators can be obtained by taking higher
derivatives, i.e.
\be
H^{(S,k)}=i\left(C_S\frac{\partial}{\partial\lambda}
\right)^k \Big|_{\lambda=0} \ln\left[\tau_{S}(\lambda)\right],
\ee
where $C_S$ are some normalization constants that can be conveniently
chosen. As a consequence of \fr{commutingS} we have
\be
[H^{(S,k)},H^{(S',k')}]=0.
\ee
Apart from the special case $S=1/2$ these conservation laws are
quasi-local. This means that (in the infinite volume) their general
structure is 
\be
H^{(S>\frac{1}{2},k)}=\sum_{j=-\infty}^\infty\sum_{k\geq 1}\sum_{\alpha_1,\dots,\alpha_k}
f^{(k)}_{\alpha_1\alpha_2\dots\alpha_k}\sigma^{\alpha_1}_j\sigma^{\alpha_2}_{j+1}
\dots\sigma^{\alpha_k}_{j+k-1}\ ,
\ee
where $\alpha_j=0,x,y,z$, and the coefficient functions
$f^{(k)}_{\alpha_1\dots\alpha_k}$ decay sufficiently fast with $k$ so
that the conservation laws are extensive. As shown in
Ref.~\onlinecite{GGE_int}, the initial data of the quantum quench can
again be encoded in suitably chosen generating functions, which are
generalizations of \fr{Omega}
\be
\Omega_{\Psi(0)}^{(S)}(\lambda)=\lim_{L\to\infty}\frac{i}{L}
\langle\Psi(0)|\tau'_S(\lambda)\tau^{-1}_S(\lambda)|\Psi(0)\rangle=
\lim_{L\to\infty}\frac{i}{L}
\left[\frac{\sinh\big(\frac{\eta}{2}(1-2S)-i\lambda\big)}
{\sinh\big(\frac{\eta}{2}(1+2S)-i\lambda\big)}\right]^L
\langle\Psi(0)|\tau'_S(\lambda)\tau_S(\lambda-\eta)|\Psi(0)\rangle.
\label{OmegaS}
\ee
The generating functionals $\Omega_{\Psi(0)}^{(S)}(\lambda)$ can be
evaluated for matrix product states by the same method discussed
above (although the computational effort increases with the value of
$S$). 

The most convenient description of the stationary state turns out to
be in terms of the generalized microcanonical ensemble discussed
above. The steady state is characterized by  the set $\{\rho^{\rm
  SS}_{n,p}(\lambda)|n=1,\dots\}$ of particle densities or the
equivalent set of hole densities $\{\rho^{\rm
  SS}_{n,h}(\lambda)|n=1,\dots\}$. Ultimately this set must be
determined by the initial conditions, which are encoded in \fr{OmegaS}.
We now use that $X_S(\lambda)=\tau'_S(\lambda)\tau^{-1}_S(\lambda)$
can be diagonalized by Algebraic Bethe Ansatz \cite{Korepinbook,spinS}. The
eigenvalues of $X_S(\lambda)$ for $M$-particle states with $M\sim L$
are of the form 
\be
\nu_S(\lambda)=\sum_{k=1}^M\frac{2\sinh(2S\eta)}{\cos(2\lambda+2\lambda_k)-\cosh(2S\eta)}+o(L)\ ,
\label{nus}
\ee
where the $\lambda_k$ are solutions to the Bethe Ansatz
equations\cite{spinS}
\be
\left(\frac{\sin(\lambda_j+i\eta S)}{\sin(\lambda_j-i\eta
  S)}\right)^L=
\prod_{k\neq j=1}^M
\frac{\sin(\lambda_j-\lambda_k+i\eta)}{\sin(\lambda_j-\lambda_k-i\eta)}\ ,\quad
j=1,\dots,M.
\label{BAE}
\ee
In the thermodynamic limit this can be simplified by following through
the usual logic of the string hypothesis and the Thermodynamic Bethe
Ansatz\cite{Takahashibook,book,Gaudin}. Rather than with solutions to
the Bethe equations \fr{BAE} one then works with macro-states
$|\boldsymbol{\rho}\rangle$, which are described by sets
$\{\rho_{n,p}(\lambda)|n=1,\dots\}$ of particle densities or the
equivalent set of hole densities
$\{\rho_{n,h}(\lambda)|n=1,\dots\}$. The eigenvalue equation \fr{nus}
then becomes\cite{GGE_int} 
\be
\lim_{L\to\infty}\frac{1}{L}\langle\boldsymbol{\rho}|X_S(\mu)|\boldsymbol{\rho}\rangle
=\sum_{k\in\mathbb{Z}} \frac{e^{-i2k\mu}}{\cosh(k\eta)} \left(
\int_{-\pi/2}^{\pi/2}\mathrm{d}\lambda \,
e^{2ik\lambda} \rho_{2S,h}(\lambda) - e^{-2S|k|\eta} \right)  . 
\label{intermediate}
\ee
For $|\boldsymbol{\rho}\rangle=|\boldsymbol{\rho}^{\rm SS}\rangle$ the
right hand side of \fr{intermediate} must agree with the initial
values after the quench \fr{OmegaS}. This is achieved by setting\cite{GGE_int}
\be
\rho_{2S,h}^{\rm SS}(\lambda)=a_{2S}(\lambda)+\frac{1}{2\pi}
\left[\Omega^{(S)}_{\Psi(0)}\big(\lambda+i\frac{\eta}{2}\big)
+\Omega^{(S)}_{\Psi(0)}\big(\lambda-i\frac{\eta}{2}\big)\right]\, ,
\label{rho2Sh}
\ee
where $a_{2S}(\lambda)$ is a function independent of the initial state
that is defined in Appendix \ref{app:TBA}.
This shows that the initial data \fr{OmegaS}, which involves both
ultra-local and quasi-local conservation laws, completely determines
the macro-state that defines the generalized microcanonical
ensemble. We note that the derivation did not invoke the maximum
entropy principle. For the particular case of quenches from the N\'eel
state \fr{rho2Sh} agrees with the one obtained by the Quench Action
Approach\cite{GGE_XXZ_Amst,XXZung,QANeel,XXZunglong}. 

The generalization of the approach discussed above for quenches to
particular values of $\Delta$ with $-1<\Delta<1$ in the Heisenberg
model \fr{HXXZ} was achieved in Ref.~\onlinecite{IQNB} .

\section{Outlook}
\label{sec:Outlook}
We have given an introduction to quantum quenches in many-particle
systems and then reviewed recent developments, focussing in particular
on the role played by conservation laws. In spite of the impressive
progress of the last few years, many important questions remain
largely open. Let us list a few of them in no particular order. 
\begin{enumerate} 
\item{}
In the spin-1/2 Heisenberg XXZ chain quasi-local conservation laws
have been shown to play a prominent role in determining the stationary
state. It is believed that this holds quite generally in interacting
integrable models. The construction used in the XXZ case can in
principle be generalized to the $sl(M|N)$ family of integrable graded
quantum ``spin'' chains, see e.g. Refs~\onlinecite{slNM}, and it
would be interesting to investigate the role of quasi-local charges in such
models. The Hubbard model is another very interesting case, but is
like to be more difficult to handle due to its non standard
structure\cite{book}. 
\item{} So far only particularly simple classes of initial states can
be accommodated, \emph{cf.} Section~\ref{ssec:initialstate}. It would
be highly desirable to have a more general method for capturing the
information on the initial state.
\item{} In interacting theories the focus has so far been on
stationary state properties. The study of the full time evolution of
observables is much less
developed\cite{sine-Gordon,rel_int,bosonXX,MS:over16}. A promising method for
analyzing the time dependence of the expectation values of local
operators after a quantum quench in an interacting integrable theory
is the Quench Action approach\cite{CE_PRL13}. So far it has been implemented
only in a very small number of cases\cite{sine-Gordon,rel_int}, and
further studies are sorely needed.
\item{} As we have seen, the non-equilibrium dynamics of integrable
and non-integrable models is quite different. This poses the question
of what happens, when one adds a small perturbation to an integrable
model. This has been investigated in a number of theoretical works
\cite{KL:BH07,MK:H08,MK:preT09,SC:intft10,KWE:preT11,MG:m-c-i11,MS:noisy12,CB:glass12,SK:13,MM:preT13,mitra13,preT0,gas14,Delfino,mau-prep,preT_entropy,BF:preT,preT_dGGE,KAM,BD:spir15,O:geo15,Fdefect,Nessi15},
and is of immediate experimental relevance\cite{KI:preT11,Langen15}
(see also the review by T. Langen, T. Gasenzer and J. Schmiedmayer in this
volume\cite{LGS16}). The generic effect of adding a
small integrability breaking term appears to be the generation of an
intermediate ``prethermalization'' time scale, below which the system
retains information about being proximate in parameter space to an
integrable model. At late times thermalization seems to set
in\cite{preT_dGGE}. So far the theoretical analyses are restricted to
weak interactions and/or short times, and it is crucial to go beyond
these limitations.

\end{enumerate}
\acknowledgments
We owe gratitude to many colleagues for collaborations and sharing
their insights on quantum quenches with us. This margin is too
small to acknowledge them all, but particular thanks are due to
Bruno Bertini, Pasquale Calabrese, John Cardy, Jean-S\'ebastien Caux,
Mario Collura, Robert Konik, Giuseppe Mussardo, Marcos Rigol,
Dirk Schuricht and Alessandro Silva. This work was supported by the
EPSRC under grant EP/N01930X/1, the Isaac Newton Institute for
Mathematical Sciences under grant EP/K032208/1, and the Agence
Nationale de la Recherche under grant LabEx ENS-ICFP:ANR-10-LABX-0010/ANR- 10-IDEX-0001- 02 PSL*.

\appendix

\section{Requirements on the initial state}
\label{app:cluster}

\subsection{Cluster decomposition}

We have defined our quench protocol such that it results in initial
states $\ket{\Psi(0)}$ that have a cluster decomposition property \fr{cluster}.
This requirement is often relaxed in both numerical and analytical
investigations, and in some cases $\ket{\Psi(0)}$ is taken to be a
Schr\"odinger cat state, see e.g. Refs
\onlinecite{QANeel,GGE_XXZ_Amst,XXZunglong,XXZung}. An example is
provided by quenches where the system is initialized in a classical 
N\'eel state $|\uparrow\downarrow\uparrow\downarrow\dots\rangle$. This
breaks translational invariance and it can be calculationally
convenient to work instead with a translationally invariant cat state
\be
\frac{1}{\sqrt{2}}\left(|\uparrow\downarrow\uparrow\downarrow\uparrow\dots\rangle+
|\down\uparrow\downarrow\uparrow\downarrow\dots\rangle\right).
\ee
While for specific calculations such replacements can be useful, they significantly
affect the steady state behaviour in general. This can be seen by
considering a $\mathbb{Z}_2$ symmetric pre-quench Hamiltonian $H_0$ with a ground
state that spontaneously breaks the $\mathbb{Z}_2$ symmetry. An
example is provided by the transverse-field Ising chain \fr{eq:HTFIC}
with $h<1$. In the thermodynamic limit $H_0$ has two ground states
$\ket{\Psi_\pm}$, both of which have a cluster decomposition
property. We now consider a general linear combination
\be\label{eq:cat}
\ket{\Psi(0)}=\cos\theta \ket{\Psi_+}+e^{i\phi} \sin\theta\ket{\Psi_-}\, .
\ee
As $|\Psi_\pm\rangle$ are macroscopically distinct (as they
lead to different order parameters), we conclude that
expectation values of local operators $\mathcal O$ are given by
\be
\braket{\Psi(0)|\mathcal O|\Psi(0)}=
\cos^2\theta\braket{\Psi_+|\mathcal
  O|\Psi_+}+\sin^2\theta\braket{\Psi_-|\mathcal O|\Psi_-}\, .
\ee
As the Hamiltonian of our system is short ranged, this decomposition
persists at all finite times, i.e. 
\be
\braket{\Psi(t)|\mathcal O|\Psi(t)}=
\cos^2\theta\braket{\Psi_+(t)|\mathcal O|\Psi_+(t)}
+\sin^2\theta\braket{\Psi_-(t)|\mathcal O|\Psi_-(t)}
\ee
To see this, we may use a result derived in Ref.~\onlinecite{bravyi06} for
the time evolution of local operators with short-ranged Hamiltonians:
restricting $\mathcal O(t)=e^{i H t}\mathcal O e^{-i H t}$ to a
subsystem $S$ of size $|S|$ gives an error that scales as $e^{(2v t-|S|)/\zeta}$,
where $\zeta$ is a constant and $v$ is the Lieb-Robinson
velocity. This means that it is possible to approximate $\mathcal
O(t)$ to a given accuracy by a local operator of a ``size'' that
scales as $2v t$. This in turn implies that
$\langle\Psi_+|{\cal O}(t)|\Psi_-\rangle=0$, because
$|\Psi_\pm\rangle$ are macroscopically distinct.

On the other hand, our system relaxes locally by construction if we
initialize it in $|\Psi_\pm(0)\rangle$, i.e.
\be
\lim_{t\to\infty}|\Psi_\pm(t)\rangle\langle\Psi_\pm(t)|=_{\rm
  loc}\rho^{\rm SS}_\pm.
\ee
Putting everything together we conclude that 
\be
\lim_{t\to\infty}|\Psi(t)\rangle\langle\Psi(t)|=_{\rm  loc}
\rho^{\rm SS}=\cos^2\theta \rho^{\rm SS}_++\sin^2\theta \rho^{\rm SS}_-\, .
\label{eq:rhoSStheta}
\ee
The problem is that this form of the stationary state can be different
from what one would expect on the basis of local relaxation to
(generalized) Gibbs ensembles. To be specific, we consider the example
of a quench to an anisotropic spin-1/2 Heisenberg chain
\be
H=\frac{J}{4}\sum_\ell
\sigma_\ell^y\sigma_{\ell+1}^y+\sigma_\ell^z\sigma_{\ell+1}^z+
\Delta \sigma_\ell^x\sigma_{\ell+1}^x +g \sigma_\ell^x\sigma_{\ell+2}^x\, .
\label{counterex}
\ee
Imposing $\Delta,g \neq 0$ renders this model non-integrable, but is
has one local conservation law
\be
Q=\sum_\ell\sigma_\ell^x\ ,\quad [H,Q]=0.
\ee
Our initial state is of the form \fr{eq:cat}, where
$|\Psi_\pm(0)\rangle$ are the two ground states of the TFIC in the
thermodynamic limit at $h<1$. We note that the energy density $e$ of
\fr{counterex} is the same in both $|\Psi_+\rangle$ and
$|\Psi_-\rangle$. As these states 
have a cluster decomposition property, it follows from our general
discussion that the respective stationary states are locally
equivalent to grand canonical ensembles
\be
\lim_{t\rightarrow\infty}|\Psi_\pm(t)\rangle\langle\Psi_\pm(t)|=_{\rm loc}
\rho^{\rm GC}_\pm=\frac{e^{-\beta_\pm H-\mu_\pm Q }}{Z_\pm}\, ,
\ee
where the values for $\beta_\pm$ and $\mu_\pm$ are obtained by fixing
the energy density $e$ and charge density $q_\pm$ to agree with their
initial values at time $t=0$. Eqn \fr{eq:rhoSStheta} tells us that the
correct stationary state in this example is then
\be
\label{eq:wrong}
\rho^{\rm SS}=_{\rm loc}\cos^2\theta \rho^{\rm GC}_++\sin^2\theta
\rho^{\rm GC}_-\ .
\ee
Crucially, while the density matrices $\rho^{\rm GC}_\pm$ have the cluster
decomposition property, $\rho^{\rm SS}$ does not. This can be seen as
follows. The local conservation law distinguishes between the two
states $|\Psi_\pm\rangle$  
\be
q_+=\braket{\Psi_+|\sigma_\ell^x|\Psi_+}\neq 
\braket{\Psi_-|\sigma_\ell^x|\Psi_-}=q_-\ .
\ee
Using the cluster decomposition property of $\rho^{\rm GC}_\pm$ we have
\be
\lim_{|n-\ell|\to\infty}{\rm Tr}\left(\rho^{\rm GC}_\pm\sigma^x_\ell\sigma^x_n\right)=q^2_\pm,
\ee
which in turn establishes that $\rho^{\rm SS}$ does not have the cluster
decomposition property
\be
\lim_{|n-\ell|\to\infty}\left[{\rm Tr}\left(\rho^{\rm
  SS}\sigma^x_\ell\sigma^x_n\right)-
{\rm Tr}\left(\rho^{\rm
  SS}\sigma^x_\ell\right)
{\rm Tr}\left(\rho^{\rm
  SS}\sigma^x_n\right)\right]=\frac{(q_+-q_-)^2}{4}\sin^2(2\theta)\neq 0.
\ee
On the other hand, if we were to apply our formalism of local
relaxation blindly to our Hamiltonian \fr{counterex}, we would
conclude that the stationary state is locally equivalent to a grand
canonical ensemble, which is expected to have a cluster decomposition property~\cite{loc_temp}.

Our discussion can be summarized as follows: \emph{If there exists at
least one integral of motion that distinguishes  $\ket{\Psi_-}$ from
$\ket{\Psi_+}$, the stationary state associated with the time
evolution of the cat state \eqref{eq:cat} does not possess the cluster
decomposition property \fr{cluster} and hence is not described by a
standard generalized Gibbs ensemble.} 

\subsection{Probability distributions of energy and conservation laws}
Let us consider a post-quench Hamiltonian $H$ with a set of local
conservation laws $I^{(n)}$, \emph{cf.} \fr{CL}, \fr{CL2}. In our basic
definition of a quantum quench we initialize the system in a pure
state $|\Psi(0)\rangle$. Then the cluster decomposition property implies
that the probability distribution of energy and all local conservation
laws approach delta-functions in the thermodynamic limit, 
e.g. 
\be
P_n(\epsilon)=
\lim_{L\to\infty}
\frac{1}{L}{\rm Tr}\left[\rho(0)\delta(I^{(n)}-L\epsilon)\right]
=\delta(\varepsilon-i^{(n)})\ ,\quad
i^{(n)}=\lim_{L\to\infty}\frac{1}{L}{\rm Tr}\left[\rho(0)I^{(n)}\right],
\label{sharpness}
\ee
where $\rho(0)=|\Psi(0)\rangle\langle\Psi(0)|$.
As we have pointed out, it is sometimes desirable to consider initial
density matrices $\rho(0)$ that are not pure states. When doing so one
must ensure that \fr{sharpness} continues to hold. In cases where it
does not it is clearly impossible for the system to locally relax
to a GGE, because there the probability distributions of all
conservation laws approach delta functions in the thermodynamic limit.
Generalizations of GGE ideas to such cases have been explored in
Ref.~\onlinecite{GGGE} .

\section{``Atypical'' macro-states in integrable models}
\label{app:typical}
Integrable models have the unusual property of having \emph{atypical}
finite entropy eigenstates at finite energy densities. This is well known
for non-interacting theories and we discuss this case first. 
\subsection{Free fermions}
Let us consider a model of free fermions with Hamiltonian
\be
H=\sum_k\epsilon(k) n(k)\ ,
\ee
where $n(k)=c^\dagger(k)c(k)$. Imposing periodic boundary conditions
quantizes the allowed momenta
\be
k_n=\frac{2\pi n}{L}\ ,\quad n=-\frac{L}{2}+1,\dots,\frac{L}{2}.
\ee
We now focus on a special class of Fock states
$\prod_{j=1}^Nc^\dagger(k_j)|0\rangle$, for which the \emph{particle
  densities}
\be
\rho_p(k_j)=\frac{1}{L(k_{j+1}-k_j)}
\ee
approach smooth functions in the thermodynamic limit $N,L\to\infty$,
$n=N/L$ fixed. For such states, the number of particles in the interval
$[k_n,k_n+\Delta k]$ for large $L$ is given by
\be
\rho_p(k_n)\Delta k\ .
\ee
It is convenient to define a \emph{hole density} by
$\rho_h(k_j)=\frac{1}{2\pi}-\rho_p(k_j)$. In the thermodynamic limit
many different choices of $\{k_j\}$ lead to the same
\emph{macro-state} described by a given particle density. To enumerate
them we note that the number of different states in the interval
$[k,k+\Delta k]$, that give rise to a given density in the
thermodynamic limit, is obtained by distributing $\rho_p(k)L\Delta k$
particles among $\big(\rho_p(k)+\rho_h(k)\big)L\Delta k$
vacancies. This follows a binomial distribution. For large $L$ the
latter may be approximated by Stirling's formula, and in the thermodynamic limit
one obtains the well-known expression for the entropy per site
\be
s[\rho_p]=\int_{-\pi}^\pi dk\left[
\big(\rho_p(k)+\rho_h(k)\big)\ln\big(\rho_p(k)+\rho_h(k)\big)
-\rho_p(k)\ln\big(\rho_p(k)\big)
-\rho_h(k)\ln\big(\rho_h(k)\big)\right]\ .
\ee
Let us now investigate what kind of macro-states exist for a given
energy density $e$. The most likely (maximum entropy) macro-state can
be constructed using equilibrium statistical mechanics. Extremizing
the free energy per site 
\be
f[\rho_p]=\int_{-\pi}^\pi\frac{dk}{2\pi}\epsilon(k)\rho_p(k)-Ts[\rho_p]
\ee
with respect to the particle density $\rho_p$ gives
\be
\rho^{\rm eq}_p(k)=\frac{1}{2\pi}\frac{1}{1+e^{\epsilon(k)/T}}.
\label{rhoeq}
\ee
The ``temperature'' $T$ is related to the energy density $e$ by
\be
e=\int_{-\pi}^\pi\frac{dk}{2\pi}\frac{\epsilon(k)}{1+e^{\epsilon(k)/T}}.
\label{eoft}
\ee
The entropy per site of this equilibrium state is
\be
s[\rho_p^{\rm eq}]=\frac{\partial}{\partial
  T}T\int_{-\pi}^\pi\frac{dk}{2\pi}
\ln\left[1+e^{-\epsilon(k)/T}\right].
\ee
By construction the macro-state \fr{rhoeq} is the \emph{typical} state
at energy density \fr{eoft}: if we randomly pick an energy eigenstate
with energy density $e$ for a very large system size $L$, the
probability for this state to have particle density \fr{rhoeq} is
exponentially close (in $L$) to one. On the other hand, there are
\emph{atypical} finite entropy macro-states characterized by their
respective particle densities $\rho_p(k)$. As a particular example we
consider our tight-binding model \fr{TBM} for $\mu=0$, which gives a
dispersion $\epsilon(k)=-2J\cos(k)$. We fix the
particle density to be $1/2$ and the energy density to be
$e=-0.405838J$, which corresponds to temperature $T=J$ in \fr{rhoeq},
\fr{eoft}, i.e.
\be
\rho_p^{\rm eq}(k)=\frac{1}{2\pi}\frac{1}{1+e^{-2\cos(k)}}.
\label{rhopeq1}
\ee
The entropy of the equilibrium state is $s_{\rm eq}=0.511571$.
Let us now consider the family of macro-states described by the
particle density 
\be
\rho^{(\lambda)}_p(k)=\frac{1}{2\pi}\frac{1}{1+e^{-4\cos(k)-\lambda\cos(3k)}}\ .
\label{rhol1l2}
\ee
Fixing $\lambda=2.43096..$ gives us the same particle and energy
densities as for the equilibrium state, i.e. $n=1/2$ and
$e=-0.405838J$. The entropy density $s=0.396781$ is of course lower
than that of the (maximum entropy) equilibrium state. This means that
if we randomly select an energy eigenstate with particle density
$n=1/2$ and energy density $e=-0.405838J$ for a very large system size
$L$, we are exponentially more likely by a factor $e^{L(s_{\rm
    eq}-s)}$ to end up with a state described by \fr{rhopeq1} than one
described by \fr{rhol1l2}. Unsurprisingly, expectation values of local
operators are generally different in the two macro-states. As an
example, let us consider
\be
{\cal O}_j=c^\dagger_jc_{j+3}+c^\dagger_{j+3}c_j\ .
\ee
Setting again $\lambda=2.43096..$ and taking the thermodynamic limit
we have 
\be
\langle \rho_p^{\rm eq}(k)|{\cal O}_j|\rho_p^{\rm eq}(k)\rangle=
-0.0271229\neq \langle \rho_p^{(\lambda)}(k)|{\cal
  O}_j|\rho_p^{(\lambda)}(k)\rangle=
0.215148\ .
\ee
Here the expectation values may be taken with regards to any
micro-state that gives rise to the appropriate macro-state in the
thermodynamic limit.

\subsection{Interacting theories: anisotropic spin-1/2 Heisenberg model}
\label{app:TBA}
The situation in integrable models is analogous to what we just
discussed for free fermions. The main difference arises from the more
complicated structure and interacting nature of the elementary
excitations in integrable models. For the sake of definiteness we
consider the particular example of the spin-1/2 Heisenberg model
\fr{HXXZ} with $\Delta\geq 1$. Energy eigenstates
$|\lambda_1,\dots,\lambda_M\rangle$ on a ring of length $L$ are
parametrized by $M$ rapidity variables $\lambda_j$, which fulfil the
quantization conditions\cite{Takahashibook} 
\be
\left(\frac{\sin(\lambda_j+i\frac{\eta}{2})}{\sin(\lambda_j-i\frac{\eta}{2})}
\right)^L= \prod_{k\neq j=1}^M
\frac{\sin(\lambda_j-\lambda_k+i\eta)}{\sin(\lambda_j-\lambda_k-i\eta)}\ ,\quad 
j=1,\dots,M.
\label{BAEhalf}
\ee
Each $\lambda_j$ is associated with an elementary ``magnon''
excitation over the ferromagnetic state with all spins up, and the
total energy and momentum of the eigenstate are additive 
$E=\sum_{j=1}^Me(\lambda_j)$, $P=\sum_{j=1}^Mp(\lambda_j)$. As a
result of interactions, magnons form \emph{bound states}. These
correspond to ``string solutions'' of the quantization conditions \fr{BAEhalf}
\be
\lambda^{n,j}_\alpha=\lambda^{n}_\alpha+i\frac{\eta}{2}(n+1-2j)+i\delta^{n,j}_\alpha\ , 
\qquad j=1,\dots,n\ ,
\ee
where $\delta^{n,j}_\alpha$ are deviations from ``ideal'' strings that
become negligible when we take the thermodynamic limit at finite
densities of magnons and bound states, \emph{cf.} 6.2.A of
Ref.~\onlinecite{TsvelickWiegmann}. As a consequence of integrability
these bound states are \emph{stable} excitations. The generalization of
particle density description of macro-states in free theories to
interacting integrable models is then clear: macro-states are
characterized by an (infinite) set of densities
$\{\rho_{n,p}(\lambda)|n=1,2,\dots\}$ for magnons $(n=1)$ and bound
states of all lengths $(n\geq 2)$. Just as in the case of free
fermions, we can define corresponding hole densities
$\rho_{n,h}(\lambda)$. The relation between particle and hole
densities is fixed by the quantization conditions \fr{BAEhalf}
\bea
\rho_{n,t}(\lambda)\equiv\rho_{n,p}(\lambda)+\rho_{n,h}(\lambda)&=&a_n(\lambda)-\sum_{m=1}^\infty
\int_{-\frac{\pi}{2}}^{\frac{\pi}{2}}d\mu\ A_{nm}(\lambda-\mu)\rho_{m,p}(\mu)\ ,
\label{particlehole}
\eea
where $A_{nm}(\lambda)=(1-\delta_{n,m})a_{|n-m|}(\lambda)+
2a_{|n-m|+2}(\lambda)+\dots+2a_{n+m-2}(\lambda)+a_{n+m}(\lambda)$ and 
$a_n(\lambda)=\frac{1}{2\pi}\frac{2\sinh(n\eta)}{\cosh(n\eta)-\cos(2\lambda)}$.
Eqn \fr{particlehole} allows one to express the hole densities in
terms of the particle densities and vice versa, but in contrast to the
non-interacting case the relationship is non-trivial. The energy and
entropy per site of a macro-state are given by
\bea
e[\{\rho_{n,p}\}]&=&-J\sinh(\eta)\sum_{n=1}^\infty
\int_{-\frac{\pi}{2}}^{\frac{\pi}{2}}
d\lambda\ \rho_{n,p}(\lambda)\ a_n(\lambda)
\ ,\nn
s[\{\rho_{n,p},\rho_{n,h}\}]&=&\sum_{n=1}^\infty
\int_{-\frac{\pi}{2}}^{\frac{\pi}{2}}
d\lambda\left[
\rho_{n,t}(\lambda)\ln\big(
\rho_{n,t}(\lambda)\big)
-\rho_{n,p}(\lambda)\ln\big(\rho_{n,p}(\lambda)\big)
-\rho_{n,h}(\lambda)\ln\big(\rho_{n,h}(\lambda)\big)
\right]\ .
\label{EandS}
\eea
The typical state at a given energy density is then obtained by
extremizing the free energy per set
$f[\{\rho_{n,p}\}]=e[\{\rho_{n,p}\}]-Ts[\{\rho_{n,p},\rho_{n,h}\}]$
with respect to particle and hole densities under the constraints
\fr{particlehole}. This results in the so-called Thermodynamic Bethe
Ansatz equations\cite{Takahashibook}. By construction this state is
\emph{thermal}, i.e. corresponds to a standard Gibbs distribution.

Atypical finite entropy states can be constructed by specifying a set
of particle densities $\{\rho_{n,p}(\lambda)\}$. The corresponding
hole densities are then obtained by solving equations
\fr{particlehole}. The entropy per site of the resulting macro-state
can be calculated from \fr{EandS}, and will by construction be smaller
than that of the maximum entropy state. 


\section{Stationary state correlators in the TFIC}
\label{app:statics}
In this appendix we summarize results for the amplitudes
$C^\alpha(\ell)$ describing the subleading behaviour of stationary
state spin-spin correlators in the TFIC, \emph{cf.} Eqn
\fr{static2point}. 

\paragraph{Transverse spin correlator.}
Here the amplitude is of the form\cite{CEF2}
\be
C^z(\ell)=\bar C^z \ell^{-\alpha^z}\ ,\qquad
\alpha^z=\left\{
\begin{array}{lcl}
1&&{\rm if\ \ } |\ln h|>|\ln h_0|\ ,\\
0&& {\rm if\ \ } h_0=1/h\ ,\\
1/2&&{\rm if\ \ } |\ln h|<|\ln h_0|\, ,
\end{array}
\right. 
\ee
where the constant is known exactly
\be
\ba
&C^z&=&\left\{
\begin{array}{lcl}
\frac{|h_0-1/h_0|}{4\pi}\frac{h_0-h}{h h_0-1}&&
{\rm if\ \ } |\ln h|>|\ln h_0|\ ,\\
-\frac{(h-1/h)^2}{2\pi} &&{\rm if\ \ } h_0=1/h\ ,\\
\frac{(h-1/h)\sqrt{|h_0-1/h_0|}(h_0-h)}{8\sqrt{\pi}
  h}\sqrt{\frac{h_0-h}{h_0(h h_0-1)}}\frac{e^{\mathrm{sgn}(\ln h)|\ln
    h_0|/2}}{\sinh\frac{|\ln h|+|\ln h_0|}{2}}&&{\rm if\ \ } |\ln
h|<|\ln h_0|\, .  
\end{array}
\right. 
\ea
\ee

\paragraph{Longitudinal spin correlator.}
Here the large-$\ell$ asymptotics of prefactor $C^x(\ell)$
are as follows.
\begin{enumerate}
\item{} Quench within the ferromagnetic phase ($h_0,h<1$).
\be\label{eq:CxFF}
C^x(\ell)\equiv{\cal C}^x_{\rm FF}=
\frac{1-h h_0+\sqrt{(1-h^2)(1-h_0^2)}}{2\sqrt{1-h 
    h_0}\sqrt[4]{1-h_0^2}}\, .
\ee
\item{} Quench from the ferromagnetic to the paramagnetic phase
($h_0<1<h$). 
\be\label{eq:CxFP}
C^x(\ell)\equiv {\cal C}^x_{\rm FP}=
\sqrt{\frac{h\sqrt{1-h_0^2}}{h+h_0}}.
\ee
\item{} Quench from the paramagnetic to the ferromagnetic phase
($h_0>1>h$). 
\be
C^x(\ell)\equiv{\cal C}^x_{\rm PF}(\ell)=
 \sqrt{\frac{\ h_0-h}{\sqrt{h_0^2-1}}} \cos\bigl(\ell
 \arctan\frac{\sqrt{(1-h^2)(h_0^2-1)}}{1+h_0 h}\bigr)\, .
\label{CPFcos}
\ee
\item{} Quench within the paramagnetic phase ($1<h_0,h$). 
\be\label{eq:CxPP}
C^x(\ell)\equiv{\cal C}^x_{\rm PP}(\ell)=
\left\{
\begin{array}{ll}
-\frac{h_0\sqrt{h}\bigl(h h_0-1+\sqrt{(h^2-1)(h_0^2-1)}\bigr)^2}{4
  \sqrt{\pi } (h_0^2-1)^{3/4} (h_0 h-1)^{3/2}(h-h_0)}\ \ell^{-3/2}
&{\rm if}\  1<h_0<h\ ,\\
 \sqrt{\frac{h(h_0-h)\sqrt{h_0^2-1}}{(h+h_0)(h h_0-1)}}&{\rm if}\  1<h<h_0.
\end{array}
\right.
\ee

\end{enumerate}


\begin{thebibliography}{299}



\bibitem{ETH}
J. M. Deutsch, \emph{Quantum statistical mechanics in a closed
  system}, Phys. Rev. A
\href{http://dx.doi.org/10.1103/PhysRevA.43.2046}{\bf 43}, 2046
(1991);   M. Srednicki, \emph{Chaos and quantum thermalization}, Phys. Rev. E
  \href{http://dx.doi.org/10.1103/PhysRevE.50.888}{\bf 50}, 888
  (1994).


\bibitem{GM:col_rev02} M. Greiner, O. Mandel, T.W. H\"ansch, and I. Bloch, \emph{Collapse and revival of the matter wave field of a Bose?Einstein condensate}, Nature \href{\doi10.1038/nature00968}{\bf 419}, 51-54 (2002). 

\bibitem{kww-06}
T. Kinoshita, T. Wenger,  D. S. Weiss, \emph{A quantum Newton's cradle},
 Nature \href{\doi10.1038/nature04693}{\bf 440}, 900 (2006).

\bibitem{HL:Bose07} S. Hofferberth, I. Lesanovsky, B. Fischer, T. Schumm, and J. Schmiedmayer, \emph{Non-equilibrium coherence dynamics in one-dimensional Bose gases}, Nature \href{doi:10.1038/nature06149}{\bf 449}, 324-327 (2007).

\bibitem{hacker10}
L. Hackermuller, U. Schneider, M. Moreno-Cardoner, T. Kitagawa,
S. Will, T. Best, E. Demler, E. Altman, I. Bloch and B. Paredes,
\emph{Anomalous Expansion of Attractively Interacting Fermionic Atoms
  in an Optical Lattice}, Science \href{\doi10.1126/science.1184565}{\bf 327}, 1621 (2010).

\bibitem{tetal-11}
S. Trotzky Y.-A. Chen, A. Flesch, I. P. McCulloch, U. Schollw\"ock,
J. Eisert, and I. Bloch, 
\emph{Probing the relaxation towards equilibrium in an isolated strongly correlated 1D Bose gas},  
Nature Phys. \href{\doi10.1038/nphys2232}{\bf 8}, 325 (2012). 

\bibitem{getal-11}
M. Gring, M. Kuhnert, T. Langen, T. Kitagawa, B. Rauer, M. Schreitl, I. Mazets, D. A. Smith, E. Demler, and J. Schmiedmayer,
\emph{Relaxation Dynamics and Pre-thermalization in an Isolated Quantum System},
Science \href{\doi10.1126/science.1224953}{\bf 337}, 1318 (2012).

\bibitem{shr-12}
U. Schneider, L. Hackerm\"uller, J. P. Ronzheimer, S. Will, S. Braun, T. Best, I. Bloch, E. Demler, S. Mandt, D. Rasch, and A. Rosch, \emph{Fermionic transport and out-of-equilibrium dynamics in a homogeneous Hubbard model with ultracold atoms}, 
Nature Phys. \href{\doi10.1038/nphys2205}{\bf 8}, 213 (2012).

\bibitem{cetal-12}
M. Cheneau, P. Barmettler, D. Poletti, M. Endres, P. Schauss, T. Fukuhara, C. Gross, I. Bloch, C. Kollath, and S. Kuhr, \emph{Light-cone-like spreading of correlations in a quantum many-body system},
Nature \href{http://dx.doi.org/10.1038/nature10748}{\bf 481}, 484 (2012).

\bibitem{langen13}
T. Langen, R. Geiger, M. Kuhnert, B. Rauer, and J. Schmiedmayer, \emph{Local emergence of thermal correlations in an isolated quantum many-body system}, 
Nature Physics \href{http://dx.doi.org/10.1038/nphys2739}{\bf 9}, 640 (2013).

\bibitem{MM:Ising13} 
F. Meinert, M.J. Mark, E. Kirilov, K. Lauber, P. Weinmann, A.J. Daley, and H.-C. N\"agerl, \emph{Quantum Quench in an Atomic One-Dimensional Ising Chain}, Phys. Rev. Lett. \href{http://dx.doi.org/10.1103/PhysRevLett.111.053003}{\bf 111}, 053003 (2013). 

\bibitem{FK:mob13} T. Fukuhara, A. Kantian, M. Endres, M. Cheneau, P. Schau{\ss},	 S. Hild,	 D. Bellem,	 U. Schollw\"ock, T. Giamarchi, C. Gross, I. Bloch, and S. Kuhr,  \emph{Quantum dynamics of a mobile spin impurity}, Nature Physics \href{\doi10.1038/nphys2561}{\bf 9}, 235 (2013).


\bibitem{FS:magn13} T. Fukuhara, P. Schau{\ss}, M. Endres, S. Hild, M. Cheneau, I. Bloch, and C. Gross, \emph{Microscopic observation of magnon bound states and their dynamics}, Nature \href{\doi10.1038/nature12541}{\bf 502}, 76 (2013).

\bibitem{ronzheimer13}
J.P. Ronzheimer, M. Schreiber, S. Braun, S.S. Hodgman, S. Langer,
I.P. McCulloch, F. Heidrich-Meisner, I. Bloch and U. Schneider,
\emph{Expansion dynamics of interacting bosons in homogeneous lattices
  in one and two dimensions}, Phys. Rev. Lett. \href{http://dx.doi.org/10.1103/PhysRevLett.110.205301}{\bf 110}, 205301
(2013).

\bibitem{zoran1}
N. Navon, A.L. Gaunt, R.P. Smith and Z. Hadzibabic,
\emph{Critical Dynamics of Spontaneous Symmetry Breaking in a
  Homogeneous Bose gas}, Science \href{\doi10.1126/science.1258676}{\bf 347}, 167 (2015).
 
 \bibitem{jurcevic14}
P. Jurcevic, B. P. Lanyon, P. Hauke, C. Hempel, P. Zoller,
R. Blatt, and C. F. Roos, \emph{Quasiparticle engineering and entanglement propagation in a quantum many-body system}, Nature \href{http://dx.doi.org/10.1038/nature13461}{\bf 511}, 202 (2014).

\bibitem{richerme14}
P. Richerme, Z.-X. Gong, A. Lee, C. Senko, J. Smith,
M. Moss-Feig, S. Michalakis, A. V. Gorshkov, and C. Monroe, \emph{Non-local propagation of correlations in quantum systems with long-range interactions}, 
Nature \href{http://dx.doi.org/10.1038/nature13450}{\bf 511}, 198 (2014).

\bibitem{LGS16} T. Langen, T. Gasenzer and J. Schmiedmayer,
\emph{Prethermalization and universal dynamics in near-integrable
  quantum systems},
arXiv:\href{http://arxiv.org/abs/1603.09385}{1603.09385} (2016). 

\bibitem{LiebLiniger}
E. H. Lieb and W. Liniger,
\emph{Exact Analysis of an Interacting Bose Gas. I. The General Solution and the Ground State} Phys. Rev. \href{http://dx.doi.org/10.1103/PhysRev.130.1605}{\bf 130}, 1605 (1963); 
E. H. Lieb, \emph{Exact Analysis of an Interacting Bose Gas. II. The
  Excitation Spectrum}, Phys. Rev. \href{http://dx.doi.org/10.1103/PhysRev.130.1616}{\bf 130}, 1616 (1963).

\bibitem{Korepinbook}
V.E. Korepin, A.G. Izergin, and N.M. Bogoliubov, {\em {Quantum Inverse
  Scattering Method, Correlation Functions and Algebraic Bethe Ansatz}}
  (Cambridge University Press, 1993).

\bibitem{davies}
B. Davies and V.E. Korepin, \emph{Higher conservation laws for the
  quantum non-linear Schroedinger equation}, arXiv:\href{http://arxiv.org/abs/1109.6604}{1109.6604} (2011).
 
 
  \bibitem{CCR_PRL11}
A.~C. Cassidy, C.~W. Clark, and M.~Rigol, \emph{Generalized Thermalization in an Integrable Lattice System}, 
Phys. Rev. Lett. \href{http://dx.doi.org/10.1103/PhysRevLett.106.140405}{\bf 106}, 140405 (2011).

\bibitem{GR:quasid12} C. Gramsch and M. Rigol, \emph{Quenches in a quasidisordered integrable lattice system: Dynamics and statistical description of observables after relaxation}, Phys. Rev. A \href{http://dx.doi.org/10.1103/PhysRevA.86.053615}{\bf 86}, 053615 (2012).

\bibitem{WRDK:anyons} 
T.M. Wright, M. Rigol, M.J. Davis, and K.V. Kheruntsyan, \emph{Nonequilibrium Dynamics of One-Dimensional Hard-Core Anyons Following a Quench: Complete Relaxation of One-Body Observables}, Phys. Rev. Lett. \href{http://dx.doi.org/10.1103/PhysRevLett.113.050601}{\bf 113}, 050601 (2014).

\bibitem{CEF1} 
P. Calabrese, F.H.L. Essler, and M. Fagotti, \emph{Quantum Quench in the Transverse-Field Ising Chain}, Phys. Rev. Lett. \href{http://dx.doi.org/10.1103/PhysRevLett.106.227203}{\bf
  106}, 227203 (2011).
\bibitem{CEF2} 
P. Calabrese, F.H.L. Essler, and M. Fagotti, \emph{Quantum quench in the transverse field Ising chain: I. Time evolution of order parameter correlators}. J. Stat. Mech. (2012) \href{http://dx.doi.org/10.1088/1742-5468/2012/07/P07016}{P07016}.
\bibitem{CEF3} 
P. Calabrese, F.H.L. Essler, and M. Fagotti, \emph{Quantum Quench in the Transverse Field Ising Chain II: Stationary State Properties}, J. Stat. Mech. (2012) \href{http://dx.doi.org/10.1088/1742-5468/2012/07/P07022}{P07022}. 

\bibitem{FE_13a}
M.~Fagotti and F.H.L.~Essler, \emph{Reduced density matrix after a quantum quench}, Phys. Rev. B \href{http://dx.doi.org/10.1103/PhysRevB.87.245107}{\bf 87}, 245107 (2013).

\bibitem{EEF:12}
F.H.L. Essler, S. Evangelisti and M. Fagotti, \emph{Dynamical Correlations After a Quantum Quench}, 
Phys. Rev. Lett. \href{http://dx.doi.org/10.1103/PhysRevLett.109.247206}{\bf 109}, 247206 (2012).

\bibitem{F:13a}
M.~Fagotti, \emph{Finite-size corrections versus relaxation after a sudden quench}, Phys. Rev. B \href{http://dx.doi.org/10.1103/PhysRevB.87.165106}{\bf 87}, 165106 (2013).

\bibitem{qentexc} M. Kormos, L. Bucciantini, and P. Calabrese, \emph{Stationary entropies after a quench from excited states in the Ising chain}, EPL \href{http://dx.doi.org/10.1209/0295-5075/107/40002}{\bf 107} 40002 (2014).

\bibitem{qexc} L. Bucciantini, M. Kormos, P. Calabrese, \emph{Quantum quenches from excited states in the Ising chain}, J. Phys. A: Math. Theor. \href{http://dx.doi.org/10.1088/1751-8113/47/17/175002}{\bf 47} 175002 (2014).

\bibitem{efftherm} D. Rossini, A. Silva, G. Mussardo, and G.E. Santoro, \emph{Effective Thermal Dynamics Following a Quantum Quench in a Spin Chain}, Phys. Rev. Lett. \href{http://dx.doi.org/10.1103/PhysRevLett.102.127204}{\bf 102}, 127204 (2009); D. Rossini, S. Suzuki, G. Mussardo, G.E. Santoro, and A. Silva, \emph{Long time dynamics following a quench in an integrable quantum spin chain: Local versus nonlocal operators and effective thermal behavior}, Phys. Rev. B \href{http://dx.doi.org/10.1103/PhysRevB.82.144302}{\bf 82}, 144302 (2010). 

\bibitem{IR:quench00} F. Ig\'oi and H. Rieger, \emph{Long-Range Correlations in the Nonequilibrium Quantum Relaxation of a Spin Chain}, Phys. Rev. Lett. \href{http://dx.doi.org/10.1103/PhysRevLett.85.3233}{\bf 85}, 3233 (2000). 

\bibitem{SPS:04} K. Sengupta, S. Powell, and S. Sachdev, \emph{Quench dynamics across quantum critical points}, Phys. Rev. A \href{http://dx.doi.org/10.1103/PhysRevA.69.053616}{\bf 69}, 053616 (2004). 

\bibitem{S:work} A. Silva, \emph{Statistics of the Work Done on a Quantum Critical System by Quenching a Control Parameter}, Phys. Rev. Lett. \href{http://dx.doi.org/10.1103/PhysRevLett.101.120603}{\bf 101}, 120603 (2008).

\bibitem{IR:quenchB} F. Igl\'oi and H. Rieger, \emph{Quantum Relaxation after a Quench in Systems with Boundaries}, Phys. Rev. Lett. \href{http://dx.doi.org/10.1103/PhysRevLett.106.035701}{\bf 106}, 035701 (2011).

\bibitem{FCG:f-d11} L. Foini, L.F. Cugliandolo, and A. Gambassi, \emph{Fluctuation-dissipation relations and critical quenches in the transverse field Ising chain}, Phys. Rev. B \href{http://dx.doi.org/10.1103/PhysRevB.84.212404}{\bf 84}, 212404 (2011).

\bibitem{RI:sc11} H. Rieger and F. Igl\'oi, \emph{Semiclassical theory for quantum quenches in finite transverse Ising chains}, Phys. Rev. B \href{http://dx.doi.org/10.1103/PhysRevB.84.165117}{\bf 84}, 165117 (2011).

\bibitem{SE:Ising} D. Schuricht and F.H.L. Essler, \emph{Dynamics in the Ising field theory after a quantum quench}, J. Stat. Mech. (2012) \href{http://dx.doi.org/10.1088/1742-5468/2012/04/P04017}{P04017}.

\bibitem{FCG:dyn12} L. Foini, L.F. Cugliandolo, and A. Gambassi, \emph{Dynamic correlations, fluctuation-dissipation relations, and effective temperatures after a quantum quench of the transverse field Ising chain}, J. Stat. Mech. (2012) \href{http://dx.doi.org/10.1088/1742-5468/2012/09/P09011}{P09011}.


\bibitem{HPL:dyn13} 
M. Heyl, A. Polkovnikov, and S. Kehrein, \emph{Dynamical Quantum Phase Transitions in the Transverse-Field Ising Model}, Phys. Rev. Lett. \href{http://dx.doi.org/10.1103/PhysRevLett.110.135704}{\bf 110}, 135704 (2013).

\bibitem{CC:GGEIs11} T. Caneva, E. Canovi, D. Rossini, G.E. Santoro, and A. Silva, \emph{Applicability of the generalized Gibbs ensemble after a quench in the quantum Ising chain}, J. Stat. Mech. (2011) \href{http://dx.doi.org/10.1088/1742-5468/2011/07/P07015}{P07015}.

\bibitem{ARRS99}
T. Antal, Z. Racz, A. Rakos, and G. M. Schutz, \emph{Transport in
  the XX chain at zero temperature: Emergence of flat magnetization
  profiles}, Phys. Rev. E \href{http://dx.doi.org/10.1103/PhysRevE.59.4912}{\bf 59}, 4912 (1999). 
  
\bibitem{BMD70}
E. Barouch, B. McCoy, and M. Dresden, \emph{Statistical Mechanics of
  the XY Model. I}, Phys. Rev. A
\href{http://dx.doi.org/10.1103/PhysRevA.2.1075}{\bf 2}, 1075 (1970).

\bibitem{BM71a}
E. Barouch and B. McCoy,  \emph{Statistical Mechanics of the XY Model. II},
Phys. Rev. A  \href{http://dx.doi.org/10.1103/PhysRevA.3.786}{\bf 3}, 786 (1971).
\bibitem{BM71b}
E. Barouch and B. McCoy, \emph{Statistical Mechanics of the XY Model. III},
Phys. Rev. A  \href{http://dx.doi.org/10.1103/PhysRevA.3.2137}{\bf 3}, 2137 (1971).

\bibitem{FC08}
M. Fagotti and P. Calabrese, \emph{Evolution of entanglement entropy following a quantum quench: Analytic results for the XY chain in a transverse magnetic field}, Phys. Rev. A \href{http://dx.doi.org/10.1103/PhysRevA.78.010306}{\bf 78}, 010306(R) (2008).

\bibitem{CIC_PRE11}
M. A. Cazalilla, A. Iucci, and M.-C. Chung, \emph{Thermalization and quantum correlations in exactly solvable models}, Phys. Rev. E \href{http://dx.doi.org/10.1103/PhysRevE.85.011133}{\bf 85},
011133 (2012). 

\bibitem{BRI:sc_XY} B. Blass, H. Rieger, and F. Igl\'oi, \emph{Quantum relaxation and finite-size effects in the XY chain in a transverse field after global quenches}, EPL \href{http://dx.doi.org/10.1209/0295-5075/99/30004}{\bf 99} 30004 (2012).

\bibitem{mau-prep}
M. Fagotti, \emph{On conservation laws, relaxation and pre-relaxation after a quantum quench}, J. Stat. Mech. (2014) \href{ttp://dx.doi.org/10.1088/1742-5468/2014/03/P03016}{P03016}.

\bibitem{CLopenXY} M. Fagotti, \emph{Local conservation laws in
  spin-1/2 XY chains with open boundary conditions},
  J. Stat. Mech. (2016) 063105
  (doi:10.1088/1742-5468/2016/06/063105). 


\bibitem{gaplessXXZ} M. Collura, P. Calabrese, and F.H.L. Essler, \emph{Quantum quench within the gapless phase of the spin-1/2 Heisenberg XXZ spin chain}, Phys. Rev. B \href{http://dx.doi.org/10.1103/PhysRevB.92.125131}{\bf 92}, 125131 (2015).

\bibitem{LM:XXZ10} J. Lancaster and A. Mitra, \emph{Quantum quenches in an XXZ spin chain from a spatially inhomogeneous initial state}, Phys. Rev. E \href{http://dx.doi.org/10.1103/PhysRevE.81.061134}{\bf 81}, 061134 (2010).

\bibitem{SM13}
T. Sabetta and G. Misguich, \emph{Nonequilibrium steady states in the
  quantum XXZ spin chain}, Phys. Rev. B \href{\doi10.1103/PhysRevB.88.245114}{\bf 88}, 245114 (2013).
  
  \bibitem{bonnes14}
L. Bonnes, F.H.L. Essler and A. L\"auchli, \emph{``Light-Cone'' Dynamics After Quantum Quenches in Spin Chains}, 
Phys. Rev. Lett. \href{http://dx.doi.org/10.1103/PhysRevLett.113.187203}{\bf 113}, 187203 (2014).

\bibitem{nonint} 
E. Canovi, D. Rossini, R. Fazio, G. Santoro, and A. Silva, \emph{Many-body localization and thermalization in the full probability distribution function of observables}, 
New J. Phys. \href{http://dx.doi.org/10.1088/1367-2630/14/9/095020}{\bf 14}, 095020 (2012).

\bibitem{FE_13b}
M.~Fagotti and F.H.L. Essler, \emph{Stationary behaviour of observables after a quantum quench in the spin-1/2 Heisenberg XXZ chain}, J. Stat. Mech. (2013) \href{http://dx.doi.org/10.1088/1742-5468/2013/07/P07012}{P07012}.

\bibitem{Pozsgay:13a}
B. Pozsgay, \emph{The generalized Gibbs ensemble for Heisenberg spin chains}, J. Stat. Mech. (2013) \href{http://dx.doi.org/10.1088/1742-5468/2013/07/P07003}{P07003}.

\bibitem{fp-13}
M. Fagotti, \emph{Dynamical Phase Transitions as Properties of the Stationary State: Analytic Results after Quantum Quenches in the Spin-1/2 XXZ Chain},   arXiv:1308.0277;
B. Pozsgay, \emph{Dynamical free energy and the Loschmidt-echo for a class of quantum quenches in the Heisenberg spin chain}, Stat. Mech. (2013) \href{http://dx.doi.org/10.1088/1742-5468/2013/10/P10028}{P10028}.

\bibitem{la-13}
W. Liu and N. Andrei, \emph{Quench Dynamics of the Anisotropic Heisenberg Model}, Phys. Rev. Lett. \href{http://dx.doi.org/10.1103/PhysRevLett.112.257204}{\bf 112}, 257204 (2014).

\bibitem{bpgda-10}
P. Barmettler, M. Punk, V. Gritsev, E. Demler, and E. Altman, 
\emph{Relaxation of antiferromagnetic order in spin-1/2 chains following a quantum quench}, 
Phys. Rev. Lett.  \href{http://dx.doi.org/10.1103/PhysRevLett.102.130603}{\bf 102}, 130603 (2009); 
P. Barmettler, M. Punk, V. Gritsev, E. Demler, and E. Altman,
\emph{Quantum quenches in the anisotropic spin-1/2 Heisenberg chain: different approaches to many-body dynamics far from equilibrium},
New J. Phys. \href{http://dx.doi.org/10.1088/1367-2630/12/5/055017}{\bf 12}, 055017 (2010).

\bibitem{FCEC_14} M, Fagotti, M, Collura, F.H.L. Essler, and P. Calabrese, \emph{Relaxation after quantum quenches in the spin-1/2 Heisenberg XXZ chain}, Phys. Rev. B \href{http://dx.doi.org/10.1103/PhysRevB.89.125101}{\bf 89}, 125101 (2014).

\bibitem{GGE_int} E. Ilievski, J. De Nardis, B. Wouters, J.-S. Caux, F.H.L. Essler, T. Prosen, \emph{Complete Generalized Gibbs Ensemble in an interacting Theory}, Phys. Rev. Lett. \href{http://dx.doi.org/10.1103/PhysRevLett.115.157201}{\bf 115}, 157201 (2015).

\bibitem{QANeel} M. Brockmann, B. Wouters, D. Fioretto, J. De Nardis, R. Vlijm and J.-S. Caux, \emph{Quench action approach for releasing the N\'eel state into the spin-1/2 XXZ chain}, Stat. Mech. (2014) \href{http://dx.doi.org/10.1088/1742-5468/2014/12/P12009}{P12009}.

\bibitem{GGE_XXZ_Amst} B. Wouters, J. De Nardis, M. Brockmann, D. Fioretto, M. Rigol, and J.-S. Caux, \emph{Quenching the Anisotropic Heisenberg Chain: Exact Solution and Generalized Gibbs Ensemble Predictions}, Phys. Rev. Lett. \href{http://dx.doi.org/10.1103/PhysRevLett.113.117202}{\bf 113}, 117202 (2014).

\bibitem{p-13a}
B. Pozsgay, \emph{Overlaps between eigenstates of the XXZ spin-1/2 chain and a class of simple product states}, J. Stat. Mech. (2014) \href{http://dx.doi.org/10.1088/1742-5468/2014/06/P06011}{P06011}.

\bibitem{XXZung} B. Pozsgay, M. Mesty\'an, M.A. Werner, M. Kormos,
  G. Zar\'and, and G. Tak\'acs, \emph{Correlations after Quantum
    Quenches in the XXZ Spin Chain: Failure of the Generalized Gibbs
    Ensemble},
  Phys. Rev. Lett. \href{http://dx.doi.org/10.1103/PhysRevLett.113.117203}{\bf
    113}, 117203 (2014).

\bibitem{XXZunglong} M. Mesty\'an, B. Pozsgay, G. Tak\'acs, and M.A. Werner, \emph{Quenching the XXZ spin chain: quench action approach versus generalized Gibbs ensemble}, J. Stat. Mech. (2015) \href{http://dx.doi.org/10.1088/1742-5468/2015/04/P04001}{P04001}.

\bibitem{R:gs14} 
M. Rigol, \emph{Quantum quenches in the thermodynamic limit. II. Initial ground states}, Phys. Rev. E \href{http://dx.doi.org/10.1103/PhysRevE.90.031301}{\bf 90}, 031301(R) (2014).


\bibitem{MC:XXZ10} J. Mossel and J.-S. Caux, \emph{Relaxation dynamics in the gapped XXZ spin-1/2 chain}, New J. Phys. \href{http://dx.doi.org/10.1088/1367-2630/12/5/055028}{\bf 12} 055028 (2010).

\bibitem{AC:qa15} V. Alba and P. Calabrese, \emph{The quench action approach in finite integrable spin chains}, arXiv:\href{http://arxiv.org/abs/1512.02213}{1512.02213} (2015).





\bibitem{KE:Mott08} 
M. Kollar and M. Eckstein, \emph{Relaxation of a one-dimensional Mott insulator after an interaction quench}, Phys. Rev. A \href{http://dx.doi.org/10.1103/PhysRevA.78.013626}{\bf 78}, 013626 (2008).

\bibitem{ES12}
T. Enss and J. Sirker, \emph{Lightcone renormalization and quantum
quenches in one-dimensional Hubbard models}, New J. Phys. \href{\doi10.1088/1367-2630/14/2/023008}{\bf 14}, 023008 (2012).

\bibitem{IMWR:14} 
D. Iyer, R. Mondaini, S. Will, and M. Rigol, \emph{Coherent quench dynamics in the one-dimensional Fermi-Hubbard model}, Phys. Rev. A \href{http://dx.doi.org/10.1103/PhysRevA.90.031602}{\bf 90}, 031602(R) (2014). 

\bibitem{ROHM15}
L. Riegger, G. Orso and F. Heidrich-Meisner,
\emph{Interaction quantum quenches in the one-dimensional
  Fermi-Hubbard model with spin imbalance}
Phys. Rev. A \href{http://dx.doi.org/10.1103/PhysRevA.91.043623}{\bf 91}, 043623 (2015).

\bibitem{BDHM15}
A. Bauer, F. Dorfner and F. Heidrich-Meisner,
\emph{Temporal decay of N\'eel order in the one-dimensional Fermi-Hubbard model},
Phys. Rev. A \href{http://dx.doi.org/10.1103/PhysRevA.91.053628}{\bf 91}, 053628 (2015).





\bibitem{Sp_th} S. Sotiriadis, P. Calabrese, and J. Cardy,
\emph{Quantum quench from a thermal initial state}, EPL
\href{http://dx.doi.org/10.1209/0295-5075/87/20002}{\bf 87} 20002 (2009). 

\bibitem{Klein-Gordon}
S. Sotiriadis, G. Martelloni, \emph{Equilibration and GGE in
  interacting-to-free Quantum Quenches in dimensions $d>1$},
J. Phys. A: Math. Theor. \href{http://dx.doi.org/10.1088/1751-8113/49/9/095002}{\bf
  49} 095002 (2016). 

\bibitem{cc-07}
P. Calabrese and J. Cardy, \emph{Quantum quenches in extended systems}, J. Stat. Mech. (2007) \href{http://dx.doi.org/10.1088/1742-5468/2007/06/P06008}{P06008}.


\bibitem{C:LL06} M.A. Cazalilla, \emph{Effect of Suddenly Turning on Interactions in the Luttinger Model}, Phys. Rev. Lett. \href{http://dx.doi.org/10.1103/PhysRevLett.97.156403}{\bf 97}, 156403 (2006).

\bibitem{IC_PRA09}
A. Iucci and M. A. Cazalilla, \emph{Quantum quench dynamics of the Luttinger model}, Phys. Rev. A \href{http://dx.doi.org/10.1103/PhysRevA.80.063619}{\bf 80}, 063619 (2009).

\bibitem{RS:T-L12} J. Rentrop, D. Schuricht and V. Meden, \emph{Quench dynamics of the Tomonaga-Luttinger model with momentum-dependent interaction}, New J. Phys. \href{http://dx.doi.org/10.1088/1367-2630/14/7/075001}{\bf14} 075001 (2012).

\bibitem{KR:LL12} C. Karrasch, J. Rentrop, D. Schuricht, and V. Meden, \emph{Luttinger-liquid universality in the time evolution after an interaction quench}, Phys. Rev. Lett. \href{http://dx.doi.org/10.1103/PhysRevLett.109.126406}{\bf 109}, 126406 (2012).

\bibitem{DBZ:GGE12} B. D\'ora, \'A. B\'acsi, and G. Zar\'and, \emph{Generalized Gibbs ensemble and work statistics of a quenched Luttinger liquid}, Phys. Rev. B \href{http://dx.doi.org/10.1103/PhysRevB.86.161109}{\bf 86}, 161109(R) (2012).

\bibitem{NI:Coul13} N. Nessi and A. Iucci, \emph{Quantum quench dynamics of the Coulomb Luttinger model}, Phys. Rev. B \href{http://dx.doi.org/10.1103/PhysRevB.87.085137}{\bf 87}, 085137 (2013).
 

\bibitem{RevCaz} M. A. Cazalilla, M.-C. Chung, \emph{Quantum Quenches in the Luttinger model and its close relatives}, arXiv:\href{http://arxiv.org/abs/1603.04252}{1603.04252} (2016).

\bibitem{CCrev} P. Calabrese and J. Cardy, \emph{Quantum quenches in 1+1 dimensional conformal field theories}, arXiv:\href{http://arxiv.org/abs/1603.02889}{1603.02889} (2016)

\bibitem{cc-05}   
P. Calabrese and  J. Cardy, \emph{Evolution of entanglement entropy in one-dimensional systems}, J. Stat. Mech. (2005) \href{http://dx.doi.org/10.1088/1742-5468/2005/04/P04010}{P04010}. 

\bibitem{CC:tcf06} 
P. Calabrese and J. Cardy, \emph{Time Dependence of Correlation Functions Following a Quantum Quench}, Phys. Rev. Lett. \href{http://dx.doi.org/10.1103/PhysRevLett.96.136801}{\bf 96}, 136801 (2006).

\bibitem{Cardy14}
J. Cardy, \emph{Thermalization and Revivals after a Quantum Quench in Conformal Field Theory}, Phys. Rev. Lett. \href{http://dx.doi.org/10.1103/PhysRevLett.112.220401}{\bf 112}, 220401 (2014).

\bibitem{non-abSpyr} 
S. Sotiriadis, \emph{Memory-preserving equilibration after a quantum quench in a 1d critical model}, arXiv:\href{http://arxiv.org/abs/1507.07915}{1507.07915} (2015).

\bibitem{Cardy_non-ab} 
J. Cardy, \emph{Quantum quenches to a critical point in one dimension: some further results}, J. Stat. Mech. (2016) \href{http://dx.doi.org/10.1088/1742-5468/2016/02/023103}{023103}.


\bibitem{GD:spectro07} 
V. Gritsev, E. Demler, M. Lukin, and A. Polkovnikov, \emph{Spectroscopy of Collective Excitations in Interacting Low-Dimensional Many-Body Systems Using Quench Dynamics}, Phys. Rev. Lett. \href{http://dx.doi.org/10.1103/PhysRevLett.99.200404}{\bf 99}, 200404 (2007).

\bibitem{sine-Gordon} 
B. Bertini, D. Schuricht, and F.H.L. Essler, \emph{Quantum quench in the sine-Gordon model}, J. Stat. Mech. (2014) \href{http://dx.doi.org/10.1088/1742-5468/2014/10/P10035}{P10035}. 

\bibitem{sc_sineG} 
M. Kormos and G. Zar\'and, \emph{Quantum quenches in the sine-Gordon model: a semiclassical approach}, arXiv:\href{http://arxiv.org/abs/1507.02708}{1507.02708} (2015).

\bibitem{FM_NJPhys10}
D. Fioretto and G. Mussardo, \emph{Quantum quenches in integrable field theories}, New J. Phys. \href{http://dx.doi.org/10.1088/1367-2630/12/5/055015}{\bf 12}, 055015 (2010). 

\bibitem{sinh-Gordon} B. Bertini, L. Piroli, P. Calabrese, \emph{Quantum quenches in the sinh-Gordon model: steady state and one point correlation functions}, arXiv:\href{http://arxiv.org/abs/1602.08269}{1602.08269} (2016).

\bibitem{Pozsgay11} 
B.~Pozsgay, \emph{Mean values of local operators in highly excited Bethe states}, J. Stat. Mech. (2011) \href{http://dx.doi.org/10.1088/1742-5468/2011/01/P01011}{P01011}.

\bibitem{muss13}
G. Mussardo, \emph{Infinite-Time Average of Local Fields in an Integrable Quantum Field Theory After a Quantum Quench}, Phys. Rev. Lett. \href{http://dx.doi.org/10.1103/PhysRevLett.111.100401}{\bf 111}, 100401 (2013).




\bibitem{sc-O3} 
S. Evangelisti, \emph{Semi-classical theory for quantum quenches in the O(3) non-linear sigma model}, J. Stat. Mech. (2013) \href{http://dx.doi.org/10.1088/1742-5468/2013/04/P04003}{P04003}. 



\bibitem{austen}
A. Lamacraft, \emph{Noise correlations in the expansion of an interacting one-dimensional Bose gas from a regular array}, Phys. Rev. A \href{http://dx.doi.org/10.1103/PhysRevA.84.043632}{\bf 84}, 043632 (2011).

\bibitem{MC_NJPhys12}
J.~Mossel and J.-S. Caux, \emph{Exact time evolution of space- and time-dependent correlation functions after an interaction quench in the one-dimensional Bose gas}, New J. Phys. \href{http://dx.doi.org/10.1088/1367-2630/14/7/075006}{\bf 14}, 075006 (2012).

\bibitem{CSC:13b}
M. Collura, S. Sotiriadis and P. Calabrese, \emph{Equilibration of a
  Tonks-Girardeau Gas Following a Trap Release},  
Phys. Rev. Lett. \href{http://dx.doi.org/10.1103/PhysRevLett.110.245301}{\bf 110}, 245301 (2013); \emph{Quench dynamics of a Tonks-Girardeau gas released from a harmonic trap}, J. Stat. Mech. (2013) \href{http://dx.doi.org/10.1088/1742-5468/2013/09/P09025}{P09025}.

\bibitem{KSCCI}
M. Kormos, A. Shashi, Y.-Z. Chou, J.-S. Caux and A. Imambekov, \emph{Interaction quenches in the one-dimensional Bose gas}, Phys. Rev. B \href{http://dx.doi.org/10.1103/PhysRevB.88.205131}{\bf 88}, 205131(2013)

\bibitem{releBose} 
M. Collura, S. Sotiriadis, and P. Calabrese, \emph{Quench dynamics of a Tonks-Girardeau gas released from a harmonic trap}, J. Stat. Mech. (2013) \href{http://dx.doi.org/10.1088/1742-5468/2013/09/P09025}{P09025}.


\bibitem{nwbc-13}
J. De Nardis, B. Wouters, M. Brockmann, and J.-S. Caux, \emph{Solution for an interaction quench in the Lieb-Liniger Bose gas}, Phys. Rev. A \href{http://dx.doi.org/10.1103/PhysRevA.89.033601}{\bf 89}, 033601 (2014).

\bibitem{trappedBose} 
P.P. Mazza, M. Collura, M. Kormos, and P. Calabrese, \emph{Interaction quench in a trapped 1D Bose gas}, J. Stat. Mech. (2014) \href{http://dx.doi.org/10.1088/1742-5468/2014/11/P11016}{P11016}.

\bibitem{enttrap} 
M. Collura, M. Kormos, and P. Calabrese, \emph{Stationary entanglement entropies following an interaction quench in 1D Bose gas}, J. Stat. Mech. (2014) \href{http://dx.doi.org/10.1088/1742-5468/2014/01/P01009}{P01009}.

\bibitem{KCC14}
M. Kormos, M. Collura and P. Calabrese, \emph{Analytic results for a quantum quench from free to hard-core one-dimensional bosons}, Phys. Rev. A \href{http://dx.doi.org/10.1103/PhysRevA.89.013609}{\bf  89}, 013609 (2014).

\bibitem{quenchLL} J. De Nardis and J.-S. Caux, \emph{Analytical expression for a post-quench time evolution of the one-body density matrix of one-dimensional hard-core bosons}, J. Stat. Mech. (2014) \href{http://dx.doi.org/10.1088/1742-5468/2014/12/P12012}{P12012}.

\bibitem{GGEqft} F.H.L. Essler, G. Mussardo, and M. Panfil, \emph{Generalized Gibbs ensembles for quantum field theories}, Phys. Rev. A \href{http://dx.doi.org/10.1103/PhysRevA.91.051602}{\bf 91}, 051602(R) (2015).

\bibitem{rel_int} 
J. De Nardis, L. Piroli and J.-S. Caux, \emph{Relaxation dynamics of local observables in integrable systems}, Phys. A: Math. Theor. \href{http://dx.doi.org/10.1088/1751-8113/48/43/43FT01}{\bf 48} 43FT01 (2015).

\bibitem{bound_Bose} L. Piroli, P. Calabrese, and F.H.L. Essler, \emph{Multiparticle Bound-State Formation following a Quantum Quench to the One-Dimensional Bose Gas with Attractive Interactions}, Phys. Rev. Lett. \href{http://dx.doi.org/10.1103/PhysRevLett.116.070408}{\bf 116}, 070408 (2016).



\bibitem{MarioRamp} M. Collura and D. Karevski, \emph{Critical Quench Dynamics in Confined Systems}, Phys. Rev. Lett. \href{http://dx.doi.org/10.1103/PhysRevLett.104.200601}{\bf 104}, 200601 (2010).

\bibitem{ramps}
J. Dziarmaga, \emph{Dynamics of a Quantum Phase Transition and
  Relaxation to a Steady State}, Adv. Phys. \href{10.1080/00018732.2010.514702}{\bf 59}, 1063 (2010);
C. De Grandi, V. Gritsev and A. Polkovnikov,
\emph{Quench dynamics near a quantum critical point: application to
the sine-Gordon model}, Phys. Rev. B \href{10.1103/PhysRevB.81.224301}{\bf 81}, 224301 (2010);
A. Dutta, G. Aeppli, B.K. Chakrabarti, U. Divakaran, T.F. Rosenbaum and
D. Sen, \emph{Quantum phase transitions in transverse field spin
  models: from statistical physics to quantum information}, Cambridge
University Press, Cambridge 2015.

\bibitem{KibbleZurek}
W.H. Zurek, \emph{Cosmological experiments in condensed matter
  systems}, Phys. Rep. \href{\doi10.1016/S0370-1573(96)00009-9}{\bf 276}, 177 (1996).

\bibitem{review} A. Polkovnikov, K. Sengupta, A. Silva, and M. Vengalattore, \emph{Colloquium: Nonequilibrium dynamics of closed interacting quantum systems}, 
Rev. Mod. Phys. \href{http://dx.doi.org/10.1103/RevModPhys.83.863}{\bf 83}, 863 (2011). 

\bibitem{Eis_rev}
C. Gogolin and J. Eisert, \emph{Equilibration, thermalisation, and the
  emergence of statistical mechanics in closed quantum systems - a
  review}, arXiv:\href{http://arxiv.org/abs/1503.07538}{1503.07538} (2015).
  
  \bibitem{Austenreview}
A. Lamacraft and  J.E. Moore, \emph{Potential insights into
non-equilibrium behavior from atomic physics}, in  \emph{Ultracold
Bosonic and Fermionic Gases}, eds A. Fetter, K. Levin and
D. Stamper-Kurn, Elseview 2012;  arXiv:\href{http://arxiv.org/abs/1106.3567}{1106.3567} (2011).

\bibitem{DKPR16}
L. D'Alessio, Y. Kafri, A. Polkovnikov and M. Rigol,
\emph{From Quantum Chaos and Eigenstate Thermalization to Statistical
  Mechanics and Thermodynamics},  arXiv:\href{http://arxiv.org/abs/1509.06411}{1509.06411} (2015).
  
  \bibitem{BCS}
R. A. Barankov, L. S. Levitov, 
\emph{Synchronization in the BCS Pairing Dynamics as a Critical Phenomenon},
Phys. Rev. Lett. \href{\doi10.1103/PhysRevLett.96.230403}{\bf96}, 230403 (2006);
E.A. Yuzbashyan and M. Dzero, \emph{Dynamical vanishing of the order
  parameter in a fermionic condensate}, Phys. Rev. Lett. \href{http://dx.doi.org/10.1103/PhysRevLett.96.230404}{\bf 96}, 230404 (2006).


\bibitem{vidal03}
G. Vidal, J.I. Latorre, E. Rico, and A. Kitaev,
  \emph{Entanglement in Quantum Critical Phenomena},
  Phys. Rev. Lett. \href{http://dx.doi.org/10.1103/PhysRevLett.90.227902}{\bf
    90}, 227902 (2003).

\bibitem{peschel}
I. Peschel and V. Eisler, \emph{Reduced density matrices and
  entanglement entropy in free lattice models}, J. Phys. A:
Math. Theor. \href{\doi10.1088/1751-8113/42/50/504003}{\bf 42} 504003 (2009).

\bibitem{arealaw} 
F. Verstraete, M.M. Wolf,  D. Perez-Garcia, and J.I. Cirac, \emph{Criticality, the Area Law, and the Computational Power of Projected Entangled Pair States}, Phys. Rev. Lett. \href{http://dx.doi.org/10.1103/PhysRevLett.96.220601}{\bf 96} 220601 (2006); M. B. Hastings, \emph{Entropy and entanglement in quantum ground states}, Phys. Rev. B \href{http://dx.doi.org/10.1103/PhysRevB.76.035114}{\bf 76}, 035114 (2007); N. de Beaudrap, T.J. Osborne, and J. Eisert, \emph{Ground states of unfrustrated spin Hamiltonians satisfy an area law}, New J. Phys. \href{http://dx.doi.org/10.1088/1367-2630/12/9/095007}{\bf 12} 095007 (2010).


\bibitem{BS_PRL08}
T. Barthel and U. Schollw\"ock, \emph{Dephasing and the Steady State in Quantum Many-Particle Systems}, 
Phys. Rev. Lett. \href{http://dx.doi.org/10.1103/PhysRevLett.100.100601}{\bf 100}, 100601 (2008).




\bibitem{preT_entropy} M. Fagotti and M. Collura, \emph{Universal prethermalization dynamics of entanglement entropies after a global quench}, arXiv:\href{http://arxiv.org/abs/1507.02678}{1507.02678} (2015).





\bibitem{clusterGGE} S. Sotiriadis and P. Calabrese, \emph{Validity of the GGE for quantum quenches from interacting to noninteracting models}, J. Stat. Mech. (2014) \href{http://dx.doi.org/10.1088/1742-5468/2014/07/P07024}{P07024}. 



\bibitem{BAW06a}
E. Bettelheim, A.G. Abanov and P. Wiegmann, \emph{Quantum Shock Waves
  - the case for non-linear effects in dynamics of electronic liquids},
Phys. Rev. Lett. \href{\doi10.1103/PhysRevLett.97.246401}{\bf97}, 246401 (2006).

\bibitem{BAW06b}
E. Bettelheim, A.G. Abanov and P. Wiegmann, \emph{Orthogonality
  catastrophe and shock waves in a non-equilibrium Fermi gas},
Phys.Rev.Lett. \href{\doi10.1103/PhysRevLett.97.246402}{\bf 97}, 246402 (2006).

\bibitem{SC08}
S. Sotiriadis and J. Cardy, \emph{Inhomogeneous Quantum Quenches},
J. Stat. Mech. \href{\doi10.1088/1742-5468/2008/11/P11003}{P11003} (2008).

\bibitem{CHL08}
P. Calabrese, C. Hagendorf, and P. Le Doussal, \emph{Time evolution of
1D gapless models from a domain-wall initial state: SLE continued?},
J. Stat. Mech. \href{\doi10.1088/1742-5468/2008/07/P07013}{P07013} (2008). 

\bibitem{BW11}
E. Bettelheim and P. Wiegmann, \emph{Fermi distribution of
  semicalssical non-eqilibrium Fermi states}, Phys. Rev. B \href{http://dx.doi.org/10.1103/PhysRevB.84.085102}{\bf 84},
085102 (2011).

\bibitem{HM13}
L. Vidmar, S. Langer, I.P. McCulloch, U. Schneider, U. Schollwoeck and
F. Heidrich-Meisner,
\emph{Sudden expansion of Mott insulators in one dimension}
Phys. Rev. B \href{\doi10.1103/PhysRevB.88.235117}{\bf 88}, 235117 (2013).

\bibitem{ER13}
V. Eisler and Z. Racz, \emph{Full Counting Statistics in a Propagating
  Quantum Front and Random Matrix Spectra}, Phys. Rev. Lett. \href{http://dx.doi.org/10.1103/PhysRevLett.110.060602}{\bf
  110}, 060602 (2013). 

\bibitem{RS:Cal14}  
M.A. Rajabpour and S. Sotiriadis, \emph{Quantum quench of the trap frequency in the harmonic Calogero model}, Phys. Rev. A \href{http://dx.doi.org/10.1103/PhysRevA.89.033620}{\bf 89}, 033620 (2014).

\bibitem{HM16}
Z. Mei, L. Vidmar, F. Heidrich-Meisner and C. J. Bolech,
\emph{Unveiling hidden structure of many-body wavefunctions of
  integrable systems via sudden expansion experiments},
Phys. Rev. A \href{\doi10.1103/PhysRevA.93.021607}{\bf 93}, 021607 (2016).


\bibitem{AP:XY03} W.H. Aschbacher and C.-A. Pillet, \emph{Non-Equilibrium Steady States of the XY Chain}, J. Stat. Phys. \href{\doi10.1023/A:1024619726273}{\bf 112}, 1153 (2003).
  
  \bibitem{AB:XY06} W.H. Aschbacher and J.-M. Barbaroux, \emph{Out of Equilibrium Correlations in the XY Chain}, Lett. Math. Phys. \href{\doi10.1007/s11005-006-0049-7}{\bf 77}, 11 (2006).
  
  \bibitem{BD:EnCFT12} D. Bernard and B. Doyon, \emph{Energy flow in non-equilibrium conformal field theory}, J. Phys. A: Math. Theor. \href{http://dx.doi.org/10.1088/1751-8113/45/36/362001}{\bf 45}, 362001 (2012).
  
  \bibitem{CK_PRL12}
J.-S. Caux and R.~M. Konik, \emph{Constructing the Generalized Gibbs Ensemble after a Quantum Quench}, Phys. Rev. Lett. \href{http://dx.doi.org/10.1103/PhysRevLett.109.175301}{\bf 109}, 175301 (2012).

  \bibitem{BD:nessCFT} D. Bernard and B. Doyon, \emph{Non-Equilibrium Steady States in Conformal Field Theory}, Ann. H. Poincar\'e \href{\doi10.1007/s00023-014-0314-8}{\bf 16}, 113 (2015).
  
  \bibitem{MS:NESSLutt} M. Mintchev and P. Sorba, \emph{Luttinger liquid in a non-equilibrium steady state}, J. Phys. A: Math. Theor. \href{http://dx.doi.org/10.1088/1751-8113/46/9/095006}{\bf 49}, 095006 (2013).
  
 \bibitem{NESSIsing} A. De Luca, J. Viti, D. Bernard, and B. Doyon, \emph{Nonequilibrium thermal transport in the quantum Ising chain}, Phys. Rev. B \href{http://dx.doi.org/10.1103/PhysRevB.88.134301}{\bf 88}, 134301 (2013).
  
  \bibitem{SM:NESSXXZ} T. Sabetta and G. Misguich, \emph{Nonequilibrium steady states in the quantum XXZ spin chain}, Phys. Rev. B \href{http://dx.doi.org/10.1103/PhysRevB.88.245114}{\bf 88}, 245114 (2013). 
  
  \bibitem{DHB:flow14} B. Doyon, M. Hoogeveen, and D. Bernard, \emph{Energy flow and fluctuations in non-equilibrium conformal field theory on star graphs}, Stat. Mech. (2014) \href{http://dx.doi.org/10.1088/1742-5468/2014/03/P03002}{P03002}. 
  
  \bibitem{C-A:QFT14} O. Castro-Alvaredo, Y. Chen, B. Doyon, and M. Hoogeveen, \emph{Thermodynamic Bethe ansatz for non-equilibrium steady states: exact energy current and fluctuations in integrable QFT}, J. Stat. Mech. (2014) \href{http://dx.doi.org/10.1088/1742-5468/2014/03/P03011}{P03011}. 
  
  \bibitem{LMV:freeqa14} A. De Luca, G. Martelloni, and J. Viti, \emph{Stationary states in a free fermionic chain from the quench action method}, Phys. Rev. A \href{http://dx.doi.org/10.1103/PhysRevA.91.021603}{\bf 91}, 021603(R) (2014).
  
  \bibitem{DLSB:K-G15} B. Doyon, A. Lucas, K. Schalm, and M.J Bhaseen, \emph{Non-equilibrium steady states in the Klein-Gordon theory}, J. Phys. A: Math. Theor. \href{http://dx.doi.org/10.1088/1751-8113/48/9/095002}{48} 095002 (2015).
  
  \bibitem{D:bal15} B. Doyon, \emph{Lower bounds for ballistic current and noise in non-equilibrium quantum steady states}, Nucl. Phys. B \href{\doi10.1016/j.nuclphysb.2015.01.007}{\bf 892}, 190 (2015). 
  
  \bibitem{VSDH:inh} J. Viti; J.-M. St\'ephan; J. Dubail; M. Haque, \emph{Inhomogeneous quenches in a fermionic chain: exact results}, arXiv:\href{http://arxiv.org/abs/1507.08132}{1507.08132} (2015). 
  
  \bibitem{PK:XX07} T. Platini and D. Karevski, \emph{Relaxation in the XX quantum chain}, J. Phys. A: Math. Theor. \href{http://dx.doi.org/10.1088/1751-8113/40/8/002}{\bf 40} 1711 (2007).
  
  \bibitem{EKPP:conn} V. Eisler, D. Karevski, T. Platini, and I. Peschel, \emph{Entanglement evolution after connecting finite to infinite quantum chains}, J. Stat. Mech. (2008) \href{http://dx.doi.org/10.1088/1742-5468/2008/01/P01023}{\bf P01023}. 
  
  \bibitem{BDrev} D. Bernard and B. Doyon, \emph{Conformal field theory out of equilibrium: a review}, arXiv:\href{http://arxiv.org/abs/1603.07765}{1603.07765} (2016).
  
  
  \bibitem{VMrev} R. Vasseur and J.E. Moore, \emph{Nonequilibrium quantum dynamics and transport: from integrability to many-body localization}, arXiv:\href{http://arxiv.org/abs/1603.06618}{1603.06618} (2016).
  
  \bibitem{CD:therm08} 
M. Cramer, C.M. Dawson, J. Eisert, and T.J. Osborne, \emph{Exact Relaxation in a Class of Nonequilibrium Quantum Lattice Systems}, Phys. Rev. Lett. \href{http://dx.doi.org/10.1103/PhysRevLett.100.030602}{\bf 100}, 030602 (2008).

\bibitem{Cramer_10}
M. Cramer and J. Eisert, \emph{A quantum central limit theorem for non-equilibrium systems: exact local relaxation of correlated states}, New J. Phys. \href{http://dx.doi.org/10.1088/1367-2630/12/5/055020}{\bf 12}, 055020 (2010).

\bibitem{SK:therm14} 
J. Sirker, N.P. Konstantinidis, F. Andraschko, and N. Sedlmayr, \emph{Locality and thermalization in closed quantum systems}, Phys. Rev. A \href{http://dx.doi.org/10.1103/PhysRevA.89.042104}{\bf 89}, 042104 (2014). 
  
\bibitem{GG15}
J.R. Garrison and T. Grover,
\emph{Does a single eigenstate encode the full Hamiltonian?},
arXiv:1503.00729.

\bibitem{RS:altET} M. Rigol and M. Srednicki, 
  \emph{Alternatives to Eigenstate Thermalization}, Phys. Rev. Lett. \href{http://dx.doi.org/10.1103/PhysRevLett.100.100601}{\bf 108}, 110601 (2012).
  
  \bibitem{RDO08}
M. Rigol, V. Dunjko, and 
M. Olshanii, \emph{Thermalization and its mechanism for generic isolated quantum systems}, Nature \href{\doi10.1038/nature06838}{\bf 452}, 854 (2008).

\bibitem{negativeT}
S. Braun, J.P. Ronzheimer, M. Schreiber, S.S. Hodgman, T. Rom, I. Bloch, U. Schneider,
\emph{Negative Absolute Temperature for Motional Degrees of Freedom},
Science \href{\doi10.1126/science.1227831
}{\bf 339}, 52 (2013).

\bibitem{CLIsing} T. Prosen, \emph{A new class of completely integrable quantum spin chains}, J. Phys. A: Math. Gen. \href{http://dx.doi.org/10.1088/0305-4470/31/21/002}{\bf 31} L397 (1998); M. Grady, \emph{Infinite set of conserved charges in the Ising model}, Phys. Rev. D \href{http://dx.doi.org/10.1103/PhysRevD.25.1103}{\bf 25}, 1103 (1981). 

\bibitem{maxent} E.T. Jaynes, \emph{Information Theory and Statistical Mechanics}, Phys. Rev. \href{http://dx.doi.org/10.1103/PhysRev.106.620}{\bf 106}, 620 (1957).
  
  \bibitem{GGE}
M. Rigol, V. Dunjko, V. Yurovsky,  and M. Olshanii, \emph{Relaxation in a Completely Integrable Many-Body Quantum System: An Ab Initio Study of the Dynamics of the Highly Excited States of 1D Lattice Hard-Core Bosons},
Phys. Rev. Lett. \href{http://dx.doi.org/10.1103/PhysRevLett.98.050405}{\bf 98}, 50405 (2007).

\bibitem{Doyon} B. Doyon, \emph{Thermalization and pseudolocality in extended quantum systems},  arXiv:\href{http://arxiv.org/abs/1512.03713}{1512.03713} (2015).





\bibitem{rigol09a}
M. Rigol, \emph{Breakdown of thermalization in finite one-dimensional systems},
Phys. Rev. Lett. \href{http://dx.doi.org/10.1103/PhysRevLett.103.100403}{\bf 103}, 100403 (2009).

\bibitem{rigol09b}
M. Rigol, \emph{Quantum quenches and thermalization in one-dimensional
  fermionic systems}, Phys. Rev. A \href{\doi10.1103/PhysRevA.80.053607}{\bf 80}, 053607 (2009).

\bibitem{BKL_PRL10} 
G. Biroli, C. Kollath, and A.M. L\"auchli, \emph{Effect of Rare Fluctuations on the Thermalization of Isolated Quantum Systems}, Phys. Rev. Lett. \href{http://dx.doi.org/10.1103/PhysRevLett.105.250401}{\bf 105},
250401 (2010). 


\bibitem{RS10a}
L.F. Santos and M. Rigol, \emph{Localization and the effects of
  symmetries in the thermalization properties of one-dimensional
  quantum systems}, Phys. Rev. E \href{http://dx.doi.org/10.1103/PhysRevE.82.031130}{\bf 82}, 031130 (2010).

\bibitem{RS10b}
M. Rigol and L.F. Santos, \emph{Quantum chaos and thermalization in
   gapped systems}, Phys. Rev. A \href{\doi10.1103/PhysRevA.82.011604}{\bf 82}, 011604(R) (2010).

\bibitem{neuenhahn12}
C. Neuenhahn and F. Marquardt, \emph{Thermalization of interacting
  fermions and delocalization in Fock space}, Phys. Rev. E \href{\doi10.1103/PhysRevE.85.060101}{\bf 85}
060101 (2012).

\bibitem{steinigeweg13}
R. Steinigeweg, J. Herbrych, and P. Prelovsek, \emph{
Eigenstate thermalization within isolated spin-chain systems},
Phys. Rev. E \href{\doi10.1103/PhysRevE.87.012118}{\bf 87}, 012118 (2013).

\bibitem{beugeling1}
W. Beugeling, R. Moessner, and M. Haque, \emph{Finite-size scaling of
  eigenstate thermalization}, Phys. Rev. E \href{\doi10.1103/PhysRevE.89.042112}{\bf 89} 042112 (2014).

\bibitem{kid14}
H. Kim, T.N. Ikeda, and D.A. Huse, \emph{Testing whether all
  eigenstates obey the Eigenstate Thermalization Hypothesis},
Phys. Rev. E \href{\doi10.1103/PhysRevE.90.052105}{\bf 90}, 052105 (2014).

\bibitem{beugeling2}
W. Beugeling, R. Moessner, and M. Haque, \emph{Off-diagonal matrix
  elements of local operators in many-body quantum systems},
Phys. Rev. E \href{\doi10.1103/PhysRevE.91.012144}{\bf 91}, 012144 (2015).

\bibitem{CE_PRL13} 
J.-S.~Caux and F.H.L.~Essler, \emph{Time Evolution of Local Observables After Quenching to an Integrable Model}, Phys. Rev. Lett. \href{http://dx.doi.org/10.1103/PhysRevLett.110.257203}{\bf 110}, 257203 (2013).

\bibitem{LSM:XY} E. Lieb, T. Schultz, D. Mattis, \emph{Two soluble models of an antiferromagnetic chain}, Ann. Phys. \href{\doi10.1016/0003-4916(61)90115-4}{\bf 16}, 407 (1961). 

\bibitem{A:GGE15} V. Alba, \emph{Simulating the Generalized Gibbs Ensemble (GGE): a Hilbert space Monte Carlo approach}, arXiv:\href{http://arxiv.org/abs/1507.06994}{1507.06994} (2015).

\bibitem{spscopy} B. Fr\"ohlich, M. Feld, E. Vogt, M. Koschorreck, W. Zwerger, and M. K\"ohl \emph{Radio-Frequency Spectroscopy of a Strongly Interacting Two-Dimensional Fermi Gas}, Phys. Rev. Lett. \href{http://dx.doi.org/10.1103/PhysRevLett.106.105301}{\bf 106}, 105301 (2011).

\bibitem{bravyi06}
S. Bravyi, M. B. Hastings, and F. Verstraete, \emph{Lieb-Robinson Bounds and the Generation of Correlations and Topological Quantum Order}, Phys. Rev. Lett.
\href{http://dx.doi.org/10.1103/PhysRevLett.97.050401}{\bf 97}, 050401 (2006).

\bibitem{KPSR:f-d} E. Khatami, G. Pupillo, M. Srednicki, and M. Rigol, \emph{Fluctuation-Dissipation Theorem in an Isolated System of Quantum Dipolar Bosons after a Quench}, Phys. Rev. Lett. \href{http://dx.doi.org/10.1103/PhysRevLett.111.050403}{\bf 111}, 050403 (2013).

\bibitem{dechiara06}
G. De Chiara, S. Montangero, P. Calabrese, and R. Fazio, \emph{Entanglement entropy dynamics of Heisenberg chains}, J. Stat. Mech. (2006) \href{http://dx.doi.org/10.1088/1742-5468/2006/03/P03001}{P03001}.

\bibitem{laeuchli08}
A. L\"auchli and C. Kollath, \emph{Spreading of correlations and entanglement after a quench in the one-dimensional Bose-Hubbard model}, J. Stat. Mech. (2008) \href{http://dx.doi.org/10.1088/1742-5468/2008/05/P05018}{P05018}.

\bibitem{manmana09}
S. R. Manmana, S. Wessel, R. M. Noack, and A. Muramatsu, \emph{Time evolution of correlations in strongly interacting fermions after a quantum quench}, 
Phys. Rev. B \href{http://dx.doi.org/10.1103/PhysRevB.79.155104}{\bf 79}, 155104 (2009).

\bibitem{carleo14}
G. Carleo, F. Becca, L. Sanchez-Palencia, S. Sorella, and
M. Fabrizio, \emph{Light-cone effect and supersonic correlations in one- and two-dimensional bosonic superfluids}. Phys. Rev. A \href{http://dx.doi.org/10.1103/PhysRevA.89.031602}{\bf 89}, 031602 (2014).

\bibitem{hauke13}
P. Hauke and L. Tagliacozzo, \emph{Spread of Correlations in Long-Range Interacting Quantum Systems}, Phys. Rev. Lett. \href{http://dx.doi.org/10.1103/PhysRevLett.111.207202}{\bf 111}, 207202
(2013).

\bibitem{schachenmayer13}
J. Schachenmayer, B. P. Lanyon, C. F. Roos, and A. J.
Daley, \emph{Entanglement Growth in Quench Dynamics with Variable Range Interactions}, Phys. Rev. X \href{http://dx.doi.org/10.1103/PhysRevX.3.031015}{\bf 3}, 031015 (2013).

\bibitem{eisert13}
J. Eisert, M. van den Worm, S. R. Manmana, and M. Kastner, \emph{Breakdown of Quasilocality in Long-Range Quantum Lattice Models}, 
Phys. Rev. Lett. \href{http://dx.doi.org/10.1103/PhysRevLett.111.260401}{\bf 111}, 260401 (2013).

\bibitem{vodola15}
D. Vodola, L. Lepori, E. Ercolessi, and G. Pupillo, \emph{Long-range Ising and Kitaev models: phases, correlations and edge modes},  2016 New J. Phys. \href{http://dx.doi.org/10.1088/1367-2630/18/1/015001}{\bf 18} 015001. 

\bibitem{regemortel15}
M. V. Regemortel, D. Sels, and M. Wouters, \emph{Information propagation and equilibration in long-range Kitaev chains}, 
arXiv:\href{http://arxiv.org/abs/1511.05459}{1511.05459} (2015).

\bibitem{buyskikh16}
A.S. Buyskikh, M. Fagotti, J. Schachenmayer, F.H.L. Essler,
A. J. Daley, \emph{Entanglement growth and correlation spreading with
  variable-range interactions in spin and fermionic tunnelling
  models}, Phys. Rev. A 93, 053620 (2016) [http://dx.doi.org/10.1103/PhysRevA.93.053620].

\bibitem{localquenches}
D.B. Abraham, E. Barouch, G. Gallavotti and A. Martin-L\"of,
\emph{Thermalization of a Magnetic Impurity in the Isotropic XY
  Model}, Phys. Rev. Lett. \href{http://dx.doi.org/10.1103/PhysRevLett.25.1449}{\bf25}, 1449 (1970); 
D.B. Abraham, E. Barouch, G. Gallavotti and A. Martin-L\"of,
\emph{Dynamics of a Local Perturbation in the XY Model. I-Approach to 
  Equilibrium}, Stud. Appl. Math. \href{\doi10.1002/sapm1971502121}{\bf 50}, 121 (1971);
C. Kollath, U. Schollw\"ock and W. Zwerger,
\emph{Spin-charge separation in cold Fermi-gases: a real time analysis}
Phys. Rev. Lett. \href{http://dx.doi.org/10.1103/PhysRevLett.95.176401}{\bf 95}, 176401 (2005);
P. Calabrese and J. Cardy, \emph{Entanglement and correlation
  functions following a local quench: a conformal field theory
  approach}, J. Stat. Mech. (2007) \href{\doi10.1088/1742-5468/2007/10/P10004}{P10004};
J.-M. St\'ephan and J\'er\^ome Dubail, \emph{Local quantum quenches in
  critical one-dimensional systems: entanglement, the Loschmidt echo,
  and light-cone effects}, J. Stat. Mech. (2011) \href{\doi	10.1088/1742-5468/2011/08/P08019}{P08019};
M. Ganahl, E. Rabel, F.H.L. Essler and H.-G. Evertz, \emph{Observation of Complex Bound States in the Spin-1/2 Heisenberg XXZ Chain Using Local Quantum Quenches}, 
Phys. Rev. Lett. \href{http://dx.doi.org/10.1103/PhysRevLett.108.077206}{\bf 108}, 077206 (2012).

\bibitem{LR72}
E. H. Lieb and D. W. Robinson, \emph{The finite group velocity of quantum spin systems}, Commun. Math. Phys. \href{http://dx.doi.org/10.1007/BF01645779}{\bf 28}, 251
(1972).

\bibitem{SN10}
R. Sims and B. Nachtergaele, \emph{Lieb-Robinson bounds in quantum
many-body physics}, edited by R. Sims and D. Ueltschi, Entropy
and the Quantum, Vol. 529 (American Mathematical Society, 2010).

\bibitem{juneman13}
J. J\"unemann, A. Cadarso, D. Perez-Garcia, A. Bermudez, and
J. J. Garc\'ia-Ripoll, \emph{Lieb-Robinson Bounds for Spin-Boson Lattice Models and Trapped Ions}, Phys. Rev. Lett. \href{http://dx.doi.org/10.1103/PhysRevLett.111.230404}{\bf 111}, 230404 (2013).


\bibitem{kliesch13}
M. Kliesch, C. Gogolin, and J. Eisert, \emph{Lieb-Robinson bounds
and the simulation of time evolution of local observables in lattice
systems}, edited by L. D. Site and Bach, Many-Electron Approaches
in Physics, Chemistry and Mathematics: A Multidisciplinary
View (Springer, 2013).

\bibitem{poulin10}
D. Poulin, \emph{Lieb-Robinson Bound and Locality for General Markovian Quantum Dynamics}, Phys. Rev. Lett. \href{http://dx.doi.org/10.1103/PhysRevLett.104.190401}{\bf 104}, 190401 (2010).

\bibitem{Bocc57}
P. Bocchieri and A. Loinger, \emph{Quantum Recurrence Theorem}, Phys. Rev. \href{http://dx.doi.org/10.1103/PhysRev.107.337}{\bf 107}, 337 (1957).

\bibitem{sachdevbook}
S. Sachdev, \emph{Quantum Phase Transitions}, Cambridge University Press,
2001.

\bibitem{VD:QED13} 
O. Viehmann, J. von Delft, and F. Marquardt, \emph{Observing the Nonequilibrium Dynamics of the Quantum Transverse-Field Ising Chain in Circuit QED}, Phys. Rev. Lett. \href{http://dx.doi.org/10.1103/PhysRevLett.110.030601}{\bf 110}, 030601 (2013); O. Viehmann, J. von Delft, and F. Marquardt, \emph{
The quantum transverse-field Ising chain in circuit quantum electrodynamics: effects of disorder on the nonequilibrium dynamics}, New J. Phys. \href{http://dx.doi.org/10.1088/1367-2630/15/3/035013}{\bf 15} 035013 (2013).


\bibitem{LSM}
E. Lieb, T. Schultz and D. Mattis, \emph{Two soluble models of an
  antiferromagnetic chain}, \href{http://dx.doi.org/j.aop.10.1016}
{\bf 16}, 407 (1961).

\bibitem{Isingusual}
B.M. McCoy, E. Barouch and D.B. Abraham, \emph{Statistical Mechanics
  of the XY Model. IV. Time-Dependent Spin-Correlation Functions},
\href{http://dx.doi.org/10.1103/PhysRevA.4.2331}{\bf 4}, 2331 (1971);
O. Derzhko and T. Krokhmalskii, \emph{Dynamic structure factor of the
  spin-1/2 transverse Ising chain},
\href{http://dx.doi.org/10.1103/PhysRevB.56.11659}{\bf 56}, 11659 (1997);
O. Derzhko and T. Krokhmalskii, \emph{Numerical approach for the study
of the spin- 1/2 xy chains dynamic properties}, Physica \href{\doi10.1002/(SICI)1521-3951(199807)208:1<221::AID-PSSB221>3.0.CO;2-E}{\bf B 208},
221 (1999).

\bibitem{Isingformfactors}
A. Bugrij, \emph{Correlation function of the two-dimensional Ising
  model on the finite lattice. I}, Theor. Math. Phys. \href{\doi10.1023/A:1010320126700}{\bf 127}, 528
(2001); 
A. Bugrij and O. Lisovyy, \emph{Spin matrix elements in 2D Ising model
  on the finite lattice}, Phys. Lett. \href{\doi10.1016/j.physleta.2003.10.039}{\bf A 319}, 390 (2003);
G. von Gehlen, N. Iorgov, S. Pakuliak, V. Shadura and Y. Tykhyy,
\emph{Form-factors in the Baxter-Bazhanov-Stroganov model II: Ising
  model on the finite lattice}, J. Phys. \href{\doi10.1088/1751-8113/41/9/095003}{\bf A 41}, 095003 (2008);
N. Iorgov, V. Shadura and Yu. Tykhyy, \emph{Spin operator matrix elements in
the quantum Ising chain: fermion approach}, J. Stat. Mech. (2011) \href{\doi10.1088/1742-5468/2011/02/P02028}{P02028}.

\bibitem{peschel03}
I. Peschel, \emph{Calculation of reduced density matrices from
  correlation functions}, J. Phys. A \href{http://dx.doi.org/10.1088/0305-4470/36/14/101}{\bf 36}, L205  (2003).

\bibitem{strongweakt} 
M. C. Ba\~nuls, J. I. Cirac, and M. B. Hastings, \emph{Strong and Weak Thermalization of Infinite Nonintegrable Quantum Systems}, Phys. Rev. Lett. \href{http://dx.doi.org/10.1103/PhysRevLett.106.050405}{\bf 106}, 050405 (2011). 

\bibitem{ISL:entdis12} F. Igl\'oi, Z. Szatm\'ari, and Y.-C. Lin, \emph{Entanglement entropy dynamics of disordered quantum spin chains}, Phys. Rev. B \href{http://dx.doi.org/10.1103/PhysRevB.85.094417}{\bf 85}, 094417 (2012). 

\bibitem{TT:spectrum14} G. Torlai, L. Tagliacozzo, and G. De Chiara, \emph{Dynamics of the entanglement spectrum in spin chains}, J. Stat. Mech. (2014) \href{http://dx.doi.org/10.1088/1742-5468/2014/06/P06001}{P06001}.

\bibitem{CTC:neg14} A. Coser, E. Tonni and P. Calabrese, \emph{Entanglement negativity after a global quantum quench}, J. Stat. Mech. (2014) \href{http://dx.doi.org/10.1088/1742-5468/2014/12/P12017}{P12017}.

\bibitem{NR:harm14} M.G. Nezhadhaghighi and M.A. Rajabpour, \emph{Entanglement dynamics in short- and long-range harmonic oscillators}, Phys. Rev. B \href{http://dx.doi.org/10.1103/PhysRevB.90.205438}{\bf 90}, 205438 (2014).

\bibitem{tDMRG}
S. R. White and A. E. Feiguin, \emph{Real-Time Evolution Using the Density Matrix Renormalization Group}, Phys. Rev. Lett. \href{http://dx.doi.org/10.1103/PhysRevLett.93.076401}{\bf 93}, 076401 (2004); 
A. J. Daley, C. Kollath, U. Schollw\"ock, and G. Vidal, \emph{Time-dependent density-matrix renormalization-group using adaptive effective Hilbert spaces}, J. Stat. Mech. (2004) \href{http://dx.doi.org/10.1088/1742-5468/2004/04/P04005}{P04005}.

\bibitem{iTEBD}
G. Vidal, \emph{Classical Simulation of Infinite-Size Quantum Lattice Systems in One Spatial Dimension}, Phys. Rev. Lett. \href{http://dx.doi.org/10.1103/PhysRevLett.98.070201}{\bf 98}, 070201 (2007).

\bibitem{G:14} V. Gurarie, \emph{Global large time dynamics and the generalized Gibbs ensemble}, J. Stat. Mech. (2013) \href{http://dx.doi.org/10.1088/1742-5468/2013/02/P02014}{P02014}.

\bibitem{book}
F.~H.~L.~Essler, H.~Frahm, F.~G{\"o}hmann, A.~Kl{\"u}mper, and
V.~E.~Korepin, {\it The One-Dimensional Hubbard Model}, Cambridge
University Press, Cambridge (2005).

\bibitem{Gaudin}
M. Gaudin, \emph{La fonction d'onde de Bethe}, Masson, Paris 1983,
English translation by J.-S. Caux, Cambridge University Press,
Cambridge 2014.

















\bibitem{SFM:Z-F12} S. Sotiriadis, D. Fioretto, and G. Mussardo,
  \emph{Zamolodchikov-Faddeev algebra and quantum quenches in
    integrable field theories}, J. Stat. Mech. (2012)
  \href{http://dx.doi.org/10.1088/1742-5468/2012/02/P02017}{P02017}.

\bibitem{ini_state} D.X. Horv\'ath, b, , S. Sotiriadis, and G. Tak\'acs,  \emph{Initial states in integrable quantum field theory quenches from an integral equation hierarchy}, Nucl. Phys. B \href{\doi10.1016/j.nuclphysb.2015.11.025}{\bf 902}, 508 (2016).

\bibitem{Crev} J.-S. Caux, \emph{The Quench Action}, arXiv:\href{http://arxiv.org/abs/1603.04689}{1603.04689} (2016).

\bibitem{Pozs_qB}  
B. Pozsgay, \emph{Quantum quenches and generalized Gibbs ensemble in a Bethe Ansatz solvable lattice model of interacting bosons}, J. Stat. Mech. (2014) \href{http://dx.doi.org/10.1088/1742-5468/2014/10/P10045}{P10045}. 

\bibitem{GA:fail14} 
G. Goldstein and N. Andrei, \emph{Failure of the GGE hypothesis for integrable models with bound states}, arXiv:\href{http://arxiv.org/abs/1405.4224}{1405.4224} (2014).

\bibitem{Zam}
A.B. Zamolodchikov and Al.B. Zamolodchikov, \emph{Factorized
  S-matrices in Two Dimensions as the Exact Solutions of Certain
  Relativistic Quantum Field Theory Models}, Ann. Phys. \href{\doi10.1016/0003-4916(79)90391-9}{\bf 120},
253 (1979). 

\bibitem{Fadd}
L.~D. Faddeev, \emph{Quantum completely integral models of field
  theory}, Sov. Sci. Rev. Math. Phys.~C {\bf 1},  107  (1980).

 


\bibitem{EK:finiteT}
F.H.L. Essler and R.M. Konik, \emph{Finite Temperature Dynamical
  Correlations in Massive Integrable Quantum Field Theories},
J. Stat. Mech. (2009) \href{http://dx.doi.org/10.1088/1742-5468/2009/09/P09018}{P09018}.


\bibitem{6vertex}
E.H. Lieb, \emph{Exact Solution of the Problem of the Entropy of Two-Dimensional Ice}, \emph{Exact Solution of the Problem of the Entropy of Two-Dimensional Ice}, Phys. Rev. Lett. \href{http://dx.doi.org/10.1103/PhysRevLett.18.692}{\bf 18}, 692 (1967);
E.H. Lieb, \emph{Exact Solution of the Two-Dimensional Slater KDP Model of a Ferroelectric}, Phys. Rev. Lett. 19, \href{http://dx.doi.org/10.1103/PhysRevLett.19.108}{\bf 108} (1967);
B. Sutherland, \emph{Exact Solution of a Two-Dimensional Model for Hydrogen-Bonded Crystals}, Phys. Rev. Lett. \href{http://dx.doi.org/10.1103/PhysRevLett.19.103}{\bf 19}, 103 (1967).

\bibitem{FT84}
L.D. Faddeev and L. Takhtadzhyan, \emph{Spectrum and scattering of excitations in the one-dimensional isotropic Heisenberg model}, J. Sov. Math. \href{http://dx.doi.org/10.1007/BF01087245}{\bf 24}, 241 (1984).


\bibitem{Takahashibook}
M. Takahashi and M. Suzuki, \emph{One-Dimensional Anisotropic
  Heisenberg Model at Finite Temperatures},
Prog. Theor. Phys. \href{\doi10.1143/PTP.48.2187}{\bf 48}, 2187
(1972);
M. Takahashi, \emph{Thermodynamics of one dimensional solvable
  models}, Cambridge University Press, Cambridge 1999.


\bibitem{higherCL}
A.M. Tsvelik, \emph{Incommensurate phases of quantum one-dimensional magnetics}, Phys. Rev. B \href{http://dx.doi.org/10.1103/PhysRevB.42.779}{\bf 42}, 779 (1990);
H. Frahm, \emph{Integrable spin-1/2 XXZ Heisenberg chain with competing interactions}, J. Phys. A: Math. Gen. \href{http://dx.doi.org/10.1088/0305-4470/25/6/005}{\bf 25}, 1417 (1992);
A.A. Zvyagin and A. Kl\"umper, \emph{Quantum phase transitions and thermodynamics of quantum antiferromagnets with next-nearest-neighbor couplings}, Phys. Rev. B \href{http://dx.doi.org/10.1103/PhysRevB.68.144426}{\bf 68}, 144426 (2003).


\bibitem{takahashi}
N. Muramoto and M. Takahashi, \emph{Integrable Magnetic Model of Two Chains Coupled by Four-Body Interactions}, J. Phys. Soc. Jpn \href{http://dx.doi.org/10.1143/JPSJ.68.2098}{\bf 68}, 2098 (1999). 

\bibitem{GM:chain} M.P. Grabowski, P. Mathieu, \emph{Structure of the Conservation Laws in Quantum Integrable Spin Chains with Short Range Interactions}, Ann. Phys. \href{\doi10.1006/aphy.1995.1101}{\bf 243}, 299 (1995).

\bibitem{KS:2002} A.~Kl\"umper and K.~Sakai, \emph{The thermal conductivity of the spin-1/2 XXZ chain at arbitrary temperature}, J. Phys. A \href{http://dx.doi.org/10.1088/0305-4470/35/9/307}{\bf35}, 2173 (2002);
K.~Sakai and A.~Kl\"umper, \emph{Non-dissipative thermal transport in the massive regimes of the XXZ chain}, J. Phys. A:  Math. Gen. \href{http://dx.doi.org/10.1088/0305-4470/36/46/006}{\bf 36} 11617 (2003). 

\bibitem{QTM}
A. Kl\"umper, \emph{Thermodynamics of the anisotropic spin-1/2 Heisenberg chain and related quantum chains}, Z. Phys. B \href{\doi10.1007/BF01316831}{\bf 91}, 507 (1993);
C. Destri and H.J. de Vega, \emph{Unified approach to Thermodynamic Bethe Ansatz and finite size corrections for lattice models and field theories}, Nucl. Phys. B \href{\doi10.1016/0550-3213(94)00547-R}{\bf 438}, 314 (1995).


\bibitem{wuppertal} 
M.~Bortz and F.~G\"ohmann, \emph{Exact thermodynamic limit of short-range correlation functions of the antiferromagnetic XXZ-chain at finite temperatures}, Eur. Phys. J. B \href{http://dx.doi.org/10.1140/epjb/e2005-00272-6}{\bf 46}, 399 (2005);
H.~E.~Boos, F.~G\"ohmann, A.~Kl\"umper, and J.~Suzuki, \emph{Factorization of the finite temperature correlation functions of the XXZ chain in a magnetic field}, J. Phys. A: Math. Theor. \href{http://dx.doi.org/10.1088/1751-8113/40/35/001}{\bf 40}, 10699 (2007);
H.E. Boos, J. Damerau, F. G\"ohmann, A. Kl\"umper, J. Suzuki, and
A. Wei{\ss}e, \emph{Short-distance thermal correlations in the XXZ chain}, J. Stat. Mech. (2008) \href{http://dx.doi.org/10.1088/1742-5468/2008/08/P08010}{P08010};
C.~Trippe, F.~G\"ohmann, and A.~Kl\"umper, \emph{Short-distance thermal correlations in the massive XXZ chain}, 
Eur. Phys. J. B \href{http://dx.doi.org/10.1140/epjb/e2009-00417-7}{\bf 73}, 253 (2010).


\bibitem{GKS:2004} F.~G\"ohmann, A.~Kl\"umper, and A.~Seel, \emph{Integral representation of the density matrix of the XXZ chain at finite temperatures}, J. Phys. A: Math Gen. \href{http://dx.doi.org/10.1088/0305-4470/38/9/001}{\bf 37}, 7625 (2004). 

\bibitem{Smirnov}
H.E.~Boos, M.~Jimbo,~T. Miwa, F.~Smirnov, and Y.~Takeyama, \emph{Algebraic Representation of Correlation Functions in Integrable Spin Chains}, Annales
Henri Poincare \href{\doi10.1007/s00023-006-0285-5}{\bf 7}, 1395 (2006);
H.E.~Boos, M.~Jimbo,~T. Miwa, F.~Smirnov, and Y.~Takeyama, \emph{Hidden Grassmann Structure in the XXZ Model}, 
Commun. Math. Phys \href{\doi10.1007/s00220-007-0202-x}{\bf 272}, 263 (2007); 
H.E.~Boos, M.~Jimbo,~T. Miwa, F.~Smirnov, and Y.~Takeyama,\emph{Hidden Grassmann Structure in the XXZ Model II: Creation Operators}, 
Commun. Math. Phys \href{\doi10.1007/s00220-008-0617-z}{\bf 286}, 875 (2009); 
M.~Jimbo,~T. Miwa, and F.~Smirnov, \emph{Hidden Grassmann structure in the XXZ model III: introducing the Matsubara direction}, J. Phys. A: Math. Theor. \href{http://dx.doi.org/10.1088/1751-8113/42/30/304018}{\bf 42},
304018. 

\bibitem{quasilocal_XXX}
E. Ilievski, M. Medenjak, and T. Prosen, \emph{Quasilocal Conserved Operators in the Isotropic Heisenberg Spin-1/2 Chain}, Phys. Rev. Lett. \href{http://dx.doi.org/10.1103/PhysRevLett.115.120601}{\bf 115}, 120601 (2015).

\bibitem{Prosrev} E. Ilievski, M. Medenjak, T. Prosen, and L. Zadnik, arXiv:\href{http://arxiv.org/abs/1603.00440}{1603.00440} (2016).

\bibitem{qlXXZ} T. Prosen, \emph{Open XXZ Spin Chain: Nonequilibrium Steady State and a Strict Bound on Ballistic Transport}, Phys. Rev. Lett. \href{http://dx.doi.org/10.1103/PhysRevLett.106.217206}{\bf 106}, 217206 (2011);
T. Prosen and E. Ilievski, \emph{Families of Quasilocal Conservation
  Laws and Quantum Spin Transport},
Phys. Rev. Lett. \href{http://dx.doi.org/10.1103/PhysRevLett.111.057203}{\bf
  111}, 057203 (2013); T. Prosen, \emph{Quasilocal conservation laws
  in XXZ spin-1/2 chains: Open, periodic and twisted boundary
  conditions}, Nucl. Phys. B
\href{\doi10.1016/j.nuclphysb.2014.07.024}{\bf 886}, 1177 (2014);
R.G. Pereira, V. Pasquier, J. Sirker, and I. Affleck, \emph{Exactly
  conserved quasilocal operators for the XXZ spin chain},
J. Stat. Mech. (2014);
\href{http://dx.doi.org/10.1088/1742-5468/2014/09/P09037}{P09037};
L. Piroli and E. Vernier, \emph{Quasi-local conserved charges and spin transport in spin-$1$ integrable chains}, arXiv:\href{http://arxiv.org/abs/1601.07289}{1601.07289} (2016).

\bibitem{qlXXZ2} J. Sirker, R.G. Pereira, I. Affleck
\emph{Conservation laws, integrability and transport in one-dimensional quantum systems}
Phys. Rev. B \href{\doi10.1103/PhysRevB.83.035115}{\bf 83}, 035115 (2011).

\bibitem{spinS}
L.A. Takhtajan, \emph{The picture of low-lying excitations in the
  isotropic Heisenberg chain of arbitrary spins}, Phys. Lett. A \href{\doi10.1016/0375-9601(82)90764-2}{\bf
  87}, 479 (1982);
H.M. Babujian, \emph{Exact solution of the isotropic Heisenberg chain
  with arbitrary spins: Thermodynamics of the model}, Nucl. Phys. \href{\doi10.1016/0550-3213(83)90668-5}{\bf
  B215}, 317 (1983);
A.N. Kirillov and N.Yu. Reshetikhin, \emph{Exact solution of the
  integrable XXZ Heisenberg models with arbitrary spin. I.
The ground state and the excitation spectrum}
J. Phys. \href{\doi10.1088/0305-4470/20/6/038}{\bf A20}, 1565 (1987); 
A.N Kirillov and N.Yu. Reshetikhin
\emph{Exact solution of the integrable XXZ Heisenberg model with
  arbitrary spin. II. Thermodynamics of the system}
J. Phys. \href{http://dx.doi.org/10.1088/0305-4470/20/6/039}{\bf A20}, 1587 (1987).

\bibitem{IQNB}
E. Ilievski, E. Quinn, J. De Nardis and M. Brockmann,
\emph{String-charge duality in integrable lattice models},
arXiv:\href{http://arxiv.org/abs/1512.04454}{1512.04454} (2015).



\bibitem{slNM}
B. Sutherland, \emph{Model for a multicomponent quantum system},
Phys. Rev. B \href{http://dx.doi.org/10.1103/PhysRevB.12.3795}{\bf 12}, 3795 (1975); 
P.P. Kulish, E.K. Sklyanin, \emph{Solutions of the Yang-Baxter equation}, J. Sov. Math. \href{\doi10.1007/BF01091463}{\bf 19}, 1596 (1982);
N. Andrei and H. Johannesson, \emph{Higher dimensional representations
  of the SU(N) Heisenberg model}, Phys. Lett. A\href{\doi10.1016/0375-9601(84)90819-3}{\bf 104}, 370 (1984);
P.P.Kulish, \emph{Integrable graded magnets}, J. Soviet Math. \href{\doi10.1007/BF01083770}{\bf
  35}, 2648 (1985); 
F.H.L. Essler and V.E. Korepin, \emph{Algebraic Bethe Ansatz and
  Higher Conservation Laws for the Supersymmetric t-J Model},
Phys. Rev. B \href{http://dx.doi.org/10.1103/PhysRevB.46.9147}{\bf 46}, 9147 (1992);
A. Foerster and M. Karowski, \emph{Algebraic properties of the Bethe
  ansatz for an $spl(2,1)$ supersymmetric t-J model},
Nucl.Phys. B \href{\doi10.1016/0550-3213(93)90665-C}{\bf 396}, 611 (1993);
F.H.L. Essler, V.E. Korepin and K. Schoutens, \emph{Exact Solution of
  an Electronic Model of Superconductivity I}, Int. J. Mod. Phys. B \href{\doi10.1142/S0217979294001354}{\bf
  8}, 3205 (1994);
F. G\"ohmann, \emph{Algebraic Bethe ansatz for the gl(1|2) generalized
  model and Lieb-Wu equations}, Nucl. Phys. B \href{\doi10.1016/S0550-3213(01)00497-7}{\bf 620}, 5-1, (2002).
S. Belliard and E. Ragoucy, \emph{Nested Bethe ansatz for 'all' closed
  spin chains}, J. Phys. \href{\doi10.1088/1751-8113/41/29/295202}{\bf A41}, 295202 (2008).


\bibitem{bosonXX} 
B. Pozsgay and V. Eisler, \emph{Real-time dynamics in a strongly interacting bosonic hopping model: Global quenches and mapping to the XX chain}, arXiv:\href{http://arxiv.org/abs/1602.03065}{1602.03065} (2016).

\bibitem{MS:over16} 
P.P. Mazza, J.-M. St\'ephan, E. Canovi, V. Alba, M. Brockmann, and M. Haque, \emph{Overlap distributions for quantum quenches in the anisotropic Heisenberg chain}, J. Stat. Mech. (2016) \href{http://dx.doi.org/10.1088/1742-5468/2016/01/013104}{013104}. 



\bibitem{KL:BH07} 
C. Kollath, A. M. L\"auchli, and E. Altman, \emph{Quench Dynamics and Nonequilibrium Phase Diagram of the Bose-Hubbard Model}, Phys. Rev. Lett. \href{http://dx.doi.org/10.1103/PhysRevLett.98.180601}{\bf 98}, 180601 (2007). 

\bibitem{MK:H08} M. Moeckel and S. Kehrein, \emph{Interaction Quench in the Hubbard Model}, Phys. Rev. Lett. \href{http://dx.doi.org/10.1103/PhysRevLett.100.175702}{\bf 100}, 175702 (2008).

\bibitem{MK:preT09} M. Moeckel and S. Kehrein, \emph{Real-time evolution for weak interaction quenches in quantum systems}, Ann. of Phys. \href{\doi10.1016/j.aop.2009.03.009}{\bf 324}, 2146 (2009).

\bibitem{SC:intft10} 
S. Sotiriadis and J. Cardy, \emph{Quantum quench in interacting field theory: A self-consistent approximation}, Phys. Rev. B \href{http://dx.doi.org/10.1103/PhysRevB.81.134305}{\bf 81}, 134305 (2010). 

\bibitem{KWE:preT11} M. Kollar, F.A. Wolf, and M. Eckstein, \emph{Generalized Gibbs ensemble prediction of prethermalization plateaus and their relation to nonthermal steady states in integrable systems}, Phys. Rev. B \href{http://dx.doi.org/10.1103/PhysRevB.84.054304}{\bf 84}, 054304 (2011).

\bibitem{MG:m-c-i11} 
A. Mitra and T. Giamarchi, \emph{Mode-Coupling-Induced Dissipative and Thermal Effects at Long Times after a Quantum Quench}, Phys. Rev. Lett. \href{http://dx.doi.org/10.1103/PhysRevLett.107.150602}{\bf 107}, 150602 (2011).

\bibitem{CB:glass12} 
G. Carleo, F. Becca, M. Schir\'o, and
  M. Fabrizio, \emph{Localization and Glassy Dynamics Of Many-Body
    Quantum Systems}, Scientific Reports
  \href{\doi10.1038/srep00243}{\bf 2} (2012). 

\bibitem{MS:noisy12} 
J. Marino and A. Silva, \emph{Relaxation, prethermalization, and diffusion in a noisy quantum Ising chain}, Phys. Rev. B \href{http://dx.doi.org/10.1103/PhysRevB.86.060408}{\bf 86}, 060408(R) (2012). 

\bibitem{SK:13}
M. Stark and M. Kollar, \emph{Kinetic description of thermalization
  dynamics in weakly interacting quantum systems}, arXiv:\href{http://arxiv.org/abs/1308.1610}{1308.1610} (2013).

\bibitem{MM:preT13} 
M. Marcuzzi, J. Marino, A. Gambassi, and A. Silva, \emph{Prethermalization in a Nonintegrable Quantum Spin Chain after a Quench}, Phys. Rev. Lett. \href{http://dx.doi.org/10.1103/PhysRevLett.111.197203}{\bf 111}, 197203 (2013).

\bibitem{mitra13}
A. Mitra, \emph{Correlation functions in the prethermalized regime
  after a quantum quench of a spin-chain}, Phys. Rev. B \href{\doi10.1103/PhysRevB.87.205109}{\bf 87}, 205109 (2013)

\bibitem{preT0} 
F.H.L. Essler, S. Kehrein, S.R. Manmana, and N.J. Robinson, \emph{Quench Dynamics in a Model with Tuneable Integrability Breaking}, Phys. Rev. B \href{http://dx.doi.org/10.1103/PhysRevB.89.165104}{\bf 89}, 165104 (2014).


\bibitem{gas14}
B. Nowak, J. Schole, T. Gasenzer, \emph{Universal dynamics on the way
  to thermalisation}, New J. Phys. \href{\doi10.1088/1367-2630/16/9/093052}{\bf 16}, 093052 (2014).


\bibitem{Delfino} G. Delfino, 
\emph{Quantum quenches with integrable pre-quench dynamics},  J. Phys. A: Math. Theor. \href{http://dx.doi.org/10.1088/1751-8113/47/40/402001}{\bf 47} 402001 (2014).


\bibitem{BF:preT} 
B. Bertini and M. Fagotti, \emph{Pre-relaxation in weakly interacting models}, J. Stat. Mech. (2015) \href{http://dx.doi.org/10.1088/1742-5468/2015/07/P07012}{P07012}.

\bibitem{preT_dGGE} 
B. Bertini, F.H.L. Essler, S. Groha, N.J. Robinson, \emph{Prethermalization and thermalization in models with weak integrability breaking}, Phys. Rev. Lett. \href{http://dx.doi.org/10.1103/PhysRevLett.115.180601}{\bf 115}, 180601 (2015).

\bibitem{KAM} G.P. Brandino, J.-S. Caux, and R.M. Konik, \emph{Glimmers of a Quantum KAM Theorem: Insights from Quantum Quenches in One-Dimensional Bose Gases}, Phys. Rev. X \href{http://dx.doi.org/10.1103/PhysRevX.5.041043}{\bf 5}, 041043 (2015). 

\bibitem{BD:spir15} M. Babadi, E. Demler, and M. Knap, \emph{Far-from-Equilibrium Field Theory of Many-Body Quantum Spin Systems: Prethermalization and Relaxation of Spin Spiral States in Three Dimensions}, Phys. Rev. X \href{http://dx.doi.org/10.1103/PhysRevX.5.041005}{\bf 5}, 041005 (2015).

\bibitem{O:geo15} 
M. Olshanii, \emph{Geometry of Quantum Observables and Thermodynamics of Small Systems}, Phys. Rev. Lett. \href{http://dx.doi.org/10.1103/PhysRevLett.114.060401}{\bf 114}, 060401 (2015). 


\bibitem{Fdefect} 
M. Fagotti, \emph{Control of global properties in a closed many-body quantum system by means of a local switch}, arXiv:\href{http://arxiv.org/abs/1508.04401}{1508.04401} (2015).

\bibitem{Nessi15}
N. Nessi and A. Iucci, \emph{Glass-like Behavior in a System of One
Dimensional Fermions after a Quantum Quench}, arXiv:\href{http://arxiv.org/abs/1503.02507}{1503.02507} (2015).

\bibitem{KI:preT11} T. Kitagawa, A. Imambekov, J. Schmiedmayer and E. Demler, \emph{The dynamics and prethermalization of one-dimensional quantum systems probed through the full distributions of quantum noise}, New J. Phys. \href{http://dx.doi.org/10.1088/1367-2630/13/7/073018}{\bf 13} 073018 (2011).

\bibitem{Langen15}
T. Langen, S. Erne, R. Geiger, B. Rauer, T. Schweigler, M. Kuhnert,
W. Rohringer, I.E. Mazets, T. Gasenzer and J. Schmiedmayer,
\emph{Experimental Observation of a Generalized Gibbs Ensemble},
Science \href{\doi10.1126/science.1257026}{\bf  348}, 207 (2015).




\bibitem{GGGE}
G. Goldstein and N. Andrei, \emph{Equilibration and Generalized GGE in
  the Lieb Liniger gas}, arXiv:\href{http://arxiv.org/abs/1309.3471}{1309.3471} (2013).


 
\bibitem{loc_temp} 
M. Kliesch, C. Gogolin, M.J. Kastoryano, A. Riera, and J. Eisert,
\emph{Locality of Temperature}, Phys. Rev. X
\href{http://dx.doi.org/10.1103/PhysRevX.4.031019}{\bf 4}, 031019
(2014). 

\bibitem{TsvelickWiegmann}
A.M. Tsvelick and P.B. Wiegmann,
\emph{Exact results in the theory of magnetic alloys}, Adv. Phys. \href{\doi10.1080/00018738300101581}{\bf
  32}, 453 (1983).

{
\color{red}





%





































 
 }
  

\end{thebibliography}
\end{document}